\RequirePackage{fix-cm}
\documentclass{svjour3}                     % onecolumn (standard format)
\smartqed  % flush right qed marks, e.g. at end of proof
\usepackage{graphicx}
\usepackage{psfrag}
\usepackage{pstool}
\usepackage{amsfonts}
\usepackage{amsmath}
\usepackage{amssymb}
\journalname{Reviews of Modern Plasma Physics}

\usepackage[authoryear]{natbib}

\newcommand\be{\begin{equation}}
\newcommand\ee{\end{equation}}
\newcommand\bea{\begin{eqnarray}}
\newcommand\eea{\end{eqnarray}}
\newcommand\nn{\nonumber}
\newcommand{\ms}{\noalign{\vspace{3pt plus2pt minus1pt}}}

\newfont{\myfont}{cmmib10}

\newcommand{\bbeta}{\hbox{\myfont \symbol{12} }}

\newcommand{\bkappa}{\hbox{\myfont \symbol{20} }}

\newcommand{\bomega}{\hbox{\myfont \symbol{33} }}

\newcommand{\bi}{\bf}
\newcommand{\half}{{\textstyle{1\over2}}}

\newcommand\lapprox{\mathrel{\hbox{\rlap{\hbox{\lower4pt\hbox{$\sim$}}}\hbox{$<$}}}}
\newcommand\gapprox{\mathrel{\hbox{\rlap{\hbox{\lower3pt\hbox{$\sim$}}}\hbox{$>$}}}}

\newcommand{\araa}{{Ann. Rev. Astron. Astrophys.} }
\newcommand{\solphys}{{Solar Phys.} }
\newcommand{\sovast}{{Sov. Astron.} }
\newcommand{\pr}{{Phys. Rev.} }
\newcommand{\prl}{{Phys. Rev. Lett.} }
\newcommand{\grl}{{Geophys. Res. Lett.} }
\newcommand{\pop}{{Phys. Plasmas} }
\newcommand{\ajp}{{Aust. J. Phys.} }
\newcommand{\apj}{{Astrophys. J.} }
\newcommand{\apjl}{{Astrophys. J. Lett.} }
\newcommand{\apjs}{{Astrophys. J. Suppl.} }
\newcommand{\jgr}{{J. Geophys. Res.} }
\newcommand{\nat}{{Nature} }
\newcommand{\aap}{{Astron. Astrophys.} }
\newcommand{\aplett}{{Astrophys. Lett.} }
\newcommand{\jetp}{{Sov. Phys. JETP} }
\newcommand{\sjpp}{{Sov. J. Plasma Phys.} }
\newcommand{\ssr}{{Space Sci. Rev.} }
\newcommand{\planss}{{Planet. Space Sci.} }
\newcommand{\icarus}{{Icarus} }
\newcommand{\mnras}{{Mon. Not. Roy. Astron. Soc.} }
\newcommand{\apss}{{Astrophys. Space Sci.} }
\newcommand{\aapr}{   {Astron. Astrophys. Rev.} }
\newcommand{\pre}{   {Phys. Rev. E} }
\newcommand{\physrep}{   {Phys. Reports} }

\newcommand{\pasj}{ {Publ. Astron. Soc. Japan} }

\usepackage{color}  

\newcommand{\br}{\null%\bf\cm
}
 
\begin{document}

\title{Coherent Emission Mechanisms in Astrophysical Plasmas
}

\titlerunning{Coherent Emission}        % if too long for running head

\author{D. B. Melrose        %\and
        %Second Author %etc.
}


\institute{D. B. Melrose \at
              SIfA, School of Physics, The University of Sydney, NSW 2006, Australia\\
              Tel.: +61-2-93514234\\
             % Fax: +123-45-678910\\
              \email{donald.melrose@sydney.edu.au} \\
              {\br ORCiD: {000-0002-6127-4545}}          %  \\
          % \and
          % S. Author \at
            %  second address
}

\date{Received: date / Accepted: date}

\maketitle

\begin{abstract}
Three known examples of coherent emission in radio astronomical sources are reviewed: plasma emission, electron cyclotron maser emission (ECME) and pulsar radio emission. 

Plasma emission is a multi-stage mechanism with the first stage being generation of Langmuir waves through a streaming instability, and subsequent stages involving partial conversion of the Langmuir turbulence into escaping radiation at the fundamental (F) and second harmonic (H) of the plasma frequency.  The early development and subsequent refinements of the theory, motivated by application to solar radio bursts, are reviewed. The driver of the instability is faster electrons outpacing slower electrons, resulting in a positive gradient ($df(v_\parallel)/dv_\parallel>0$) at the front of the beam. Despite many successes of the theory, there is no widely accepted explanation for type~I bursts and various radio continua.

The earliest models for ECME were purely theoretical, and the theory was later adapted and applied to Jupiter (DAM), the Earth (AKR),  solar spike bursts and flare stars. ECME strongly favors the x~mode, whereas plasma emission favors the o~mode. Two drivers for ECME are a ring feature (implying $df(v)/dv>0$) and a loss-cone feature. Loss-cone driven ECME was initially favored for all applications. The now favored driver for AKR is the ring-feature in a horseshoe distribution, which results from acceleration by a parallel electric on converging magnetic field lines. The driver in DAM and solar and stellar applications is uncertain.

The pulsar radio emission mechanism remains an enigma. Ingredients needed in discussing possible mechanisms are reviewed: general properties of pulsars, pulsar electrodynamics, the properties of pulsar plasma and wave dispersion in such plasma. Four specific emission mechanisms (curvature emission, linear acceleration emission, relativistic plasma emission and anomalous Doppler emission) are discussed and it is argued that all encounter difficulties. 

Coherent radio emission from extensive air showers in the Earth's atmosphere is reviewed briefly. The difference in theoretical approach from astrophysical theories is pointed out and discussed. 

Fine structures in DAM and in pulsar radio emission are discussed, and it is suggested that trapping in a large-amplitude wave, as in a model for discrete VLF emission, provides a plausible explanation. A possible direct measure of coherence is pointed out.

\end{abstract}

\tableofcontents

\section{Introduction}
\label{intro}
Radio astronomy began in the 1930s, through the pioneering work of Jansky in the USA. The field grew rapidly after WWII, when radar groups, in Australia and England, redirected their interests to the cosmos.  By the late 1940s, a variety of radio sources had been recognized, including the Sun, the Milky Way,  supernova remnants and radio galaxies. Most of these sources are highly nonthermal, in the sense that the brightness temperature, $T_B$, of the emission is much greater than any plausible thermal temperature, $T_e$, of the electrons in the source. By the early 1950s it was recognized that most non-solar emission is due to synchrotron radiation, that is, to highly relativistic electrons in magnetic fields. Such emission is incoherent, in the sense that each electron radiates independently of the others, and the total emission from a collection of electrons is found by summing over the emission by a distribution of electrons. Two mechanisms restrict $T_B$ in a synchrotron source: self-absorption implies $T_B<\varepsilon/k_B$, where $\varepsilon$ is the energy of the synchrotron-emitting electrons and $k_B$ is Boltzmann's constant, and inverse Compton scattering restricts $T_B$ to $<10^{12}\,$K \citep{KP-T69}. 

Radio bursts from the Sun are not due to synchrotron emission and cannot be explained by any other incoherent emission mechanism. The emission frequency is associated with the local plasma frequency {\br $\omega_p=(e^2n_e/\varepsilon_0m_e)^{1/2}$, where $n_e$ is the electron number density} in the source, and this led to the emission mechanism being referred to as ``plasma emission''. Plasma emission is one example of a ``coherent'' emission mechanism, where ``coherent'' means ``not incoherent''. All coherent emission mechanisms involve some plasma instability, and an alternative description of them is ``collective plasma radiation processes'' \citep{Me91}. Three distinct classes of coherent emission are now well established. {\br Early theories for two of these were developed in the late 1950s: plasma emission and electron cyclotron maser emission (ECME), which occurs near the cyclotron frequency $\Omega_e=eB/m_e$ where $B$ is the magnetic field in the source.} The third coherent emission mechanism is involved in pulsar radio emission, which is extremely bright, $T_B\approx 10^{25}$--$10^{30}\,$K \citep{LK04}, but the specific mechanism remains uncertain. A fourth class of coherent emission is from extensive air showers in the Earth's atmosphere; this is included here, although the medium is (un-ionized) air rather than a plasma.

Any theory for a coherent emission involves particles emitting in phase with each other. Theoretical models invoke one of three forms of coherence: (a) emission by bunches, (b) a reactive instability, and (c) a kinetic instability. Idealized limits of these can be defined by assuming that the distribution of emitting particles is described by its distribution function, $f({\bi x},{\bi p},t)$, in the 6-dimensional ${\bi x}$--${\bi p}$ phase space. There are two versions of (a) that differ according to whether the distribution function is regarded as (i) a continuum, assumed here, or (ii) a collection of individual particles, cf.\ Equation (\ref{Pcoh6}) below. In either case, $N$ particles in a bunch are assumed to be separated from each other by less than a wavelength (of the emitted radiation) and to move along nearly identical orbits, so that they act like a single macro-charge, emitting a power $N^2$ times the power emitted by a single charge. An idealized continuum model for (a) corresponds to $f({\bi x},{\bi p},t)$ proportional to $\delta$-functions in both ${\bi x}$ and ${\bi p}$. In (b) the particles are highly localized in momentum space, idealized by a $\delta$-function in ${\bi p}$, but not in coordinate space. In this case, a wave with a specific phase grows due to a feedback mechanism involving self-bunching of the particles in the wave fields. In (c) there is no localization in ${\bi x}$ and the momentum distribution is ``inverted'' in the sense that there is available free energy that leads to negative absorption, as in a maser or laser. Maser mechanisms are favored in most astrophysical applications. One reason is that if (a) were to develop,  the back reaction can be shown to broaden the distribution in ${\bi x}$ so that the bunch spreads out and the emission evolves into (b). The back reaction to (b) can be shown to broaden the distribution in ${\bi p}$ so that the reactive instability suppresses itself, and evolves into a kinetic instability (c). The back reaction to (c), which can be described by kinetic theory (the quasilinear equations), tends to reduce the growth rate of the instability until a marginally stable state is approached. Over the relatively large volumes and long times required to produce observable emission from an astrophysical source, one expects that marginally stable maser growth should determine the average properties of any coherent emission. 

Plasma emission is reviewed and discussed in Section~\ref{sect:plasma}, with emphasis on radio bursts in the solar corona. ECME is discussed in  Section~\ref{sect:ECME}, starting from the early theoretical ideas, which preceded the applications to planetary, solar and stellar applications. The pulsar radio emission mechanism remains uncertain, and the discussion in Section~\ref{sect:pulsar} is aimed at explaining why this is the case. Radio emission from extensive air showers is discussed briefly in Section~\ref{sect:EAS}. A more general discussion of coherence and its role in these applications is given in Section~\ref{sect:coherent}. Some concluding remarks are made in Section~\ref{sect:conclusion}.

\section{Plasma Emission}
\label{sect:plasma}

The first theory for plasma emission \citep{GZ58} was motivated by the known observational properties of solar radio bursts, established over the preceding decade.

\subsection{Properties of solar radio bursts}

\begin{figure}[t]
 \begin{center}    
     \includegraphics[width=1.0\hsize]{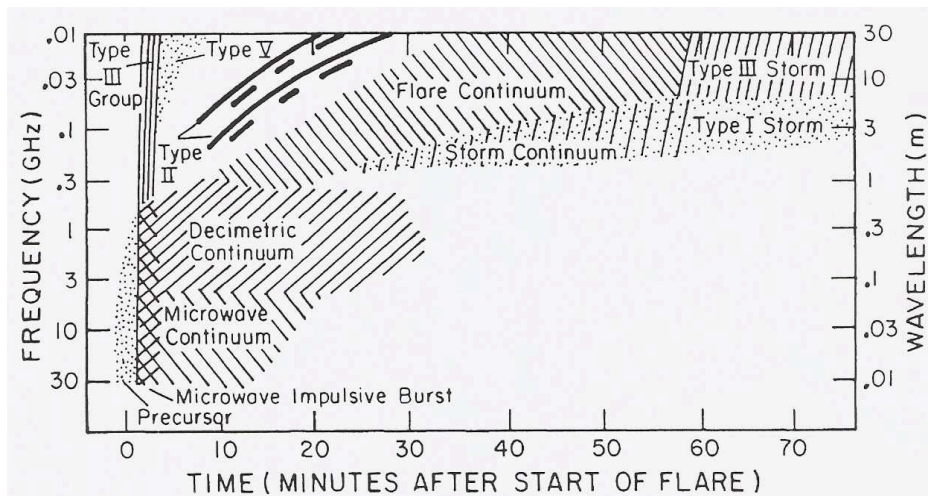}
     \caption{Schematic of the dynamic spectrum of radio emission following a flare, shown in terms of decreasing frequency (corresponding to increasing height) as a function of time following the flare [from \citet{McLL85}]
     }
 \label{1fig:schematic}    
      \end{center}
      \end{figure}

The initial definitive classification of solar radio bursts, into  types~I, II and III, was made in 1950 based on the appearance of the bursts in dynamic spectra \citep{WM50,W50a,W50b}:
 \begin{itemize}
    
    \item Type I bursts (or storm bursts)  have durations 1--$20\,$s and a bandwidth, $\Delta f$, of a few megahertz.

    \item Type II are slow drift, with $df/dt\approx-0.25\,\rm MHz\,s^{-1}$.
    
    \item Type III are fast drift with $df/dt\approx-20\,\rm MHz\,s^{-1}$.
  
    \end{itemize}
Two other types of bursts were added later in the 1950s: type~IV bursts by \citet{B57} and type~V by \citet{typeV59}. A schematic showing meter-wavelength solar radio emission during and after a solar flare is shown in Figure~\ref{1fig:schematic}. 
    
The exciting agency for type~III bursts was identified, as a stream of mildly relativistic electrons, by Ruby Payne-Scott in her research notes circa 1947 \citep{GossM10}, but evidently she considered the idea too radical to publish. The exciting agency for a type~II burst was also recognized, early in the development of the field, a shock wave propagating at close to the Alfv\'en speed, $v_{\rm A}$. There is still no consensus on the exciting agency for type~I bursts. 

An important observation, in defining the properties of plasma emission, was that of harmonic structure in type~III and type~II bursts \citep{1WMR53,1WMR54}. The implication is that plasma emission occurs near both the fundamental (F) and second harmonic (H) of the plasma frequency, $\omega_p$. (Angular and cyclic frequencies are used when discussing theory and observation, respectively, with $\omega=2\pi f$.)  However, type~I bursts are seen only in F emission with no evidence for H emission.

Plasma emission, in types~I, II and III bursts, is partly circularly polarized. F emission in type~III bursts \citep{SD85} can be highly (e.g., 70\%) but never completely polarized, with H emission weakly polarized ($<10$\%). Type~I emission, which has no H component, is typically nearly 100\% polarized for sources near the central meridian, with the polarization decreasing from day to day as a storm approaches the solar limb \citep{Z75}.

\subsection{Theories for plasma emission}

The theory of \citet{GZ58} was proposed before {\br the development of plasma kinetic theory,  as reviewed by \citet{T67}, cf.\ also \citet{T72,T72U,KT73}. Important details of the theory of \citet{GZ58} needed to be} updated \citep{M70a,M70b,ZZ70a,ZZ70b} without changing the overall concept of the theory. Confirmation of the essential features of the theory occurred later, when observations from spacecraft provided detailed information on type~III bursts in the interplanetary medium (IPM).

\subsubsection{Stages in plasma emission}

\begin{figure}[t]
\begin{center}
\includegraphics[scale=0.8, angle=0]{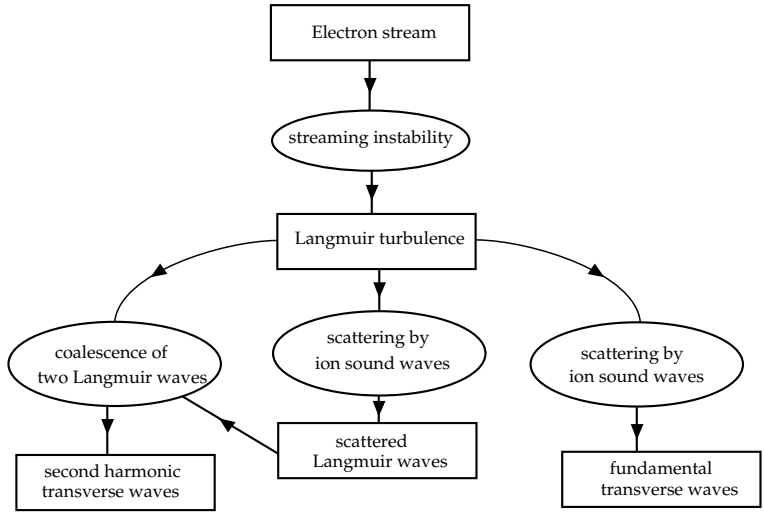}
\caption{Flow diagram for a variant of the theory of \citet{GZ58} for the generation of plasma
emission; in other variants the processes indicated that involve ion sound waves are replaced by other nonlinear plasma
processes [from \citet{Me91}]}
\label{fig:annrev1}
\end{center}
\end{figure}

The original theory involves three stages, as indicated schematically in Figure~\ref{fig:annrev1}. The first stage is the generation of Langmuir turbulence through a beam instability. The second stage is production of F emission due to {\br scattering} off fluctuations associated with the ions, producing transverse waves with little change in frequency. The third stage is the production of H emission through coalescence of two Langmuir waves, one from the beam-generated distribution and the other from the thermal distribution of Langmuir waves. In later versions of plasma emission all three stages were modified. The beam instability was treated using quasilinear theory, F emission was attributed to induced scattering of Langmuir waves into transverse or to coalescence of Langmuir and ion sound waves, and H emission was attributed to coalescence with a secondary nonthermal distribution of Langmuir waves produced from the primary (beam-generated) waves by a nonlinear process. 

\subsubsection{Nonlinear conversion mechanisms}

To produce escaping radiation, the energy in Langmuir turbulence must be partially converted into energy in escaping transverse waves. Two nonlinear processes are relevant. One involves three-wave interactions and the other is induced scattering. In most examples of plasma emission it is unclear which of these two processes dominates. 

A three-wave interaction involves two waves beating to generate a third wave. Let the three waves be in modes $M$, $P$, $Q$, with frequencies $\omega_M({\bi k})$, $\omega_P({\bi k}')$, $\omega_Q({\bi k}'')$ and wave vectors ${\bi k}$, ${\bi k}'$, ${\bi k}''$, respectively. In a coalescence process $P+Q\to M$ waves $P$ and $Q$ beat to form wave $M$. The inverse $M\to P+Q$ is referred to as a decay process. Both satisfy the beat conditions, also called {Manley-Rowe} conditions,
\be
\omega_P({\bi k}')+\omega_Q({\bi k}'')=\omega_M({\bi k}),
\quad
{\bi k}'+{\bi k}''={\bi k}.
\label{3.1}
\ee
One process that leads to F emission is the coalescence $L+S\to T$, where $L$ refers to a Langmuir wave, $S$ to an ion sound wave and $T$ to a transverse wave. The relevant dispersion relations are $\omega=\omega_L(k)$, $\omega_S(k)$, $\omega_T(k)$, with
\be
\omega_L^2(k)=\omega_p^2+3k^2V_e^2,
\qquad
\omega_S(k)=kv_{\rm s},
\qquad
\omega_T^2(k)=\omega_p^2+k^2c^2,
\label{dr1}
\ee
where $v_{\rm s}=V_e/43$ is the ion sound speed and $V_e=(k_BT_e/m)^{1/2}$ is the thermal speed of electrons with $T_e$ the electron temperature. The decay, $L\to T+S$ also produces F emission. Analogous processes $L+S\to L'$ and $L\to S+L'$ produce scattered Langmuir waves, denoted by $L'$. H emission results from the coalescence process $L+L'\to T$. The wavenumber, $k$, of a $T$ wave is much smaller than the wavenumber, $k'$, of a beam-generated $L$ wave, requiring ${\bi k}''\approx-{\bi k}'$, that is, requiring that $L'$ be from a back-scattered distribution. 

Qualitatively, one may regard induced scattering as analogous to a three-wave interaction in which the $S$ waves is replaced by a fluctuation associated with Debye screening of ions, with the relevant frequency and wave vector not satisfying any dispersion relation. Induced scattering is related to the more familiar ``spontaneous'' scattering by a single particle (in this case an ion) in the same way as absorption is related to spontaneous emission of a wave, with the wave replaced by the beat between two waves. Induced scattering causes the lower-frequency wave to grow and the higher frequency wave to damp. 

An alternative conversion process is mode coupling due to inhomogeneities in the plasma.  Although this may be significant under some conditions, it seems implausible as the basic conversion mechanism for type~III and type~II bursts.

\subsection{Generation of Langmuir waves}

Plasma instabilities can have both reactive and kinetic versions. A reactive instability applies when the imaginary part of the response function can be neglected. The dispersion equation is then a polynomial equation in $\omega$ and $k$ with real coefficients, so that its complex solutions appear in complex conjugate pairs. Reactive growth requires that the growth rate exceed the bandwidth of the growing waves, which determined the rate of phase mixing. When phase mixing is unimportant, the growing wave has a well-defined phase. A kinetic (or maser) instability applies when the imaginary part of the response function is included and leads to negative absorption. Maser growth requires that the growth rate be less than the bandwidth of the growing waves. Phase mixing then occurs faster than wave growth, implying that the random phase approximation (RPA) applies. As in a maser or laser, the RPA does not imply that the growing waves necessarily have random phases; rather the RPA implies that the phase is irrelevant when considering wave growth.

\subsubsection{Reactive growth of Langmuir waves}

A model for a weak beam corresponds to an electron density $n_e=n_0+n_{\rm b}$, with $n_0$ the number density of background electrons, assumed to be at rest, and $n_{\rm b}\ll n_0$ the number density of beam electrons with velocity ${\bi v}_{\rm b}$. The dispersion equation for the Langmuir waves is found by setting the longitudinal dielectric constant to zero. This gives
\be
K^{\rm L}(\omega,{\bi k})=1-{\omega_{p0}^2\over\omega^2}-{\omega_{pb}^2\over(\omega-{\bi k}\cdot{\bi v}_{\rm b})^2}
=0,
\label{KLb}
\ee
with $\omega_{p0}^2=e^2n_0/\varepsilon_0m$, $\omega_{pb}^2=e^2n_{\rm b}/\varepsilon_0m$. The dispersion equation (\ref{KLb}) may be written as a quartic equation in $\omega$. For $n_{\rm b}\ll n_0$ and most values of ${\bi k}\cdot{\bi v}_{\rm b}$, the four solutions are all real, $\omega\approx\pm\omega_{p0}$ and $\omega\approx{\bi k}\cdot{\bi v}_{\rm b}\pm\omega_{pb}$, with the latter two solutions called beam modes. When $(\omega-{\bi k}\cdot{\bi v}_{\rm b})^2$ is sufficiently close to zero equation (\ref{KLb}) may be approximated by the cubic equation 
\be
\Delta\omega(\Delta\omega+\Delta\omega_0)^2-\half\omega_{p0}\omega_{pb}^2=0,
\label{rbi3}
\ee
with $\Delta\omega=\omega-\omega_{p0}$, $\Delta\omega_0=\omega_{p0}-{\bi k}\cdot{\bi v}_{\rm b}$. For $\Delta\omega_0\ll(\half\omega_{p0}\omega_{pb}^2)^{1/3}$ one may neglect $\Delta\omega_0$, and the three solutions become
\be
\Delta\omega=\alpha(\half\omega_{p0}\omega_{pb}^2)^{1/3},
\qquad
\alpha=1^{1/3}=1,(-1\pm i\sqrt{3})/2.
\label{rbi4}
\ee
The growth rate for the reactive instability is identified as the imaginary (Im) part of the frequency:
\be
{\rm Im}(\Delta\omega)=\frac{\sqrt{3}}{2}\left({n_{\rm b}\over2n_0}\right)^{1/3}\omega_{p0}.
\label{rbi5}
\ee

\subsubsection{Phase bunching and wave trapping}

The mechanism that drives reactive growth is  bunching of particles in phase with the wave. This is a general feature of reactive instabilities, and one may regard the following model for growing Langmuir waves as a prototype for other reactive instabilities. 

\begin{figure}[t]
\begin{center}
\vspace{-4cm}
\includegraphics[scale=0.4, angle=0]{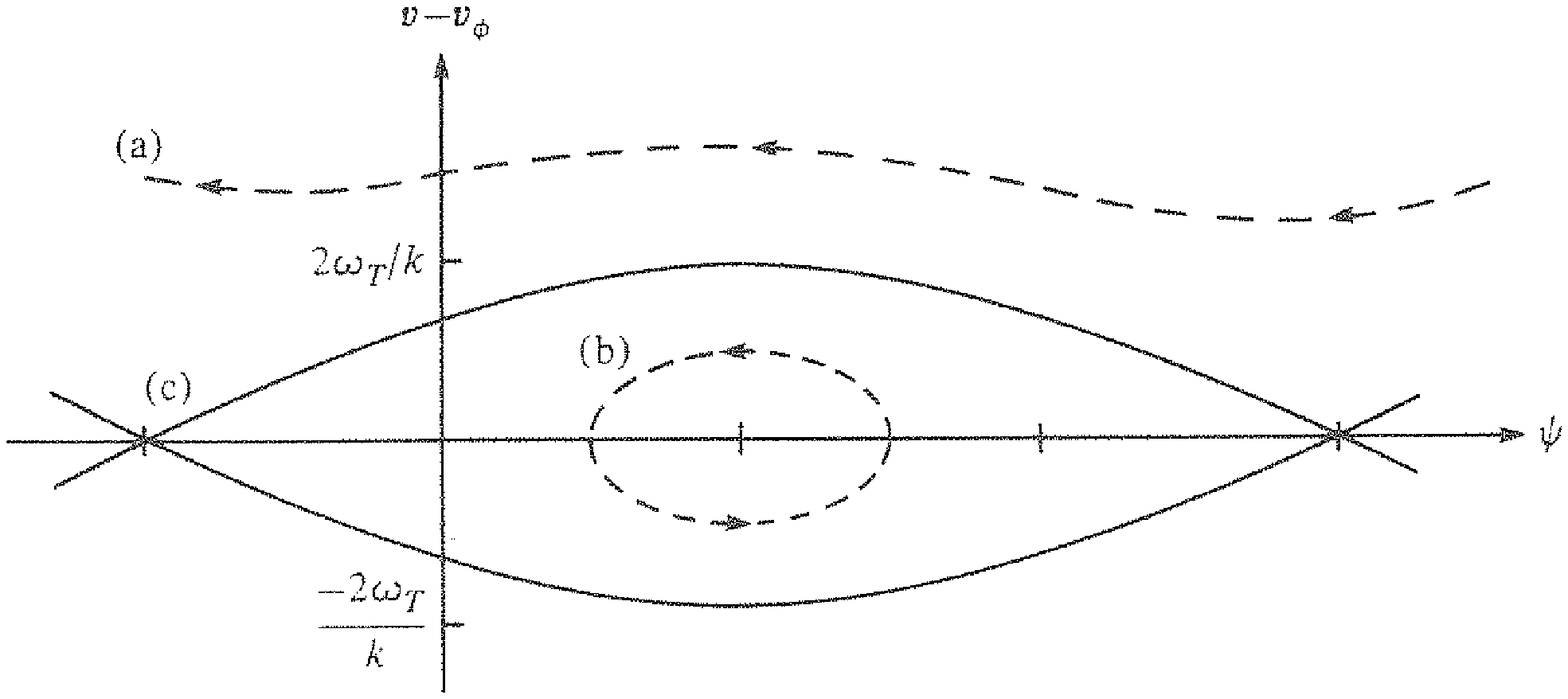}
\vspace{-4cm}
\caption{The velocity component of a particle along the direction of propagation of a finite amplitude wave ($\alpha=1$)  is illustrated schematically as a function of the phase $\psi$ of the wave for $v\approx v_\phi$ by the dashed curves, (a) for an untrapped particle and (b) for a trapped particle; (c) the solid curves denoted the separatrix between trapped and untrapped particles [from \citet{M86}]}
\label{Fig:closed}
\end{center}
\end{figure}

Let $\psi=\omega t-kz$ be the phase of the wave, such that the electric field varies as $E=E_0\cos\psi$. The first and second derivatives of the phase are
\be
{d\psi\over dt}=\omega-kv=-k(v-v_\phi),
\qquad
{d^2\psi\over dt^2}={eE_0k\over m}\,\cos\psi, 
\label{wt1}
\ee
where Newton's equation $mdv/dt=-eE$ is used. Electrons trapped in the wave oscillate about the phase velocity $v_\phi\approx v_{\rm b}$:
\be
{d^2\over dt^2}(v-v_\phi)=-(v-v_\phi)\,\omega_{\rm t}^2\sin\psi,
\qquad
\omega_{\rm t}^2={eE_0k\over m},
\label{wt2}
\ee
where $\omega_{\rm t}$ is the trapping (or bounce) frequency. To describe the oscillations, it is convenient to consider the frame moving with the wave, at velocity $v_\phi$. Noting that the electric field, $E=E_0\cos\psi$, may be written as $E=-{\bf\nabla}\Phi$, the electrostatic potential is $\Phi=(E_0/k)\sin\psi$. The total energy of an electron, $\varepsilon'$, is the sum of the kinetic energy, $\half m(v-v_\phi)^2$, and potential energy, $-e\Phi$, and is a constant of the motion. Equation (\ref{wt1}) implies
\be
\left({d\zeta\over dt}\right)^2=\omega_{\rm t}^2(\alpha^2-\sin^2\zeta),
\qquad
\zeta=\half\left(\psi-\frac{\pi}{2}\right),
\qquad
(\alpha^2-\half)\omega_{\rm t}^2=\frac{\varepsilon'k^2}{2m},
\label{wt3}
\ee
which may be solved in terms of elliptic integrals. Closed orbits occur for $\alpha^2<1$, as illustrated in Figure~\ref{Fig:closed}. The range $|v-v_\phi|<2\omega_{\rm t}/k$ of velocities corresponding to trapped electrons increases as the amplitude of the wave (and hence $\omega_{\rm t}^2$) increases.

In a reactive instability, as the phase-coherent wave grows, more and more electrons become trapped in it, and the bounce frequency of the trapped electrons increases. It follows that such wave trapping causes a spread in electron velocity, around the resonant velocity, which increases as the amplitude, $E_0$, of the wave increases. The reactive instability is derived under the assumption that there is no spread in the velocity of the electrons, and this is justified only if the spread associated with the bounce motion of the trapped electrons remains smaller that the growth rate. This suggests that the reactive instability suppresses itself when the wave amplitude and associated spread in velocities reaches a threshold. Ignoring factors of order unity, the threshold corresponds to a wave energy density 
\be
W_L=\varepsilon_0E_0^2=\left({n_{\rm b}\over n_0}\right)^{2/3}\,n_{\rm b}mv_{\rm b}^2.
\label{wt5}
\ee
As this threshold is approached, phase mixing becomes important, and the growth passes over from the phase-coherent reactive form to the phase-random kinetic form of the instability. This transition between reactive and kinetic growth may be treated by including a velocity spread, {\br $\Delta v_{\rm b}$}, in the distribution function of the beam. The condition for this transition to occur then corresponds to
\be
\left({n_{\rm b}\over n_0}\right)^{1/3}\gapprox {v_{\rm b}\over\Delta  v_{\rm b}}.
\label{rit3}
\ee

\subsubsection{Kinetic version of beam instability}

The kinetic version of the beam instability may be attributed to negative Landau damping. A kinetic theory derivation of the absorption coefficient for Langmuir waves resonating with electrons gives
\be
\gamma_L({\bi k})=-{\pi e^2\omega_p\over\varepsilon_0 k^2}
\int d^3{\bi p}\,\delta(\omega_p-{\bi k}\cdot{\bi v})\,{\bi k}\cdot{\partial f({\bi p})\over\partial{\bi p}}.
\label{kbi3}
\ee
The simplest example corresponds to a one dimensional (1D) model in which only Langmuir waves propagating along the streaming direction are considered.

%\subsubsection{Weak-beam instability in 1D}

Let the streaming direction, and hence the 1D direction, be the $z$-axis. A reduced distribution function for the electrons is defined by integrating over the momentum components perpendicular to this axis:
\be
F(v_z)={1\over n_{\rm b}}\int d^3{\bi p}\,\delta(v_z-\bkappa\cdot{\bi v})\,F({\bi p}),
\label{bit1}
\ee
with $\bkappa={\bi k}/k$. In the 1D model the absorption coefficient (\ref{kbi3}) reduces to
\be
\gamma_L(v_\phi)=-\pi{\omega_{pb}^2\over\omega_p}\,v_\phi^2{dF(v_\phi)\over dv_\phi},
\label{bit2}
\ee
with $v_\phi=\omega_p/k$. The waves may be described by their energy density per unit range of $v_\phi$, such that the total energy density in the waves is
\be
W_L=\int dv_\phi\,W(v_\phi).
\label{Wvphi}
\ee
The kinetic equation for the waves is
\be
{d W(v_\phi)\over d t}=-\gamma_L(v_\phi)W(v_\phi),
\qquad
\gamma_L(v_\phi) =-{\pi\omega_p\over n_e}\,v_\phi^2
{d F(v_\phi)\over d v_\phi}.
\label{1Dql1}
\ee
The evolution of the distribution of electrons is described by a quasilinear equation, which corresponds to diffusion in momentum space. In the 1D approximation, the evolution of the reduced distribution is described by
\be
{d F(v)\over d t}
={\partial\over\partial v}\,D(v) 
{\partial F(v)\over\partial v},
\qquad
D(v)={\pi\omega_p\over n_e m}\,vW(v),
\label{1Dql2}
\ee
which corresponds to (1D) diffusion in velocity space with diffusion coefficient $D(v)$. The 1D equations (\ref{1Dql1}) and (\ref{1Dql2}) conserve both energy and momentum, as well as the number of electrons. These conservation laws correspond to
\be
n_{\rm b}\int dv\,\left[
\begin{array}{c}
1\\mv\\\half mv^2
\end{array}
\right]\,{d F(v)\over d t}
+\int dv_\phi\,\left[
\begin{array}{c}
0\\1/v_\phi\\1
\end{array}\right]\,{d W(v_\phi)\over d t}=0.
\label{1Dql3}
\ee

\begin{figure}[t]
\begin{center}
\includegraphics[scale=0.6, angle=0]{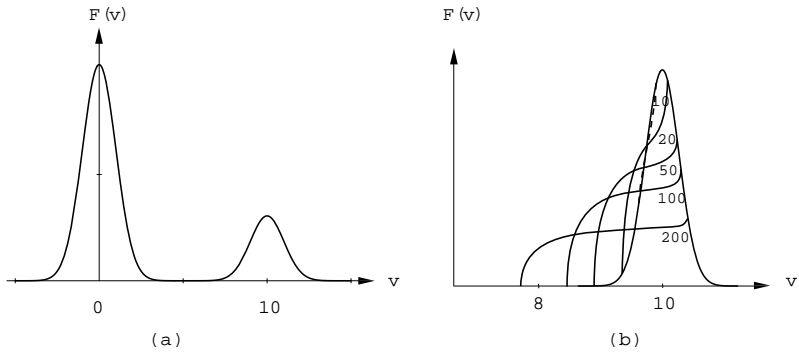}
\caption{The evolution of the bump-in-tail instability: (a)~the initial distribution causes
waves to grow for $v_\phi=v$ in the range where $dF(v)/dv$ is
positive; (b)~as the growth proceeds, the bump in the tail {\br (shown on an expanded scale)} is eaten
away to form a plateau $dF(v)/dv\approx0$ that extends to lower $v$
with time (denoted by numbers in units of the initial growth time) [from  \citet{G75}]}
\label{lindaufig1}
\end{center}
\end{figure}  

An example of the 1D evolution of the distribution of particles is illustrated in Figure~\ref{lindaufig1}. The distribution function relaxes towards a plateau, $dF(v)/dv=0$, in velocity space. The asymptotic solution corresponds to $F_{\infty}=$ constant for $0<v<v_{\rm b}$.  Suppose the initial beam has a small velocity spread, and hence an energy density $\half n_{\rm b}mv_{\rm b}^2$. The energy density for a plateau distribution with the same $n_{\rm b}$ is only a third of this initial energy density, implying that the remaining two thirds of the energy has been transferred to the Langmuir waves. 

\subsubsection{Sturrock's dilemma}

A large loss of energy to Langmuir waves leads to an inconsistency called ``Sturrock's dilemma'' \citep{S64}: the beam would slow down before propagating any significant distance from the source. For example, assume a plasma frequency of $100\,$MHz and a beam with number density $n_{\rm b}=10^{-6}n_0$. The growth rate is then of order $10^{-6}$ times the plasma frequency, and the asymptotic state is approached after about 100 growth times. These numbers suggest that a plateau should form on a time scale of order $0.1\,$s, so that the beam must slow down on this time scale. A beam with speed $v_{\rm b}$ of order $10^8\rm\,m\,s^{-1}$ would lose a large fraction of its energy to Langmuir waves after propagating about $10^7\,$m. This clearly does not occur.  There is no evidence for systematic slowing down of the electrons that generate type~III bursts in the corona, and the electrons are known to propagate through the IPM, to well beyond the orbit of the Earth, apparently without slowing down. Evidently, the loss of energy to the Langmuir waves is never as catastrophic as simple theory implies. 

One early suggestion on how this dilemma might be overcome is to appeal to an inhomogeneous beam, in which faster electrons continually outpace slower electrons, causing the positive slope at the front of the beam to be continually regenerated \citep{ZMR72}. To avoid the catastrophic energy loss it was suggested that slower electrons towards the back of the beam absorb the Langmuir waves generated by the faster electrons near the front of the beam, so that the energy is recycled. However, this recycling would need to occur with an impossibly high efficiency to resolve the dilemma. The dilemma was only resolved when type~III bursts in the IPM were studied in detail (\S\ref{sect:IPM}). 

\subsection{Three-wave interactions}

The three-wave interaction, satisfying the beat conditions (\ref{3.1}), can be described by a set of three kinetic equations, one for each mode $M,P,Q$. These equations are particularly useful in determining the saturation conditions, and hence the effective temperature of the F and H emission. An effective temperature, $T_M({\bi k})$, for waves in the mode $M$, corresponds to an energy density $k_BT_M({\bi k})$ in the range $d^3{\bi k}/(2\pi)^3$, where $k_B$ is Boltzmann's constant. It is convenient to introduce the wave action, $N_M({\bi k})=k_BT_M({\bi k})/\omega_M({\bi k})$.

\subsubsection{Kinetic equations for three-wave processes}

The kinetic equations for the three waves involved in the coalescence, $P+Q\to M$, and decay, $M\to P+Q$, processes
are 
\begin{align} 
{ d N_M({\bi k})\over  d t}
&=\int{ d^3{\bi k}'\over(2\pi)^3}\,
{ d^3{\bi k}''\over(2\pi)^3}\,
u_{MPQ}({\bi k},{\bi k}',{\bi k}')\,\big\{N_{P}({\bi k}')N_{Q}({\bi k}'')
\nn
\\
&\qquad\qquad\qquad\qquad\qquad
-N_M({\bi k})\big[N_{P}({\bi k}')+N_{Q}({\bi k}'')\big]
\big\},
\nn
\\
\ms
{ d N_{P}({\bi k}')\over  d t}&=
-\int{ d^3{\bi k}\over(2\pi)^3}\,
{ d^3{\bi k}''\over(2\pi)^3}\,
u_{MPQ}({\bi k},{\bi k}',{\bi k}')\,\big\{N_{P}({\bi k}')N_{Q}({\bi k}'')
\nn
\\
&\qquad\qquad\qquad\qquad\qquad
-N_M({\bi k})\big[N_{P}({\bi k}')+N_{Q}({\bi k}'')\big]
\big\},
\nn
\\
\ms
{ d N_{Q}({\bi k}'')\over  d t}&=
-\int{ d^3{\bi k}\over(2\pi)^3}\,
{ d^3{\bi k}'\over(2\pi)^3}\,
u_{MPQ}({\bi k},{\bi k}',{\bi k}')\,\big\{N_{P}({\bi k}')N_{Q}({\bi k}'')
\nn
\\
&\qquad\qquad\qquad\qquad\qquad
-N_M({\bi k})\big[N_{P}({\bi k}')+N_{Q}({\bi k}'')\big]
\big\}.
\label{5e.9}
\end{align}
The quantity $u_{MPQ}({\bi k},{\bi k}',{\bi k}')\propto\delta[\omega_M({\bi k})-\omega_P({\bi k}')-\omega_Q({\bi k}'')]\delta^3({\bi k}-{\bi k}'-{\bi k}')$ ensures that the beat conditions (\ref{3.1}) are satisfied, but its explicit form is not important in the present discussion. Together the set of equations (\ref{5e.9}) ensures that the energy and the momentum summed over the three wave distributions are conserved.

In a plasma emission model based on evolution of the Langmuir turbulence due to ion-sound or other low-frequency waves, one of the three modes is the beam-generated Langmuir waves, denoted $L$ say, and the other modes are denoted $L'$, $S$, and $T$, corresponding to scattered Langmuir waves, ion-sound waves and transverse waves, which include fundamental, $T\to{\rm F}$, and harmonic, $T\to{\rm H}$, emissions.

\subsubsection{Saturation of the three-wave interactions}

The three-wave interactions saturate when the condition
\be
N_{P}({\bi k}')N_{Q}({\bi k}'')=N_M({\bi k})\big[N_{P}({\bi k}')+N_{Q}({\bi k}'')\big]
\label{saturation1}
\ee
is satisfied. The condition (\ref{saturation1}) may be rewritten as
\be
\frac{T_{P}({\bi k}')}{\omega_{P}({\bi k}')}\,
\frac{T_{Q}({\bi k}'')}{\omega_{Q}({\bi k}'')}=
\frac{T_M({\bi k})}{\omega_M({\bi k})}\left[\frac{T_{P}({\bi k}')}{\omega_{P}({\bi k}')}
+\frac{T_{Q}({\bi k}'')}{\omega_{Q}({\bi k}'')}\right].
\label{saturation2}
\ee
In particular, in thermal equilibrium each of the effective temperatures is equal to $T_e$, and Equation (\ref{saturation2}) is satisfied in view of $\omega_M({\bi k})=\omega_P({\bi k}')+\omega_Q({\bi k}'')$.

The saturation condition (\ref{saturation2}) provides significant constraints on a model for plasma emission. Two of the waves must be from nonthermal distributions to produce a nonthermal distribution of the third wave. For example, consider the original suggestion by \citet{GZ58} that H emission is due to coalescence of nonthermal beam-generated Langmuir waves with thermal Langmuir waves; the identifications $P\to L$, $Q\to L'$, $M\to{\rm H}$, with $T_L\gg T_{L'}=T_e$ imply $T_{\rm H}=2T_e$, and hence that the mechanism cannot explain nonthermal H emission \citep{M70a}. For the processes involving ion sound waves to be effective in F and H emission, one must have $T_S\gg T_e$, and such waves cannot be generated effectively from a single nonthermal distribution of $L$ waves.  The assumption that these three-wave processes saturate in plasma emission offers a natural explanation for the observation that the brightness temperatures for F and H emission are often comparable: both are limited to $<T_L$.

\subsection{Modified forms of plasma emission}
\label{sect:modified}

The original theory of plasma emission was developed to explain type~III emission. Type~II emission was assumed to be analogous to type~III, due to electrons streaming away from a shock front. The original theory has been modified in various ways, to explain specific fine structures in observed emission, and to include the effect of the magnetic field, notably in connection with the polarization. Some examples of these various modifications are discussed here. Also discussed are two poorly understood phenomena that are clearly due to plasma emission.
Type~I emission has properties that are qualitatively different from type~III emission, notably the absence of a harmonic and high polarization, and why this should be the case is inadequately understood. Type~I emission also includes a continuum, the ``storm continuum in Figure~\ref{1fig:schematic}, and there are other continua that are not associated with type~I bursts. There is no widely accepted model for such continua in terms of plasma emission.

\subsubsection{Fine structures}

There have been extensive observations of various fine structures in solar radio bursts, cf.\ the review by \citet{Chernov11}. An early observation was of drift pair bursts \citep{R58}, and more detailed observations \citep{EM66,E69} led to the identification of three classes of fine structure:
(a) fast drift storm bursts with a mean frequency-time slope of $1·9\rm\,MHz\,s^{-1}$, a mean bandwidth $\Delta f\approx 0·03\,$MHz and a mean duration $\Delta t=0·6\,$s; 
(b) drift pair bursts with $df/dt = 1·2\rm\,MHz\,s^{-1}$ and $\Delta f= 0·45\,$MHz; 
(c) split pair bursts with $df/dt = 0·08\rm\,MHz\,s^{-1}$, $\Delta f = 0·05\,$MHz, and $\Delta t = 1·4\,$s. 
Another type of fine structure occurs in type~IIIb  bursts \citep{dLNB72}, which consist of ``striae'' within an envelope typical of a type~III burst. These bursts are observed at metric wavelengths. A different class of fine structures is observed in the decimetric band, including zebra patterns, fiber bursts and tadpoles \citep{E61,S72,Chernov06, Chernov16}.

Various modifications of plasma emission have been suggested to account for these fine structures. Some of these suggestions involve inclusion of the magnetic field, but others do not. Amongst the latter are models for drift pairs and type~IIIb bursts. One early model for drift pair bursts \citep{R58} involves two rays reaching the observer from the same source; one is the direct ray, and the other is a ray initially directed downward and then refracted strongly so that it is redirected upward (a ``reflected'' ray). Another model involves the frequency difference resulting from two 3-wave processes that result in fundamental emission: $L+S\to T$ and $L\to T+S$, where $S$ is any appropriate low-frequency fluctuation \citep{MS71,M83}. A more favored interpretation is in terms of propagation through a medium with filamentary irregularities \citep{TY75}. The irregularities can lead to favored locations where the growth factor for the Langmuir waves is large resulting in intermittency in the Langmuir turbulence governed by the irregularities \citep{LCR11b,LCR11a,LCR12,LCL14}.

{\br Another modification of plasma emission, which is not discussed here, involves including a third harmonic \citep{TY74,Zetal98,Yetal02,FYC13}.}

\subsubsection{Inclusion of the magnetic field in plasma emission}

A magnetic field affects all three stages of plasma emission: the properties of the Langmuir waves and of the instability that generates them, the properties of the nonlinear processes, the properties of possible counterparts of the ion sound waves in the three-wave interactions, and the properties of the escaping waves, which are in either the o~mode or the x~mode of magnetoionic theory. The formal theory for wave dispersion and the wave properties of most relevance here can be summarized as follows.

In a coordinate system with the magnetic field along the $z$~axis and the wave vector in the $x$-$z$ plane at an angle $\theta$ to it, the wave equation may be written in the form
\be
\Lambda_{ij}E_j=-\frac{i}{\varepsilon\omega}J_{{\rm ext}\,i},
\qquad
\Lambda_{ij}=n^2\kappa_i\kappa_j-n^2\delta_{ij}+K_{ij},
\label{we1}
\ee
where ${\bi J}_{\rm ext}$ is an extraneous current, regarded as a source term, $n=kc/\omega$  is the refractive index, $\bkappa={\bi k}/k=(\sin\theta,0,\cos\theta)$ is the unit vector along the wave-normal direction, and $K_{ij}$ is the dielectric tensor. The polarization vector is ${\bi e}={\bi E}/|{\bi E}|$. The dispersion equation is obtained by neglecting the source term and setting the determinant of the coefficients on the left hand side of Equation (\ref{we1}) to zero: $|\Lambda_{ij}|=0$. Different approximations are made in treating the Langmuir-like waves, the low-frequency waves and the o~and x~mode waves.

\subsubsection{Magnetoionic modes}

Radio-wave propagation in the solar corona is well described by the magnetoionic theory, in which the plasma is treated as a cold magnetized electron gas,  corresponding to the magnetoionic theory, with $\omega_p$ and $\Omega_e$ incorporated into the two magnetoionic parameters $X=\omega_p^2/\omega^2$, $Y=\Omega_e/\omega$. Far from the source, the o~and x~modes are oppositely circularly polarized, and the degree of polarization of observed plasma emission is defined in terms of the degree of circular polarization. Plasma emission is polarized in the sense of the o~mode.

For a cold electron gas $K_{ij}$ has the form \citep{Stix62}
\be
K_{ij}=
\left(
\begin{array}{ccc}
S&-i D& 0\cr
i D& S& 0\cr
0& 0& P
\end{array}
\right),
\qquad
S=\frac{1-X-Y^2}{1-Y^2},
\quad
D=\frac{XY}{1-Y^2},
\quad
P=1-X.
\label{cp1}
\ee
The dispersion equation, $|\Lambda_{ij}|=0$, may be written as a quadratic equation for $n^2$. The two solutions are called the ordinary and extraordinary wave modes. Each mode has a lower frequency branch (the whistler mode and the z~mode) and higher frequency branch (the o~mode and the x~mode), separated by a region of evanescence (where the solution for $k$ is imaginary). Only waves on the higher frequency branches can escape. The handedness of the o~mode  (x~mode) is opposite (the same) as the sense of electron gyration in the magnetic field.

In the approximation in which the modes are assumed circularly polarized, their dispersion relations reduce to
\be
n^2_\sigma=1-\frac{X}{1+\sigma Y|\cos\theta|}\approx1-X(1-\sigma Y|\cos\theta|),
\label{n2ox}
\ee
where the approximate forms apply for the o~and x~modes with $\sigma=\pm1$, respectively. For $Y|\cos\theta|, X/Y\gg1$, the general form with $\sigma=-1$ gives the approximate dispersion relation, The approximate dispersion relation, $n^2=X/Y|\cos\theta|$, for the whistler mode follows from the general form (\ref{n2ox}) for $Y|\cos\theta|, X/Y\gg1$ and $\sigma=-1$.

\subsubsection{Longitudinal and low-frequency waves}

The inclusion of both a magnetic field and thermal motions leads to a rich variety of waves that could play the role of the Langmuir waves or the ion sound waves in plasma emission.

For the Langmuir-like waves, the general dispersion relation is replaced by the longitudinal dispersion relation $\kappa_i\kappa_jK_{ij}=0$, with solution $\omega=\omega_L(k,\theta)$. This solution has a simple form for $\omega_e^2\ll\omega_p^2$,
\be
\omega_L^2(k,\theta)=\omega_p^2+3k^2V_e^2+\Omega_e^2\sin^2\theta,
\label{dr2}
\ee
which reduces to $\omega_L(k)$ for $\Omega_e^2\sin^2\theta=0$. For perpendicular propagation, $\sin\theta=1$, the dispersion relation (\ref{dr2}) describes upper-hybrid waves. For perpendicular propagation, there are other weakly damped longitudinal waves near harmonics of  $\Omega_e$, referred to as (electron) Bernstein modes. 

The low-frequency waves in a magnetized plasma include Alfv\'en waves, magneto-acoustic waves and whistler waves. The first two correspond to MHD waves with dispersion relations approximated by $\omega=|k_\parallel|v_{\rm A}$ and $\omega=kv_{\rm A}$, respectively. The wavenumbers for these waves are too small to be relevant to plasma emission, which requires $k$ approximately equal to that of the Langmuir mode to satisfy the beat conditions (\ref{3.1}). The whistler mode is the low-frequency branch of the ordinary mode of magnetoionic theory; it has a resonance at $\omega=\Omega_e|\cos\theta|$, and has large $k$ near the resonance. Far from the resonance, the dispersion relation may be approximated by $n^2\approx\omega_p^2/\omega\Omega_e|\cos\theta|$.

Several different plasma emission mechanisms have been proposed for {\br fiber bursts, zebra patterns and spike bursts}. One involves whistler waves \citep{K75}, and another involves upper hybrid waves and Bernstein modes \citep{R72,CGR73,ZZ75a,ZZ75b,ZZ75c,BKY17}. One suggestion for the generation of the upper hybrid waves is through a loss-cone instability \citep{ZS83}. {\br A plasma-emission model for microwave spike bursts was proposed by \citet{CFL01}.}

These examples illustrate some of the possible variants of plasma emission that have been invoked to explain specific features in the metric and decimetric bands. From about the mid-1970s, detailed data on type~III and type~II emission in the interplanetary medium became available, and the emphasis in modeling plasma emission shifted to bursts in the IPM, as discussed briefly below.

\subsubsection{Polarization of plasma emission}

Simple theory suggests that F emission should be 100\% polarized in the o~mode. The argument is based on the frequency of emission, which is above the cutoff frequency, at $\omega_p$, of the o~mode, and below the cutoff frequency of the x~mode, at
\be
\omega_{\rm x}=\half\Omega_e+\half(\Omega_e^2+4\omega_p^2)^{1/2}\approx\omega_p+\half\Omega_e,
\label{omegax}
\ee 
where the approximate form applies for $\omega_p\gg\Omega_e$. The frequency of the beam-generated Langmuir waves,
\be
\omega_L(k)=(\omega_p^2+3k^2V_e^2)^{1/2}\approx\omega_p\left(1+\frac{3V_e^2}{2v_\phi^2}\right),
\label{omegaL}
\ee
with $v_\phi\approx v_{\rm b}$, is between the cutoff frequencies for the o~and x~modes. If  F emission is due to the processes $L+S\to T$, $L\to T+S$, its frequency differs from $\omega_L$ by the ion-sound frequency, $\omega_S(k')\approx k'v_{\rm s}\approx(\omega_p/43)(V_e/v_{\rm b})$, and the change in frequency is similarly small if F emission is due to induced scattering.  Even for a weak magnetic field, $\Omega_e/\omega_p\approx0.1$ and a low beam speed, $v_{\rm b}/V_e\approx10$, F emission cannot exceed $\omega_{\rm x}$. Hence F emission is allowed only in the o~mode. The fact that type~I emission can be 100\% polarized in the o~mode is consistent with this simple theory. The accepted explanation for why not all F emission is 100\% polarized is that depolarization occurs as a propagation effect. The argument for this is somewhat different for type~I emission than for type~III and type~II emission.

Detailed calculations of the polarization of H emission imply that it should be weak and generally favor the o~mode \citep{MDS78}. Although this is consistent with observations, there are too many uncertainties in the details of the processes leading to H emission for a quantitative comparison.

\subsubsection{Propagation effects on plasma emission}

Refraction has a major effect on the propagation of $F$ emission. The refractive index of F emission at its source may be estimated by assuming that its frequency is $\omega_L$, and then equation (\ref{omegaL}) and $n^2\approx1-\omega_p^2/\omega_L^2$ implies $n\approx\sqrt{3}V_e/v_{\rm b}\ll1$. Along the ray path, $n$ increases towards unity, and Snell's law implies that the emission (including nearly backward emission) is refracted into a cone with half-angle $\approx\sqrt{3}V_e/v_{\rm b}$. The average density gradient is in the radial direction, suggesting that only a source within this angle of the central meridian could be observed.  However, F emission is observed from sources anywhere between the central meridian and the solar limb. For F emission from near the limb to be observed, it must be scattered through a large angle.

For type~I emission, the visibility of a source near the limb is attributed to a single large-angle scattering \citep{BS77,WZM86}, rather than many small-angle scatterings. The model requires that the corona be locally inhomogeneous with sharply bounded overdense and underdense regions, referred to as fibers \citep{BS77}, elongated along the magnetic field lines. Such reflection-like scattering of an incident o~mode wave leads to reflected waves in both o~and x~modes, and hence to a net depolarization. For example, scattering at the Brewster angle would result in linearly polarized emission, which corresponds to equal mixtures of the o~and x~modes which would be observed as unpolarized emission.

Radioheliograph images of type~III and type~II bursts suggest that what is seen is an apparent source that is much larger and at a greater height than the actual source. The interpretation is in terms of ducting \citep{D79}. The idea is similar to light being guided along a collection of optical fibers. In this model the ducts (or fibers) are sharply-bounded underdense elongated regions that extends over a large distance (or order the solar radius), so that the emission is guided along the magnetic field until the plasma frequency outside the duct has decreased to well below the wave frequency. Reflections off the walls of the duct lead to a systematic depolarization of initially 100\% o~mode emission \citep{M06}. This is also consistent with a systematic partial depolarization of  type~III and type~II bursts. 

Ducting can also account for the effective temperature of thermal radio emission being of order a factor ten smaller than the known electron temperature of the corona. In the ducting model, the apparent source has an area of order ten times larger than the area of the actual source, implying that its (average) brightness temperature is of order ten times smaller than the coronal temperature at the actual source.

\subsubsection{Type I emission}

Type~I emission includes type~I bursts and an associated type~I continuum \citep{E77}, and neither is adequately understood. Early theories 
\citep{T63,Z65,T66} for type~I bursts were type~III-like, in that they invoked a streaming instability. An obvious problem that arises with any type~III-like model is why there is no second harmonic emission. Another obvious difference between type~I and type~III is the polarization. Assuming that type~III F emission is depolarized due to ducting suggests that type~I emission is generated in a region where ducting does not occur. An indication that this is the case is provided by chains of type~I bursts \citep{H66,EU70}. In a type~I--III storm, there is a frequency separation between type~I bursts at higher frequency and type~III bursts at lower frequency. In a chain, type~III and type~I bursts can appear to be correlated, suggesting that a MHD-like disturbance excites both, perhaps due to localized regions of reconnection at the boundary between open and closed field lines.

Type~I continuum and some other radio continua, cf. Figure~\ref{1fig:schematic}, that are attributed to plasma emission have no harmonic component. The lack of frequency-time structure suggests that the Langmuir waves are not generated through a streaming instability. Langmuir waves can be generated through a loss-cone instability \citep{HM85}, but this generates waves at large angles to the magnetic field, leading one to expect that coalescence to produce harmonic emission should occur. An alternative idea is that no instability is involved, and that the Langmuir waves are generated through spontaneous emission by trapped supra-thermal electrons \citep{M80typeI}. The model requires that Landau damping by the supra-thermal electrons be suppressed, and this is possible if the distribution function has a gap between the thermal and supra-thermal electrons, where there are too few electrons to cause significant Landau damping. The model also requires a high level of ion sound waves so that the conversion processes $L+S\to T$, $L\to T+S$ lead to saturation, so that $T_B$ is equal to the effective temperature of the $L$~waves. Even with these assumptions, it is difficult to account for the absence of H emission.

\subsection{\label{sect:IPM}Plasma emission in the interplanetary plasma (IPM)}

One of the original motivations (pre-1970) for spacecraft to carry radio receivers was to study the extension of type~III bursts from frequencies $>10\,$MHz, corresponding to emission from the corona, to lower frequencies that cannot be observed using ground-based instruments. The plasma frequency falls off roughly proportional to $1/r$ in the IPM, implying that a spacecraft at $r=r_0$ can detect F plasma emission only from $r<r_0$, and H emission only from $r<2r_0$. Early spacecraft were at $r\approx1\,$AU, where the plasma frequency is $\approx30\,$kHz. Later spacecraft (Orbiter and Voyager) went to Jupiter and beyond, and it is now known that type~III bursts continue to much lower frequencies, all the way to the heliospheric termination shock. 

In the discussion here,  emphasis is placed on the observations in the 1970s that helped clarify questions associated with plasma emission from the solar corona, including confirmation of the basic theory, the ratio of F to H emission,  Sturrock's dilemma and the role of ion sound waves. 

\begin{figure}[t]
\centerline{
\includegraphics[scale=0.4, angle=0]{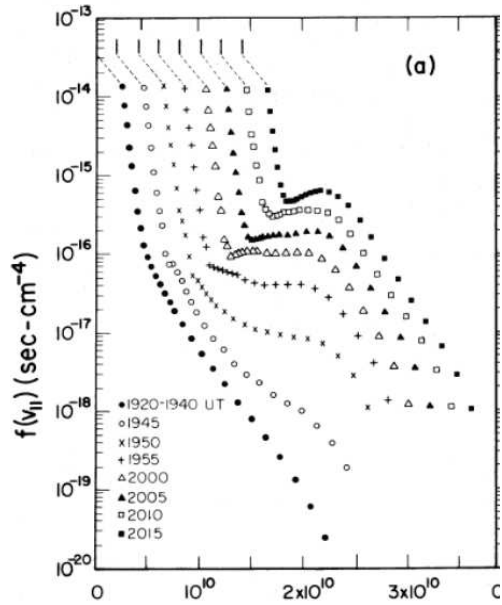}
}
\caption{The 1D distribution function in a type~III event in the IPM showing an increasing distribution function at early times; for clarity, the curves at different times are displaced from each other; the short vertical lines at the top of the figure coincide in the absence of the displacement [from \citet{Letal81}]}
\label{Lin_etal}
\end{figure}

\subsubsection{Type~III bursts in the IPM}

Confirmation of the theory for type~III emission required that the F and H radio emission, the electrons and the Langmuir waves all be observed simultaneously, and that electron distribution be unstable to the growth of Langmuir waves. As expected,  type~III emission was found to extend  down to $30\rm\, kHz$. However, the interpretation of these early results led to uncertainty as to whether the emission is F or H \citep{HA73}. This uncertainty was later resolved by identifying the emission as predominantly F at early times, before the emission peaks, and predominantly H at later times \citep{DSH84}. The early data on the electrons confirmed the direct link with the radio emission. Later measurements of the electron distribution, cf.\ Figure~\ref{Lin_etal}, suggested that it is consistent with that expected when the effect of faster electrons outpacing slower electrons, tending to increase $df(v)/dv>0$, is balanced by quasilinear relaxation \citep{G84}. 

The early observations did not confirm the presence of the Langmuir waves, and there was a hiatus of several year before this discrepancy was resolved. It was realized that the Langmuir waves are extremely intermittent, appearing only in isolated clumps whose filling factor is only a tiny fraction of the volume occupied by the type~III electrons.The energy lost by the electrons to the Langmuir waves is only this tiny fraction of that implied by the (homogeneous) quasilinear model. This resolved Sturrock's dilemma, but raised the obvious question of why the wave growth is so intermittent. One suggestion is that the clumps of Langmuir waves arise from a modulation instability \citep{Z72}, but this is not favored by observations \citep{CR95,CR98}. Another interpretation is that the $df(v)/dv>0$ is sufficiently small that the distribution in only marginally unstable, so that the Langmuir waves grow only under particularly favorable conditions. In a statistical model, called stochastic growth theory  \citep{RCG92,RCG93,RC93}, it is assumed that the growth factor is a random variable, implying that the electric field in a distribution of clumps should satisfy log-normal statistics, and the data are found to be consistent with this prediction. The intermittency also requires that the models for F and H emission and for quasilinear relaxation of the electron distribution be based on  a statistical distribution of clumps of Langmuir waves \citep{MDC86,RCG92,RCG93,RC93}. Quasilinear relaxation due to clumpy Langmuir waves has the same form as for a homogeneous distribution of Langmuir waves \citep{MC89}.

The suggestion that nonthermal $S$ waves play a central role in plasma emission is at best an over-simplication for type~III bursts in the IPM. There are nonthermal density fluctuations with appropriately low frequencies, but their properties are not well described in terms of an ion sound wave. The turbulence in the IPM is consistent with a Kolmogorov spectrum \citep{MV11,Cetal12}, which involves a turbulent cascade from larger to smaller scales. Despite density fluctuations requiring compressibility, the original theory \citep{K41} and later astrophysical models were for an incompressible fluid and  Alfv\'en turbulence \citep{GS95}, respectively. There are more specific models for the density structures, including structures that act like potential wells in the sense that Langmuir eigenmodes get trapped in them \citep{Eetal08,GC13}.

In summary, spacecraft observations of type~III events in the IPM, on the one hand, have confirmed the early theory of plasma emission in a general sense and, on the other hand, have raised many theoretical challenges that are being addressed in the field of space plasma physics. From the point of view of coherent emission, perhaps the most important lesson is the extreme intermittency of the wave growth and the interpretation in terms of marginal stability. In maser terminology, marginal instability applies when the relaxation due to the maser emission occurs on a much faster timescale than the pump that causes the energy inversion. In the present case, the ``pump'' is faster electrons outpacing slower electrons, and this occurs on time and space scales that are very much larger than the scales involved in the instability itself. This is plausibly a general feature of all examples of coherent emission in astrophysical and space plasmas.

\section{Electron Cyclotron Maser Emission (ECME)}
\label{sect:ECME}

Gyromagnetic emission from nonrelativistic electrons is referred to as cyclotron emission. It occurs near harmonics of the cyclotron frequency, $\omega\approx s\Omega_e$, with the intensity decreasing  rapidly with increasing harmonic number $s$. Cyclotron absorption can be negative, and this is the basis for ECME. Early (late 1950s) theories for ECME assumed fundamental ($s=1$) emission in vacuo. Gyromagnetic emission is modified by the presence of a plasma. For plasmas with $\omega_p>s\Omega_e$ cyclotron emission at the $s$th harmonic cannot escape directly. Cyclotron emission at the fundamental is possible in principle only for $\omega_p<\Omega_e$, and the wave dispersion in the plasma further restricts the conditions for effective emission to $\omega_p\ll\Omega_e$.

\subsection{Early versions of ECME}

The earliest theory for ECME was proposed by \citet{T58}, who considered the possibility of negative absorption for three radio emission mechanisms. Negative cyclotron absorption was considered in a model in which the emission is at the relativistic electron gyrofrequency, $\omega=\Omega_e/\gamma\approx\Omega_e(1-v^2/2c^2)$. Other early maser theories \citep{S59,BHB61} were similar to Twiss's theory in three notable ways. First, the intrinsic role played by this relativistic effect, which implies a one-to-one correspondence between the resonant frequency and the speed, $v$, of the electron. Second, the driving term for the maser is an inverted energy population, which in the isotropic case corresponds to a ring distribution, with $d f(v)/d v>0$  below a maximum at $v=v_0$, and $d f(v)/d v<0$  for $v>v_0$. Third, the effect of the plasma on the wave dispersion was ignored. It is interesting that a reactive version of the cyclotron instability was also recognized at about the same time by \citet{G59a}. In an addendum \citet{G59a} included the relativistic term in the gyrofrequency, and this paper became the basis for the subsequent development of the laboratory gyrotron. 

\subsubsection{Effect of plasma dispersion}

Although these early theories for ECME were not motivated by any specific astrophysical application, at about the same time it was recognized that Jupiter's decametric radio emission (DAM) is emitted at the cyclotron frequency. The interpretation of DAM requires a coherent form of cyclotron emission, and in the astrophysical literature this was initially attributed to emission by bunches. Cyclotron emission strongly favors the x~mode over the o~mode, and it was also recognized that it is essential to take the properties of the wave modes into account in a cyclotron model for DAM \citep{E62}. Specifically, in order for cyclotron emission to escape  it needs to be Doppler shifted to above the cutoff frequency for the x~mode \citep{E62,E65}. The cutoff frequency of the x~mode is given by equation (\ref{omegax}), which may be approximated by
\be
\omega_{\rm x}\approx\Omega_e+{\omega_p^2/\Omega_e}
\label{omegax1}
\ee 
for $\omega_p\ll\Omega_e$. An upward Doppler shift by $>\omega_p^2/\Omega_e$ is required. This effect was included in the model for cyclotron emission by bunches, but was neglected in early ECME models for DAM. Specifically, \citet{HB63} suggested that the ECME model of \citet{BHB61} applies to DAM, but this and some later cyclotron maser models \citep{GL-B69} assumed vacuum conditions, and hence ignored the requirement that the emission be at $\omega>\omega_{\rm x}$ in order for it to escape from a cold plasma. Before discussing ECME theories that overcome this difficulty it is relevant to outline the formal theory for ECME and to summarize the properties of DAM and AKR that a theory needs to explain.

\subsection{Absorption coefficient}

Both the early theories of ECME and the later developments of the theory can be treated as special cases of a general theory for the gyromagnetic absorption coefficient. Consider waves in a wave mode $M$, with dispersion relation $\omega=\omega_M({\bi k})$, polarization vector ${\bi e}_M({\bi k})$ and ratio of electric to total energy $R_M({\bi k})$, and electrons with a distribution function $f(p_\parallel,p_\perp)$. The absorption coefficient involves a sum over harmonics, $s$, of the gyrofrequency $\Omega=\Omega_e/\gamma$: 
\bea
\gamma_M({\bi k})&=&
-{2\pi e^2R_M({\bi k})\over\varepsilon_0\omega_M({\bi k})}\sum_s
\int d^3{\bi p}\,
\big|{\bi e}_M^*({\bi k})\cdot
{\bi V}({\bi k},{\bi p};s)\big|^2\nn
\\
&&\qquad\times
\delta[\omega_M({\bi k})-s\Omega-k_\parallel v_\parallel]\,{\hat D}_sf(p_\parallel,p_\perp),
\label{gammaM}
\eea
with $p_\perp=\gamma mv_\perp$, $p_\parallel=\gamma mv_\parallel$ and
\bea
{\widehat D}_s&=&\frac{s\Omega}{v_\perp}{\partial\over\partial p_\perp}+k_\parallel{\partial\over\partial p_\parallel}
=\frac{\omega}{v}\,{\partial\over\partial p}+\frac{\omega\cos\alpha-k_\parallel v}{pv\sin\alpha}
{\partial\over\partial \alpha},
\nn
\\
\ms
{\bf V}({\bf k},{\bf p};s)&=&
\left(
v_\perp{s\over z}J_s(z),iv_\perp J'_s(z),v_\parallel J_s(z)\right),
\qquad
z=\frac{k_\perp v_\perp}{\Omega},
\label{DV}
\eea
where $J_s(z)$ is a Bessel function.

\subsubsection{ECME in vacuo}

In the early theories, the effect of the plasma on the wave properties was ignored, the emission was assumed to be at the fundamental, $s=1$, and the electrons were assumed nonrelativistic. Vacuum dispersion corresponds to replacing $\omega_M({\bi k})$ by $\omega=kc$, $R_M({\bi k})$ by $1/2$, and ${\bi e}_M({\bi k})$ by an arbitrary transverse polarization. A convenient choice of transverse polarizations corresponds to the directions ${\bi k}\times{\bi B}$ and ${\bi k}\times({\bi k}\times{\bi B})$; \citet{BHB61} described these as modes, but they do not correspond to the natural modes of a cold plasma. For emission in vacuo by nonrelativistic electrons the argument of the Bessel functions is small, $z\ll1$, and only the leading term in an expansion of the Bessel functions in $z$ need be retained. With these assumption, the absorption  coefficient for the dominant polarization (denoted x) has the form
\be
\gamma_{\rm x}\propto-\int d^3{\bi v}\,
v_\perp^2\,\delta(\omega-\Omega_e/\gamma-k_\parallel v_\parallel)
{\widehat D}_1f(v_\parallel,v_\perp),
\label{gammax}
\ee
where a nonrelativistic notation for the distribution function is used.

\subsubsection{Drivers for ECME}

Maser action corresponds to negative absorption, and a necessary condition for this is ${\widehat D}_sf>0$. \citet{T58} assumed $k_\parallel=0$, and then the only possible driving term for the maser is $\partial f/\partial p_\perp>0$. The other early authors assumed an isotropic distribution, and then the only possible driving term is $\partial f/\partial p>0$. 

A subtle point is that it is important to include the Lorentz factor $\gamma\ne1$ in the resonance condition,
\be
\omega-s\Omega_e/\gamma-k_\parallel v_\parallel=0,
\label{gyroresonance}
\ee
with the harmonic number $s=1$ here. If one sets $\gamma=1$ in the resonance condition,  then it is trivial to partially integrate with respect to $p_\perp\to mv_\perp$  in equation (\ref{gammax}), and to show that the term $\partial f/\partial p_\perp$ leads only to positive absorption. An interpretation is that even if one has $\partial f/\partial v_\perp>0$ over some range of $v_\perp$, one must have $\partial f/\partial v_\perp<0$ at higher $v_\perp$ in order for $f$ to be normalizable. The implication is that the contribution to positive absorption from $\partial f/\partial v_\perp<0$ at higher $v_\perp$ always dominates any contribution to negative absorption from a region with  $\partial f/\partial v_\perp>0$ at lower $v_\perp$.

A quantum mechanical treatment provides further insight into the need to retain $\gamma\ne1$ in treating negative absorption. The energy eigenstates of an electron in a magnetic field are $\varepsilon_n=(m^2c^4+p_\parallel^2c^2+2neBc^2\hbar)^{1/2}$ in a relativistically correct theory and  $E_n(=\varepsilon_n-mc^2)=\half m v_\parallel^2+n\hbar\Omega_e$ in a nonrelativistic theory, with $n=0,1,\ldots$ the Landau quantum number. The separation between two neighboring states, $n$ and $n-1$ say, is equal for $\hbar\Omega_e$ independent of $n$ in the nonrelativistic case, but depends on $n$ in the relativistically-correct case. The latter effect is sometimes referred to as anharmonicity. In a maser model, it is convenient to re-interpret the electron distribution function in terms of the occupation number $N_n(p_\parallel)$ for the quantum states. A transition between neigboring states $n$ and $n-1$ gives a contribution to the absorption coefficient proportional to $-[N_n(p_\parallel)-N_{n-1}(p_\parallel-\hbar k_\parallel)]$, with negative absorption requiring an inverted energy population, $N_n>N_{n-1}$. In the nonrelativistic approximation, the transitions between any two neighboring states has the same frequency, $\omega=\Omega_e$, and the net absorption coefficient involves a sum over $n$. The transition rate is independent of $n$ and the sum from $n=n_1$ to $n=n_2$ is proportional to $N_{n_1}-N_{n_2}$, which is strictly positive for $n_1\to0$, $n_2\to\infty$. In contrast, in the relativistically correct theory, the transition frequency between $n$ and $n-1$ depends on $n$, due to the anharmonicity, and there is a contribution to negative absorption at this particular frequency for $N_n>N_{n-1}$, with all other neighboring states contributing to absorption at (slightly) different frequencies. The classical counterpart is negative absorption due to $\partial f/\partial v_\perp>0$ at the relevant relativistically-correct gyrofrequency. 

When the strictly nonrelativistic approximation, $\gamma\to1$, is made a maser can be driven only by the terms $\partial f/\partial p_\parallel$, referred to as parallel driven. Such an instability may also be attributed to an anisotropic pitch-angle distribution \citep{SS61}.

\begin{figure}[t]
\begin{center}
\includegraphics[scale=0.4, angle=0]{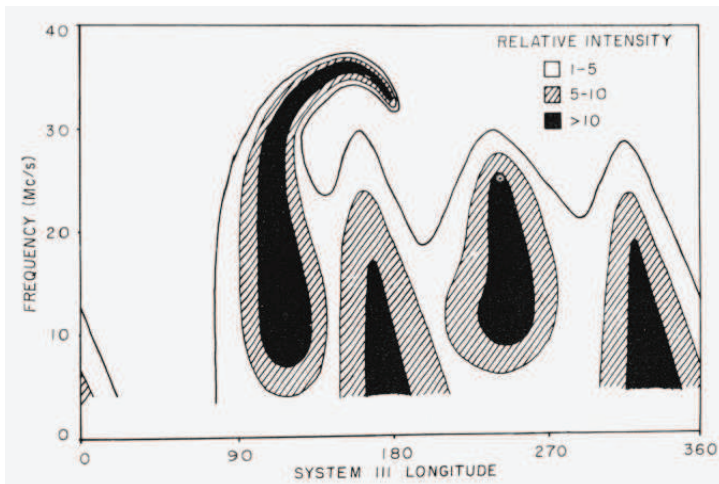}
\caption{A plot showing the average rate of occurrence of DAM as a function of frequency and system~III longitude; the regions of high occurrence rate are identified as four separate ``sources'' of DAM [from \citet{E65}]}
\label{fig:Ellis65}
\end{center}
\end{figure}

\subsection{DAM}

Jupiter was identified  as an intermittent source at decametric wavelengths (DAM) by  \citet{1BF55}, over a decade after the early investigations of solar radio bursts. The early observations and interpretations of Jupiter's radio emission were reviewed by \citet{1W64}. 

\subsubsection{Properties of DAM}

DAM is observed only below about $40\,$MHz. Individual bursts have durations of a few tenths to several seconds, generally increasing with decreasing frequency. DAM is highly circularly polarized, and this suggested electron cyclotron emission \citep{E62,1W64}, rather than plasma emission from the Jovian ionosphere \citep{1Z58}. The identification of both right- and left-hand polarization at frequencies below about $20\,$MHz led to the suggestion that the opposite polarizations originate from opposite hemispheres \citep{1D63}. The probability of observing a DAM burst varies with the central meridian longitude (CML). Four regions of enhanced probability are identified  as DAM ``sources'', as shown schematically in Figure~\ref{fig:Ellis65}. The sources centered around $\lambda_{\rm III}\approx130^\circ$ and $\approx240^\circ$, respectively, where $\lambda_{\rm III}$ is the system III longitude, are associated with the northern and southern magnetic poles. The maximum frequency is interpreted as the cyclotron frequency at the pole, implying that the northern pole has $B\approx1.4\times10^{-3}\,$T, which is nearly twice that at the southern pole.

DAM bursts with two quite different  time scales, referred to as L (long) and S (short) bursts.  L bursts mostly have dura­tions of typically 1--10$\,$s, sometimes extending to 100$\,$s. S burst have time scales $\lesssim1\,$ms. 

\subsubsection{Correlation with Io}

The discovery by \citet{B64} that the probability of observing DAM bursts correlates with the orbital phase, $\phi_{\rm Io}$, of Io (the innermost Galilean satellite) had a major impact on the subsequent observational and theoretical study of DAM. The probability is maximum in two ranges, both around $20^\circ$ wide centered on $\phi_{\rm Io}\approx90^\circ$ and $240^\circ$. The sensitivity to $\phi_{\rm Io}$ led to sources being classified as Io-related and non-Io-related. The Io effect also depends on frequency, with higher frequencies being more strongly Io-related than lower frequencies.

A physical interpretation of the Io effect was provided in the late 1960s \citep{PD68,GL-B69}. The idea is that because Io is a good conductor the ``Io flux tube'', defined by the magnetic field lines from Jupiter that intersect Io, becomes frozen in and moves at the angular frequency corresponding to Io's Keplerian motion. This implies that the Io flux tube is dragged backwards through the corotating Jovian magnetosphere. This leads to an electric field, equal to $-{\bi v}\times{\bi B}$, where ${\bi v}$ is the velocity of the Io flux tube relative to the corotating magnetosphere and ${\bi B}$ is the Jovian magnetic field at Io, leading to a potential difference of about 2$\,$MV across Io. It is assumed that electrons are accelerated along the magnetic field lines by this potential, in a manner analogous to that for auroral electrons in the terrestrial magnetosphere, and that the emission process for DAM is due to ECME by these electrons. The general features of this model for the Io effect were later confirmed by observations during Pioneer and Voyager flybys of Jupiter.

\begin{figure}[t]
\begin{center}
\includegraphics[scale=0.4, angle=0]{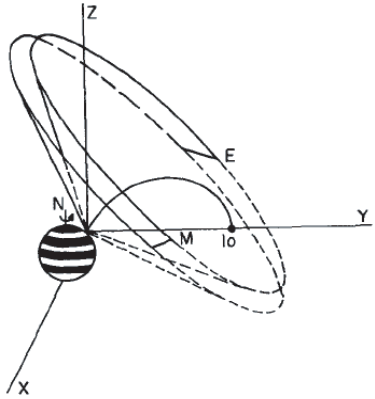}
\caption{Schematic drawing of the conical sheet of Io-related radiation leaving Jupiter's vicinity  [from  \citet{Dulk67}]}
\label{fig:dulk67}
\end{center}
\end{figure}

\subsubsection{Emission pattern of DAM}

The preferred ranges of CML and of $\phi_{\rm Io}$ provide strong geometric constraints on an acceptable model for DAM. An early model that accounts for most of the geometric features was proposed by \citet{Dulk67}. The interpretation of the geometry implies a seemingly bizarre emission pattern, confined to the narrow surface of a wide cone with its axis along the magnetic field direction, as illustrated in Figure~\ref{fig:dulk67}.  

\begin{figure}[t]
\begin{center}
\includegraphics[scale=0.4, angle=0]{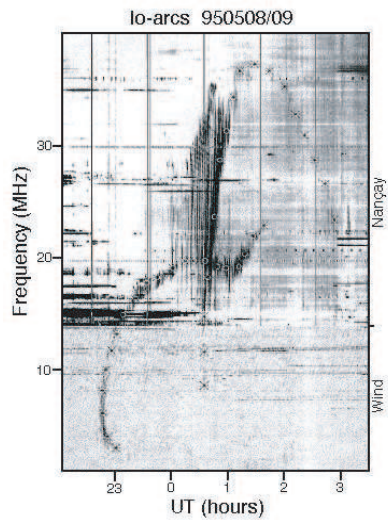}
\caption{Io-related emission in DAM: lower half from the Wind spacecraft, and upper half from ground-based observations (Nan\c cay); the lower arc is from the southern hemisphere, and the upper arc extending to 38 MHz is from the northern hemisphere [from \citet{1QZ98}]}
\label{figIorelated}
\end{center}
\end{figure}

Further evidence of the emission pattern of DAM was provided during Pioneer and Voyager flybys. A prominent feature in the observed radio emission is that enhanced emission occurs in arc patterns in the frequency-time plane. The interpretation of these Jovian decametric arcs requires highly structured emission, similar to the pattern in Figure~\ref{fig:dulk67}.  Figure~\ref{figIorelated} shows a combination of  spacecraft data at lower frequencies and ground-based data at higher frequencies; the arc-like structure at lower frequencies joins on continuously to a high-frequency feature that is known to be Io related.

\subsection{The Earth's AKR}

The early spacecraft launched to observe solar radio bursts found two classes of ``Earth noise'', one of which was identified as plasma emission from electrons accelerated at the Earth's bow shock. The other component, e.g. as reviewed by \citet{T06}, was studied in detail by \citet{G74} and called terrestrial kilometric radiation; the name was subsequently changed \citep{Ketal75} to auroral kilometric radiation (AKR). AKR correlates with ``inverted-V'' electron precipitation events, and it is assumed that the radiation is generated by these electrons. There are obvious analogies and differences between AKR and DAM, and the understanding of both has been enhanced by comparing them. 

\begin{figure}[t]
\begin{center}
\includegraphics[scale=0.3, angle=-90]{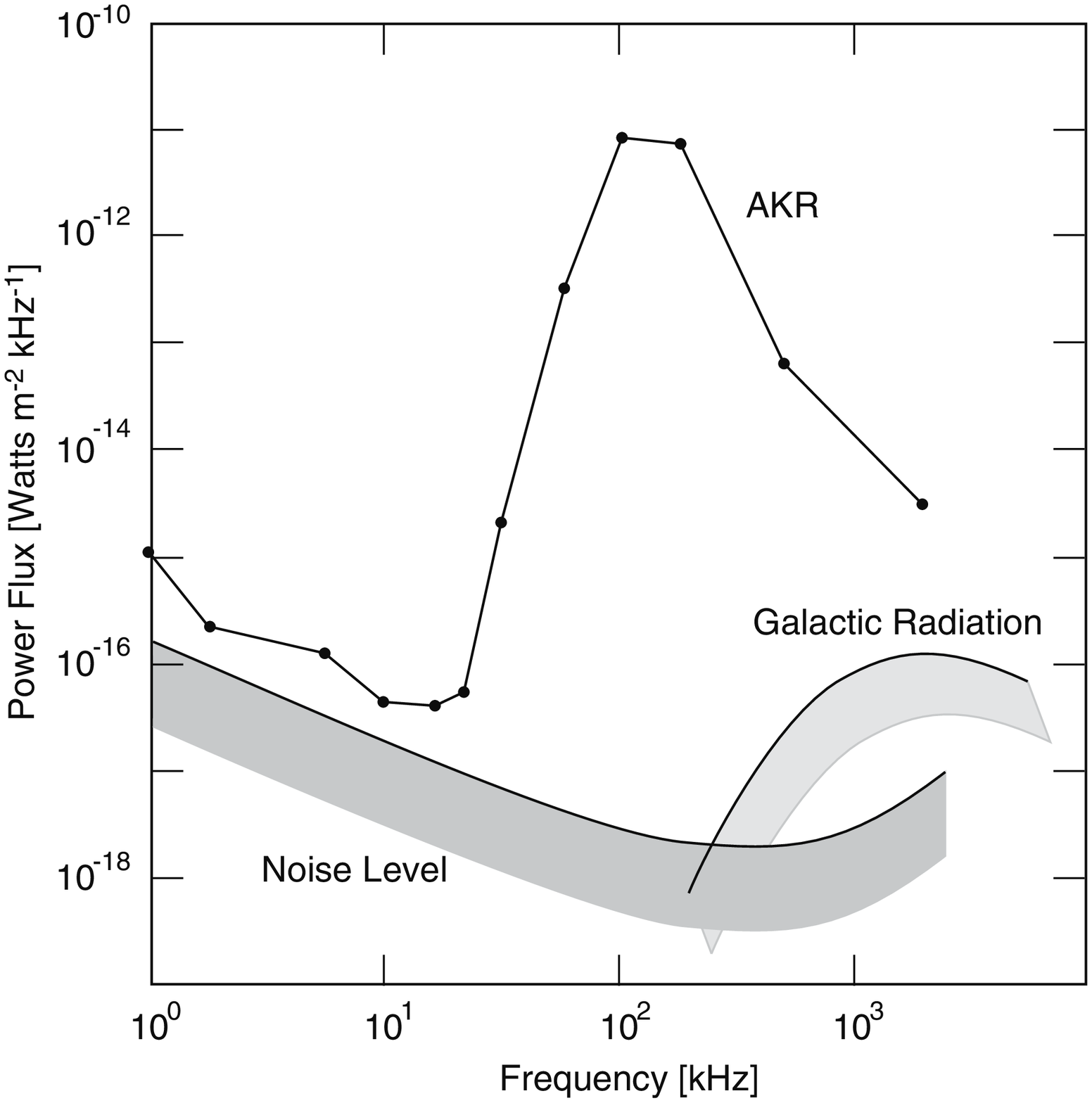}
\caption{Spectrum of AKR [from \citet{G74} modified by \citet{T06}]}
\label{fig:AKRspectrum}
\end{center}
\end{figure}

\subsubsection{Properties of AKR} 

AKR is one signature of a substorm; substorms occur  during geomagnetic storms, which can last for days, and are powered by energy transfer from the solar wind to the geomagnetic field. During a substorm, of duration around $10^3\,$s, the total power released, of order $10^{11}\,$W, can be attributed to the rate work is done by a current of a few times $10^6\,$A against an electric field with a potential of a few times $10^4\,$V. A small fraction of this power, $10^7\,$W to $10^9\,$W, appears in AKR. The main energy release involves magnetic reconnection and redirection of current in the Earth's magnetotail, at $\gtrsim10\,R_E$, where $R_E$ is the Earth's radius. Energy is transported Alfv\'enically to the source region for AKR at 2--4$\times10^3\,$km above the Earth on auroral field lines.

AKR has a frequency range of about 50 to 500$\,$kHz, as illustrated in Figure~\ref{fig:AKRspectrum}. The highest frequency corresponds to the cyclotron frequency of electrons above the auroral region. The dominant polarization corresponds to the x~mode in the source region, as expected for ECME, although there is a small admixture of o~mode. The brightness temperature is high, $T_B\gg10^{10}\,$K, with very high values implied by fine structures from very small regions.

\begin{figure}[t]
\begin{center}
\includegraphics[scale=0.3, angle=0]{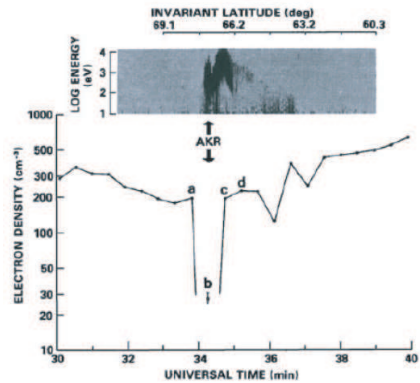}
\caption{Auroral density cavity: data from the ISIS 1 spacecraft showing that the plasma density decreases sharply by a large factor as the spacecraft enters the region where inverted-V electrons and AKR are observed [from \citet{BCK80}]}
\label{fig:cavity}
\end{center}
\end{figure}

\subsubsection{Auroral density cavity}

A surprising feature of these early observations was that AKR generation occurs within an auroral density cavity  \citep{BC79,C81}, as illustrated in Figure~\ref{fig:cavity}. The subsequent interpretation is that an upward-directed field-aligned electric field, $E_\parallel$, with a total potential drop of a few kV, creates the cavity by removing all the thermal electrons. The only electrons remaining in the cavity are those accelerated to a few keV by $E_\parallel$.  During a substorm, the auroral plasma consists of many cavities with a range of widths, generally extending further in longitude than in latitude, confined latitudinally by dense plasma walls. AKR occurs only in regions with $\omega_p/\Omega_e<0.14$  \citep{H92}. The density depletions can extend up to several Earth radii \citep{Aetal15}.

\begin{figure}  
\begin{center}
   \includegraphics[scale=0.6, angle=0]{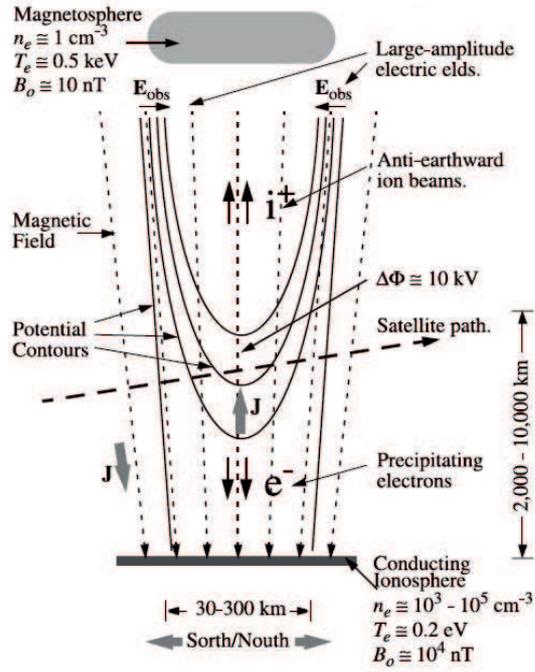}
\caption{Idealized model  for acceleration of auroral electrons by a parallel electric field [from  \citet{Eetal00}]}
\label{Fig:ergun}
\end{center}
\end{figure}

The inverted-V spectrum is interpreted in terms of the potential drop having its maximum in the center of the cavity, so that the energy of the precipitating electrons is maximum at the center, and minimum at the edges of the cavity, as illustrated in Figure~\ref{Fig:ergun}.

Reflection of the ECME from the cavity walls has been invoked for several different reasons.  \citet{C82} suggested that the reflection could act like that at the partially reflecting ends of a laboratory laser, so that a wave propagates backwards and forwards many times through the amplifying region before escaping.  \citet{HM86} argued that partial reflection and transmission at oblique angles for incident x~mode radiation results in a mixture of the two modes, and suggested that observed o~mode emission could be produced in this way. \citet{Eetal00} invoked reflections to account for ducting along the field lines to heights where the x~mode emission is above $\omega_{\rm x}$ in the surrounding medium. A detailed discussion of ducting in various wave modes was given by  \citet{C95}.

\subsubsection{Distribution of electrons generating AKR}

The correlation between AKR and ``inverted-V'' electron precipitation events allowed a direct test of models for ECME. The properties of inverted-V electrons were studied with improving resolution  over several decades. A notable feature of the electron distribution had been recognized from ground-based observations prior to the discovery of AKR: the electron distribution is approximately mono-energetic \citep{E68}, more specifically the electron distribution has a relatively sharp peak, at an energy $\varepsilon_m$ say.  ``Inverted-V'' describes the shape of the dynamic spectrum (electron energy versus time) as a spacecraft passes through the region where the electrons are precipitating: the peak at $\varepsilon_m$ is a function of position (and hence of time at the spacecraft) increasing from the edge of the density cavity to a maximum in the center of the region. Subsequently more detailed data led to the electron distribution function being described as a ``shell'' or ``ring'' reflecting the shape of the contours in velocity space, cf.\ Figure~\ref{Fig:ellipse}. Early spacecraft data on the electron distribution also indicated a one-sided loss-cone feature \citep{1EHR79}, in the sense that there is an absence of upward directed electrons with small pitch angles. Early theories for ECME in AKR assumed that the maser is driven by the loss-cone feature. However, during the 1990s it became increasingly evident that the shell- or ring-type feature can drive a maser, and this driving term seems to be the more important in AKR \citep{Eetal00,BC00}. This led to the horseshoe-driven ECME models discussed below.

It is helpful to separate cyclotron instabilities into two classes, referred to as parallel-driven and perpendicular-driven \citep{M86}, depending on whether the relativistic term in the gyrofrequency is neglected or not. The absorption coefficient (\ref{gammaM}) depends on both $\partial f/\partial v_\parallel$ and $\partial f/\partial v_\perp$. When the relativistic term is neglected, the resonance condition does not depend on $v_\perp$, and one may partially integrate the latter term with respect to $v_\perp$ and show that it contributes positively to the absorption coefficient. An instability is still possible, and is usually attributed to an anisotropy $v_\perp\partial f/\partial v_\parallel\ne v_\parallel\partial f/\partial v_\perp$ \citep{SS61}. When the relativistic term is included, negative absorption can be driven by $\partial f/\partial v_\perp>0$, as in the early models for ECME discussed above. An analogous separation applies to the two reactive forms of cyclotron instability identified by \citet{G59a}, which may be attributed to  axial and azimuthal self-bunching which do and do not, respectively, depend on the relativistic term \citep{W83}.

\subsection{Applications of ECME}

Early suggestions that DAM is due to ECME \citep{HB63,G73,M73,M76} invoked forms of ECME that did not readily account for the observed features of DAM. A loss-cone driven model \citep{WL79} became widely accepted for DAM and for AKR. Subsequent observations of the electron distribution that drives AKR strongly favored a horseshoe-like distribution,  leading to a horseshoe-driven version of ECME becoming the favored interpretation for AKR.

\subsubsection{Resonance ellipse}

A useful concept in discussing ECME is a graphical interpretation of the gyro\-resonance condition
\be
\omega-s\Omega_e/\gamma-k_\parallel v_\parallel=0,
\label{gres}
\ee 
which is the condition for an electron, with given $v_\perp,v_\parallel$, to resonate with a wave, with given $\omega,k_\parallel$ at the $s$th harmonic. When plotted in $v_\perp$-$v_\parallel$ space for given $\omega,k_\parallel,s$, equation (\ref{gres}) defines a resonance ellipse \citep{OG82,MRH82,M86}. The ellipse (actually a semi-ellipse with the region $v_\perp<0$ unphysical) is centered on the $v_\parallel$-axis, at a point $\propto k_\parallel$, with its major axis along the $v_\perp$-axis. The physical significance of the ellipse is that the absorption coefficient, which must be negative for ECME to occur, can be written as a line-integral around the ellipse. For a given distribution function, this allows one to identify the most favorable ellipse as the one that maximizes the negative contribution to the absorption coefficient. For cases of relevance here, the dominant driving term is $\propto\partial f/\partial v_\perp>0$, and the largest growth rate corresponds to the ellipse that maximizes the (weighted) contribution from this term.

The maximum growth rate for a ring distribution corresponds to the ellipse reducing to a circle centered on the origin, which corresponds to $k_\parallel=0$. The maximum contribution from $\partial f/\partial v>0$, sampled around the circle, corresponds to a speed $v$ slightly less than $v_0$. It follows that the most favorable case for ECME driven by a ring distribution is for emission perpendicular to the field lines, $\theta=\pi/2$, at $\omega\approx\Omega_e(1-v_0^2/2c^2)$. {\br Such ECME is possible at slightly backward angles \citep{Setal14}.} In this case the line-integral reduces to the integral over pitch angle, with all pitch angles contributing.

\begin{figure} [t]
\begin{center}
\includegraphics[width=0.5\textwidth]{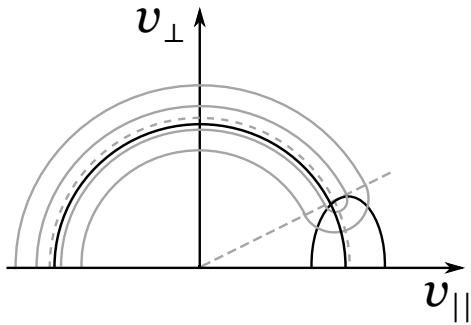}
\caption{Two resonance ellipses are illustrated for a horseshoe distribution; one is circular, corresponding to perpendicular emission, through the region just below $v=v_0$ where the ring distribution has its maximum positive value of $\partial f/\partial v$; the other indicates the ellipse that gives the maximum contribution from $\partial f/\partial v_\perp$ associated with the loss-cone feature [from \citet{MW16}]}
   \label{Fig:ellipse}
   \end{center}
   \end{figure}

\subsubsection{ECME theories for DAM and AKR}

\citet{M73} proposed a parallel-driven version of ECME and \citet{M76} applied this to both DAM and AKR. The model relies on an anisotropic distribution as the driver \citep{SS61} and this requires $\omega_p\ll\Omega_e$ and  an extreme form of anisotropy. There was little evidence for $\omega_p\ll\Omega_e$, but this condition was later confirmed  for AKR \citep{GG78} and is well satisfied within auroral cavities \citep{BC79,C81}. However, the required extreme form of anisotropy was not confirmed. \citet{WL79} proposed a perpendicular-driven version for ECME that includes both the relativistic correction to the cyclotron frequency and the Doppler shift to $>\omega_{\rm x}$.  This version is driven by $\partial f/\partial v_\perp>0$ due to a loss-cone anisotropy. An attractive feature of the loss-cone-driven maser is that upward-directed electrons cause upward-directed waves to grow, and this implies a positive Doppler shift to above $\omega_{\rm x}$. This loss-cone-driven maser became the preferred version of ECME for well over a decade. It was applied not only to DAM and AKR, but also to solar spike bursts \citep{HEK80}{\br, type~V bursts \citep{WD86}} and to radio emission from flare stars \citep{MD82a}. 

\subsubsection{Horseshoe-driven ECME}

Although early data on the electron distribution of the inverted-V electrons gave general support to the loss-cone driven model for AKR,  later data implied a horseshoe electron distribution \citep{Eetal00,BC00}, which may be regarded as a ring distribution with a one-sided loss cone. Models for ECME were modified to take this into account, and horseshoe-driven ECME became the favored mechanism for AKR. Negative absorption due to  a horseshoe distribution includes contributions with $\partial f/\partial v_\perp>0$ from both the ring-type feature and the loss-cone feature. As indicated in Figure~\ref{Fig:ellipse}, the most favorable ellipse is a circle for a ring distribution, corresponding to perpendicular emission, $\theta=\pi/2$, and the most favorable ellipse for the loss-cone feaure is for an oblique angle $\theta\ne\pi/2$.  This is also the case for the ring-type feature in a horseshoe distribution.  Maximum growth for ECME due to the ring-type feature is for $\theta\approx\pi/2$ and at a frequency $\omega\approx\Omega_e(1-v_0^2/2c^2)$. The contribution to the growth rate from the loss-cone feature is smaller, and is usually neglected. However, it is relevant to note the loss-cone feature leads to growth of qualitatively different radiation from the ring feature, notably emission at an angle $\theta$ significantly different from $\pi/2$ and Doppler shifted to $\omega>\omega_{\rm x}$, as in the \citet{WL79} model.

In treating horseshoe-driven ECME the requirement $\omega>\omega_{\rm x}$ is assumed not to be relevant. The argument is that when $\omega_p$ is sufficiently low (and the plasma is sufficiently hot) vacuum-like wave dispersion applies. Specifically, the stop band between magnetoionic z~and x~modes is assumed to be washed out, so that emission below the cyclotron frequency can escape. The observational evidence is that the auroral cavity is essentially devoid of thermal plasma so that vacuum-like dispersion applies \citep{Petal02,Setal14}. The emitted radiation is assumed to be ducted upward, by reflection from the cavity walls, until it reaches a height where it can escape. Escape is possible at a height were the cutoff frequency, $\omega_{\rm x}$, of the x~mode outside the flux tube is below the wave frequency.

\subsection{Formation of a horseshoe distribution}

The  formation of a horseshoe distribution is attributed to acceleration of electrons by $E_\parallel$ along converging magnetic field lines \citep{T06}.

\subsubsection{Generation of $E_\parallel$} 

A qualitative description of how acceleration of auroral electrons occurs is implicit in Figure~\ref{Fig:ergun}: the electrons experience a potential drop, $-\Phi$ which increases their energy by $e\Phi$. The energy source for this acceleration is associated with magnetic reconnection in the Earth's magnetotail, resulting in the released energy propagating downward as an Alfv\'enic Poynting flux. The energy transport and the acceleration of precipitating electrons occurs in an upward current region, with the current generating a magnetic field, ${\bi B}_\perp$ perpendicular to the Earth's magnetic field ${\bi B}_0$. A cross-field potential is imposed in the source region, and this produces the ${\bi E}_\perp$ such that ${\bi E}_\perp\times{\bi B}_\perp/\mu_0$ is the Poynting flux. The ionosphere is a good (cross-field) conductor with a high mass density such that the magnetic field is line-tied. This requires that the field-aligned potential surfaces high in the magnetosphere close across field lines somewhere above the ionosphere, implying that the cross-field potential becomes a field-aligned potential. The electron acceleration is attributed to the resulting $E_\parallel$.

This simple model suggests $E_\parallel$ that varies only over the relatively large scales shown in Figure~\ref{Fig:ergun}. However, the observed $E_\parallel$ consists of localized propagating structures \citep{Metal80,Betal88}. The interpretation of the observed localized structures is a long-standing problem \citep{B93}. Suggested interpretations include electrostatic shocks \citep{Metal80}, double layers \citep{B72,R89} and phase-space holes \citep{Schamel86,Netal01,Tetal11}. There is strong evidence for an association of $E_\parallel$ with  dispersive Alfv\'en waves \citep{Cetal15}.

 \begin{figure}    
\begin{center}
\includegraphics[width=0.5\textwidth]{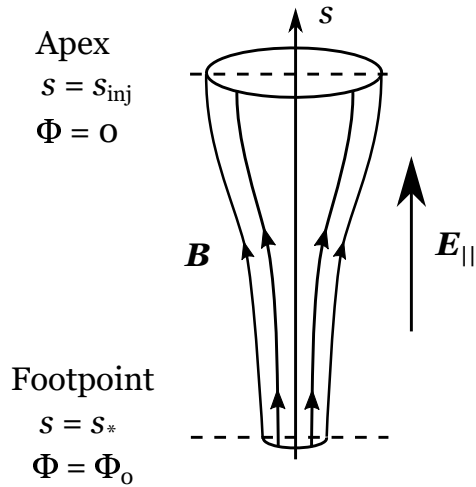}
 \caption{The 1D model is illustrated as a vertical flux tube between the chromosphere (bottom) and the apex of the flux tube (top), magnetic field lines are shown to diverge, as $B$ decreases from $s=s_*$ to $s=s_{\rm inj}$ [from \citet{MW16}]}
   \label{Fig:geometry}
   \end{center}
   \end{figure}

\subsubsection{1D Model}

\citet{MW16} showed that a one-dimensional (1D) model for the electron motion in a flux loop leads to a horseshoe distribution when the following conditions are satisfied, cf.\ Figure~\ref{Fig:geometry}. (a) In the injection region for the electrons, assumed to be around the apex of the flux loop (where $B$ is minimum),  pitch-angle scattering maintains an isotropic distribution. (b) Outside the injection region, no pitch-angle scattering occurs so that the magnetic moment, $\mu=mv_\perp^2/2B$, of an electron is conserved. (c) The total energy, ${\cal E}$, which is the sum of the kinetic energy, $\half m(v_\perp^2+v_\parallel^2)$, and the potential energy, $-e\Phi(s)$, is conserved as $-\Phi(s)$ changes from its value (assumed zero) in the injection region, $s=s_{\rm inj}$ say, to its maximum, $\Phi_0=\half mv_0^2$, just above the ionosphere, at $s=s_*$ say. (d) The speed $v_0$ is assumed much greater than the typical speed of the electrons in the injection region. 

The electron distribution function in this model can be inferred using Liouville's theorem, which becomes trivial when the distribution function can be written in terms of the two constants of the motions, ${\cal E}$ and $\mu$. The distribution function, $f(v_\perp,v_\parallel,s)$, at the injection point, $s=s_{\rm inj}$ say, is written as a function, $F({\cal E},\mu)$ say, of the two constants, and the distribution function at any other point  is equal to $F({\cal E},\mu)$, with ${\cal E}$ and $\mu$ re-expressed in term of $v_\parallel$ and $v_\perp$ at $s$. The assumptions that the only source of electrons is at $s=s_{\rm inj}$ and that the distribution function is isotropic there implies that $F({\cal E})$ does not depend on $\mu$. The argument is that ${\cal E}$ does not depend on pitch angle whereas $\mu$ does. It  follows that the distribution function, $f(v,s)$ say, at any other point does not depend on pitch angle. This corresponds to an isotropic ring distribution, peaked around the speed $v=[-2e\Phi(s)/m]^{1/2}$ with a spread determined by the spread in the injection region.

A minor complication is that there are two solutions for $v_\parallel$ in terms of ${\cal E}$ and $\mu$ and one needs to introduce separate functions, $F_\pm({\cal E},\mu)$, corresponding to downgoing and upgoing electrons, respectively. Moreover, downgoing electrons with $\mu>\mu_{\rm L}$ mirror and become upgoing electrons, whereas downgoing electrons with $\mu<\mu_{\rm L}$ precipitate into the atmosphere and are lost, where $\mu=\mu_{\rm L}$ corresponds to the loss cone. This corresponds to a sink, or negative source term, at the precipitation point.  Following \citet{MW16}, we separate $F$ into $F_\pm$ corresponding to down- and up-going electrons and make a further separation by writing
\be
F({\cal E},\mu)=F_+({\cal E},\mu)+F_-({\cal E},\mu),
\qquad 
F_\pm({\cal E},\mu)=F^>_\pm({\cal E},\mu)+F^<_\pm({\cal E},\mu),
\label{Hmu3b}
\ee
where the superscripts indicate $\mu>\mu_{\rm L}$ and $\mu<\mu_{\rm L}$, respectively. The downgoing electrons have $F^>_+=F^>_-$ and there are no upward propagating electrons in the loss cone, $F^<_+=0$.

This model implies an idealized horseshoe  distribution that is independent of pitch angle except for a one-sided loss cone. The observed one-sidedness of the loss cone implies that the electrons are not bouncing back and forth between mirror points in the two hemispheres, and this is built into the model through the assumed efficient pitch-angle scattering in the injection region. Electrons that return to this region are isotropized before leaving it again.

\subsection{Pump in horseshoe-driven ECME}

The foregoing 1D model leads to a simple interpretation of the ``pump'' for (the ring-driven component in) horseshoe-driven ECME. Acceleration by $E_\parallel$ tends to cause the ring-like feature to develop, and the (quasilinear) back reaction to the ECME tends to smooth out the ring feature. The energy in the ECME then comes from the acceleration by $E_\parallel$.

The back reaction to maser emission may be described using quasi-linear theory. Numerical treatments for AKR \citep{P86,Petal02,KV12} show that the back reaction tends to drive the electrons to lower energy, tending to decrease the positive values of the distribution function in velocity space and hence to suppress the instability. If suppression did occur, the back reaction would lead to substantial modification of the distribution function.  However, observation of a horseshoe distribution in the magnetosphere implies that any modification that is due to the back reaction is only small. As in the case of type-III bursts, it is plausible that highly intermittent wave growth occurs and that the back reaction to the statistically large number of localized bursts of growth maintains the distribution close to the marginally stable state.

The evidence that AKR occurs in a density cavity and is due to horseshoe-driven ECME is very strong. Specifically, the growth appears to be driven by the positive gradient of the distribution function at $v<v_0$ associated with the ring feature, rather than by the gradient in $v_\perp$ at $\alpha<\alpha_{\rm L}$ associated with the loss-cone feature, cf.\ Figure~\ref{Fig:ellipse}. However, whether this form of ECME also applies to other accepted and suggested applications of ECME is unclear.

\subsubsection{Is DAM horseshoe driven?}

It is plausible that the acceleration of the electrons that generate DAM is due to an $E_\parallel$ associated with kinetic Alfv\'en waves \citep{G83}, analogous to the acceleration of inverted-V electrons. This would appear to favor horseshoe-driven ECME. However, there are two observational features that suggest that DAM is due to loss-cone-driven ECME rather than ring-driven ECME.

The bizarre radiation pattern in Io-related DAM \citep{Dulk67}, cf.\ Figures~\ref{fig:dulk67} and~\ref{figIorelated}, seems to be consistent with loss-cone-driven ECME \citep{HMR81}. It is not consistent with ring-driven ECME, which leads to emission at $\theta\approx\pi/2$. Observation by \citet{DLL92} showed the polarization to be intrinsically elliptical, with an axial ratio consistent with cyclotron emission in vacuo at the angle implied by the radiation pattern. This suggests that the emission occurs in a region where the plasma density is intrinsically very low, unlike the density cavity in the AKR source region.Assuming loss-cone driven ECME does operate in DAM, as these observations suggest, raises the questions as to whether the distribution has a horseshoe form and, if so, why the loss-cone feature appears to dominate in DAM, whereas the ring feature dominates in AKR. 

\subsubsection{Application for solar and stellar emissions}

{\br As already remarked, ECME has been proposed as the emission mechanism of solar spike bursts, e.g., the review by \citet{FM98}. Assuming the ECME interpretation is correct, the question arises as to which form of ECME operates.} Are solar spike bursts (and coherent emission from flare stars) due to horseshoe-driven ECME? This question was discussed by \citet{MW16}, who concluded that there is no compelling argument against horseshoe-driven ECME in a solar or stellar flare. Acceleration by $E_\parallel$, that occurs in inverted-V events associated with AKR, is plausible as the acceleration mechanism for the precipitating electrons that produce solar hard X-ray bursts and spike bursts. Moreover, this acceleration can also plausibly result in a density cavity similar to that in the AKR source region. However, these suggestions are relatively new, and need further critical discussion. {\br Similarly, in other suggested astrophysical applications of ECME \citep{BER05,Betal13} it is important to distinguish between ring-driven and horseshoe-driven versions, with only the former requiring an extreme density cavity and the latter requiring only the weaker condition $\omega_p\ll\Omega_e$.
}

\section{Pulsar Radio Emission}
\label{sect:pulsar}

Pulsar radio emission is a third form of coherent emission, but unlike plasma emission and ECME there is no consensus on what the radio emission mechanism is. One can identify several reasons why the emission mechanism remains poorly understood. However, before discussing such reasons, it is necessary to understand the general theoretical framework, including pulsar electrodynamics, the properties of the ``pulsar plasma'' that populates the relevant regions of a pulsar magnetosphere, and the properties of the wave modes of a pulsar plasma. {\br It is of particular interest from the plasma-physics viewpoint to understand how familiar plasma physics concepts and methods need to be modified and adapted to the extreme environment of a pulsar magnetosphere.}

\subsection{Background on pulsars}

Pulsars were discovered in 1967, and there are now over 2000 known radio pulsars. Pulsars are strongly magnetized, rapidly rotating neutron stars  created as the compact remnants of supernova explosions \citep{MT77,M91,GBI93,M99,LK04,LG-S06}. However, important details needed to interpret the radio data are poorly determined. These include the obliquity angle, $\alpha$, between the rotation and magnetic axes, the angle, $\zeta$, between the line of sight and the rotation axis, and the location of the source region of the radio emission. Another important detail that is clear for plasma emission and ECME but not for pulsar emission is the relation between the source location and the frequency of the emission: is there a tight ``radius-to-frequency mapping'' or is the emission from one location relatively broad band?  There is a dilemma: one needs to understand the radio emission mechanism in order to use the radio data to determine these and other parameters, but one needs to have a reliable model that includes these parameters in order to identify the radio emission mechanism.

\subsubsection{Classification of pulsars}

\begin{figure}[t]
\begin{center}
\includegraphics[scale=0.5, angle=-90]{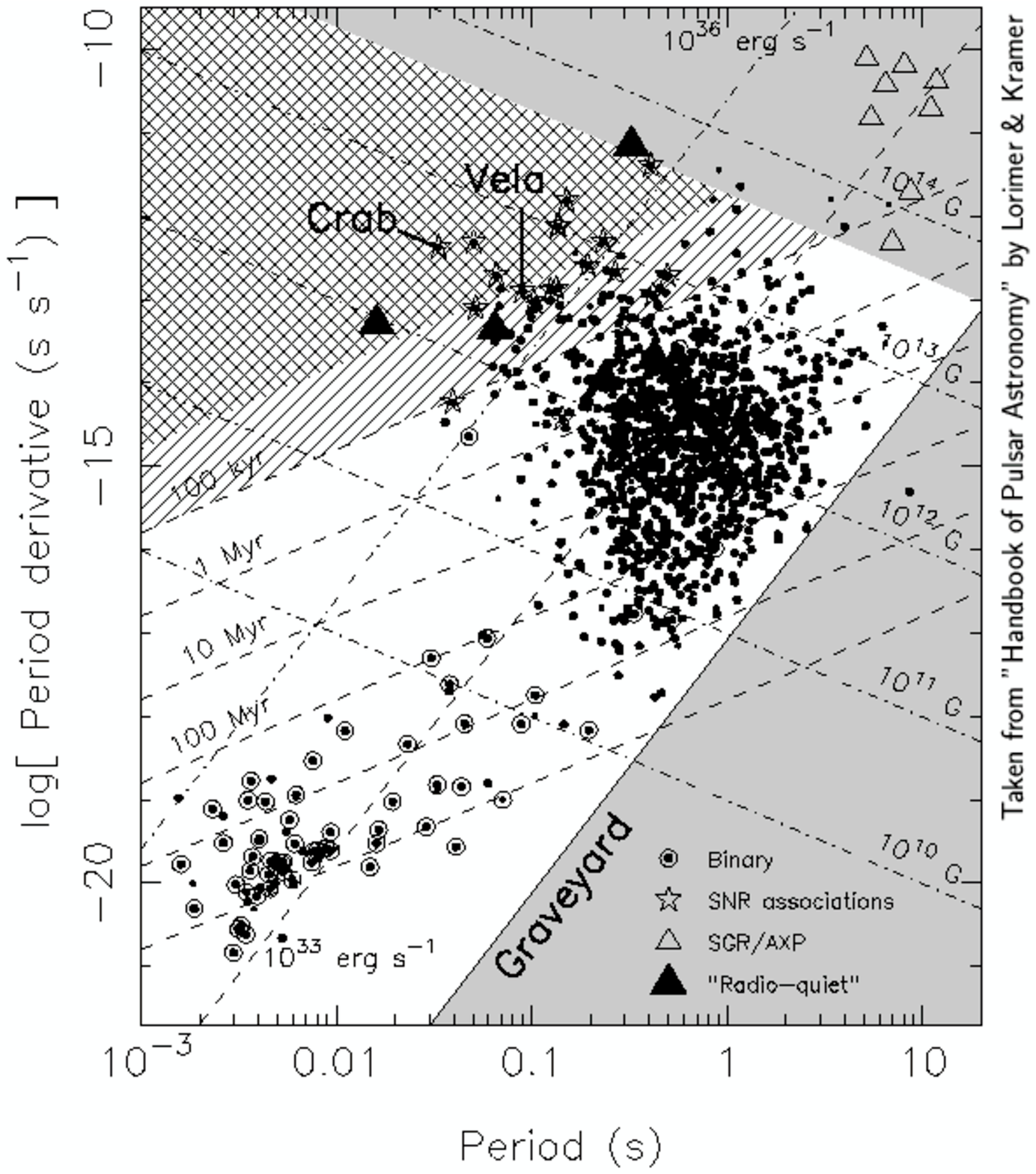}
\caption{On a $P$--$\dot P$ plot, young pulsars (including Crab and Vela) are on the upper left, old pulsars  are in the middle moving towards the ``graveyard'' as they age, recycled (millisecond) pulsars are on the lower left, with a circle indicating a companion star, and magnetars are on the upper right; lines corresponding to the indicated values of the surface magnetic field, $B_*\sin\alpha=3.2\times10^{19}(P{\dot P})^{1/2}\rm G$, and the characteristic ages, $P/2{\dot P}$ [from \citet{LK04}] }
\label{PdotP}
\end{center}
\end{figure}

Two basic parameters are measured for each pulsar: the pulse period, $P$, and its rate of change, ${\dot P}$, which determine the rotation frequency, $\omega_*=2\pi/P$ and the slowing down rate, ${\dot\nu}=-{\dot P}/P^2$, of the star.  In the vacuum-dipole model, as discussed below, the spin-down power (the rate of loss of rotation energy) is equated to the power in magnetic dipole emission, and this provides the basis for the interpretation of the distribution of pulsars in the $P$--${\dot P}$ plane, cf.\ Figure~\ref{PdotP}, in terms of the surface magnetic field, $B_*\propto(P{\dot P})^{1/2}$, and the age of the pulsar, $P/2{\dot P}$. Pulsars separate into three classes. Normal pulsars spin rapidly when they are young ($\lesssim10^4\,$years), slowing down as they age, until they disappear from view (into the ``graveyard'') after about $10^7$ years. Recycled (or millisecond) pulsars are old pulsars with weak magnetic fields that have been spun up in a binary system. Magnetars are pulsars with exceptionally strong magnetic fields, rotating relatively slowly.  

Although there are differences in the properties of the radio emission from the different classes of pulsars, the similarities are more remarkable than the differences. For example, despite the very wide ranges of the parameters $P$ and ${\dot P}$, most radio pulsars are observed over a relatively narrow frequency range, between about $100\,$MHz and $10\,$GHz. Whatever parameter determines the natural frequency of pulsar radio emission, it cannot be a strong function of either $P$ or ${\dot P}$. 

A subset of pulsars are observed to have pulsed high-energy (X-ray and gamma-ray) emission, which is due to incoherent emission by highly relativistic electrons.  There is an observed square-root relation between the power in high-energy emission and the spin-down power (there is no analogous relation for the radio emission). The best available estimates of $\alpha$ and $\zeta$ are obtained by combining models for both the high-energy emission and for the radio emission \citep{Petal15}, but the resulting values remain subject to considerable uncertainties. 

\subsubsection{General description of pulsar emission}

Some general features relevant to the interpretation of pulsar radio emission include the following.
\begin{itemize}
\item Beaming: The pulsing is interpreted in terms of a ``lighthouse'' model: relativistic beaming restricts the emission to nearly tangent to magnetic field lines such that a pulse (in radio or high-energy emission) is observed each time the beam sweeps across the line of sight to the observer \citep{RC69}. The ``pulse window'' is the range of (rotational) phase during which an observer can see emission.  
\item Integrated pulse profile: Most pulsars are not bright enough for individual pulses to be studied in detail, and the pulse profile is built up by folding many pulses together. The resulting (integrated) pulse profile is generally very stable.  
\item Polar Cap: The magnetic field is assumed to be approximately dipolar, with the (``open'') field lines that extend beyond the light cylinder, at $r_{\rm L}=c/\omega_*$, defining polar caps around the two poles. The radio emission is assumed to come from within the polar cap, maybe near its boundary defined by the last closed field lines. Some young pulsars have an inter-pulse, usually interpreted as emission from the conjugate polar-cap region. 
\item Linear polarization: In the early literature it was assumed that the position angle (PA) of the linear polarization is determined by the direction of the magnetic field in the source region, called the rotating vector model, predicting a characteristic S-shaped sweep of the PA through the pulse \citep{RC69,K70}. While the PAs of some pulsars obey this rule, for many pulsars the polarization is  more complicated than this simple model suggests. 
\item Radius-to-frequency mapping: A relation between the emission frequency and the height (denoted by the radial distance) is widely assumed, in the sense that lower frequencies are emitted at greater heights. However, the range of frequencies emitted at a given height is not known. 
\item Pulse to pulse variations: For sufficiently bright pulsars, the radio emission in individual pulses can be resolved. There are large pulse-to-pulse variations, with the integrated pulse profile being an envelope within which these variations occur.
\item Subpulses and micropulses: Those pulsars for which individual pulsars can be observed show a rich variety of features, including subpulses and micro\-pulses in the emission. 
\item Drifting subpulses: In some pulsars, the subpulses drift through the pulse window in a systematic way, and much emphasis has been placed on the interpretation of such drifting subpulses, notably in terms of a carousel model \citep{RS75,DR99}.
\item Mode changing: In young pulsars the pulse profile tends to be simple, with a single broad peak, while in older pulsars multiple peaks are common. Some pulsars have two or more quasi-stable pulse profiles, between which they jump abruptly, referred to a ``mode changing''.
\item Nulling: Some pulsars can turn off and on abruptly, with the off-state referred to as a ``null''. The rate of change, ${\dot\nu}$, of the slowing down changes abruptly when a pulsar turns off or on \citep{Ketal06}, implying a link between slowing down and radio emission.
\item Circular polarization: Observations of individual pulses show that they can be highly elliptically polarized, with large pulse-to-pulse variations \citep{MS00,J04,ES04}, requiring a statistical interpretation \citep{Metal06}. The circular polarization averages to a small value in the integrated pulse profile.
\item Orthogonally polarized modes:  Even in cases where there is a steady average swing in the PA, it can jump by 90$^\circ$ at specific phases, referred to as jumps between orthogonal polarizations \citep{Setal84,MS00,McK02}. The sign of the circular polarization also reverses, indicating that the jumps are between elliptically polarized natural modes of a birefringent medium \citep{PL00,WLH10,BP12}. 
\item Timing noise: Pulsars are extremely accurate clocks, and when all known effects (e.g., the motion of the Earth around the Sun) are taken into account, ``timing noise'' is a residual unexplained randomness in pulse arrival times. A now-favored explanation is in terms of changes in $\dot\nu$ associated with nulling or mode changing \citep{Letal10}.

\end{itemize}

\subsection{Pulsar electrodynamics}

Pulsar electrodynamics involves models for the electromagnetic field and the distributions of charges around a rapidly rotating, strongly magnetized neutron star \citep{MY16}. Two early models are referred to here as the the vacuum dipole mode (VDM) and the rotating magnetosphere model (RMM). Both models existed before the discovery of pulsars, and were modified in the application to pulsars: the VDM was developed for rotating magnetized stars \citep{D47,D55} and applied to pulsars by \citet{P68}, and the RMM was developed for rotating planetary magnetospheres \citep{HB65} and applied to pulsars by \citet{GJ69}. 

In the VDM the plasma in the magnetosphere is neglected. The VDM is used to relate the surface magnetic field on the star, and other properties, to the pulsar period, $P$, and the period derivative, ${\dot P}$. In a RMM the basic assumption is that the magnetospheric plasma is corotating with the star. In most detailed versions of the RMM, additional simplifying assumptions are made to reduce the electrodynamics to electrostatics, with the simplest such assumption being that the magnetic and rotation axes are aligned \citep{GJ69}. A later class of models is based on force-free electrodynamics (FFE), which is an MHD-like theory in which the plasma inertia and non-electromagnetic forces are neglected, and the displacement current is retained: the electromagnetic force $\rho{\bi E}+{\bi J}\times{\bi B}$ on the plasma is assumed to be negligibly small. A stationary, axisymmetric version of the FFE implies the so-called pulsar equation, whose solution provides a global model that extends from the stellar surface, through the light cylinder, where the corotation speed would be equal to $c$, into the pulsar wind zone \citep{Michel91}.

\subsubsection{Vacuum dipole model}

The magnetic field in a pulsar magnetosphere is usually approximated by that due to a rotating dipole at the center of the star. In a vacuum dipole model (VDM) the magnetospheric plasma is neglected, the magnetic field has its familiar dipolar form, $\propto1/r^3$, near the stellar surface, at $r=R_*$, and is modified by retardation effects, specifically an inductive term $\propto1/r^2r_{\rm L}$ and a radiative term $\propto1/rr_{\rm L}^2$, that become substantial near and beyond the light cylinder, at $r=r_{\rm L}=c/\omega_*$. The electric field induced by a rotating magnetic dipole includes an inductive term $\propto1/r^2$ and a radiative component $\propto1/r$. The radiative terms imply an Poynting flux $\propto1/r^2$ that corresponds to magnetic dipole radiation.

The power in magnetic dipole radiation in vacuo is
\be
P_{\rm rad}={\mu_0\omega_*^2|\bomega_*\times{\bi m}|^2\over6\pi c^3}={\mu_0m^2\omega_*^4\sin^2\alpha\over6\pi c^3},
\label{radf4}
\ee
where ${\bi m}$ is the magnetic dipole moment and $\alpha$ is the angle between the magnetic and rotational axes. The rotational energy, $\half I_*\omega_*^2$, where $I_*$ is the moment of inertia, decreases at the rate $-I_*\omega_*{\dot\omega}_*$, with $\omega_*=2\pi/P$ Equating  the equality $P_{\rm rad}$ to $-I_*\omega_*{\dot\omega}_*$ gives
\be
{I_*(2\pi)^2{\dot P}\over P^3}={(2\pi)^5R_*^6\over3c^3P^4}\,{B_*^2\sin\alpha^2\over\mu_0},
\label{calP2}
\ee
where  $m$ is expressed in terms of the magnetic field, $B_*=(\mu_0/4\pi)(2m/R_*^3)$, at the magnetic pole of the star. Assuming a characteristic values for the moment of inertia of a neutron star $I_*=10^{38}\rm\,kg\,m^2$, equation (\ref{calP2}) implies 
\be
B_*\sin\alpha=
3.2\times10^{15}\,(P{\dot P})^{1/2}\rm\,T.
\label{Bstar1}
\ee
The relation (\ref{Bstar1}) with $\sin\alpha=1$ defines the quantity identified as the surface magnetic field of the neutron star. The characteristic age of a pulsar is identified by assuming that all quantities in (\ref{calP2}) except $P$ and ${\dot P}$ are constant, integrating $\half dP^2/dt=P{\dot P}=$ const., and assuming that the value of $P$ at $t=0$ is negligible compared with the value at $t$. This gives a characteristic age\footnote{The assumption that $\sin\alpha$ is constant, made in the derivation of this age, is inconsistent with the VDM; the emission of magnetic dipole radiation exerts a torque that slows down the star and a torque that tends to cause alignment, implying that $\sin\alpha$ decreases on the slowing down timescale \citep{DG70}.}  $P/2{\dot P}$. Lines corresponding to given values of $B_*$ (in gauss, $\rm 1\,G=10^{-4}\,T$) and this age are drawn on Figure~\ref{PdotP}.

The dipolar magnetic field is also modified by currents flowing in the magnetospheric plasma at  $r\gtrsim r_{\rm L}$, where outflowing plasma forms a pulsar wind. It is widely assumed that equation (\ref{Bstar1}) with $\sin\alpha=1$ provides a plausible estimate of $B_*$ even when the loss of angular momentum is due to the wind, rather than magnetic dipole radiation.

There are two sources of electric field in the VDM: the inductive (plus radiative) electric field due to the rotating magnetic dipole, and an electric field due to a surface charge on the star when the point dipole is replaced by a conducting sphere in vacuo. Inside the star, infinite conductivity implies a corotational electric field, ${\bi E}_{\rm cor}=-(\bomega_*\times{\bi x})\times{\bi B}$. The boundary conditions at the stellar surface then imply a quadrupolar electric field at $r>R_*$ due to the surface charge distribution \citep{D47,D55}. In general, both the inductive and quadrupolar electric fields in vacuo have components, $E_\parallel$, parallel to the magnetic field that can accelerate electrons to highly relativistic energies. Such acceleration of charges from the stellar surface should trigger a pair cascade, populating the magnetosphere with plasma and invalidating the assumption of vacuum conditions.

\subsubsection{Rotating magnetosphere model}

In the RMM the magnetosphere (like the stellar interior) is assumed to be perfectly conducting. There is then no surface charge on the star, the quadrupolar electric field due to the surface charge is absent, and the corotation electric field, ${\bi E}_{\rm cor}$, is present throughout the corotating region of the magnetosphere. The divergence of ${\bf E}_{\rm cor}$ implies the Goldreich-Julian charge density
\be
\rho_{\rm GJ}(t,{\bi x})=
-2\varepsilon_0\bomega\cdot{\bi B}(t,{\bi x})
+\varepsilon_0(\bomega\times{\bi x})\cdot{\bf\nabla}\times{\bi B}(t,{\bi x}).
\label{rhoGJ}
\ee
In early models it was assumed that $\rho_{\rm GJ}$ is provided by charges of a single sign drawn from the stellar surface. With the surface the only source of charge, it is not possible to satisfy (\ref{rhoGJ}) at greater heights. An additional source of charge is needed, and this is provided by a pair cascade, which are assumed to occur in regions, called gaps, where $E_\parallel\ne0$ accelerates charges to sufficiently high energies for them to emit gamma rays that decay into electron-positron pairs. 

Force-free electrodynamics (FFE) is a modified form of magnetohydrodynamics (MHD) in which relativistic effects and the displacement current are included,  and the inertia of the plasma is neglected, corresponding to $v_{\rm A}\to\infty$. As in MHD, the assumption $E_\parallel=0$ is made in FFE. FFE is used widely to model the global electrodynamics, particularly for the region $r\gtrsim r_{\rm L}$ covering the transition from the inner magnetosphere to the pulsar wind \citep{CKF99,G05,K06,LST12}.

\subsection{Properties of pulsar plasma}

The magnetosphere of a pulsar is populated by plasma that is either drawn from the surface of the star (``primary'' particles) or produced in a pair cascade (``secondary'' particles). A distinction is drawn between the polar-cap regions, from which the plasma escapes along the open field lines, ultimately forming a pulsar wind, and the closed-field region. It is widely accepted that the plasma in the polar caps, which needs to be continuously replaced, is the source of the radio emission. The properties of this ``pulsar plasma'' must be important in any radio emission mechanism. However, these properties are poorly determined.

\subsubsection{Pair creation in gaps}

The polar-cap regions are assumed to be populated by the secondary pair plasma \citep{S71} created in gaps. Suggested locations of gaps include an inner gap \citep{RS75} near the stellar surface, a slot gap \citep{A83} near the last closed field line and an outer gap \citep{CHR86}. The high-energy photons are produced through curvature emission by primary particles, or by resonant and non-resonant Compton scattering \citep{HMZ02}.

A pair is produced through one-photon decay, which is allowed in a magnetic field provided that the photon energy perpendicular to the magnetic field satisfies the threshold condition 
\be
\varepsilon_{\rm ph}\sin\theta>2mc^2,
\qquad
\varepsilon_{\rm ph}\cos\theta=(p_\parallel+p'_\parallel)c,
\label{pct}
\ee 
where the latter condition expresses conservation of parallel momentum with $p_\parallel,p'_\parallel$, the parallel momenta of the electron and positrons. The pairs are generated by outward-propagating high-energy photons, and hence are propagating outwards. The photon decays spontaneously into an electron and a positron in Landau levels $n,n'$ satisfying
\be
\varepsilon_{\rm ph}=\varepsilon_{n}(p_\parallel)+\varepsilon_{n'}(p'_\parallel), 
\qquad
\varepsilon_{n}(p_\parallel)=(m^2c^2+p_\parallel^2+2neB\hbar c)^{1/2}.
\label{pce}
\ee
Curvature photons initially emitted by a primary particle with Lorentz factor $\gamma$ are strongly beamed along the magnetic field lines, $\theta\lesssim1/\gamma$, and $\theta$ increases as the angle between the ray path and the curved magnetic field line increases until the threshold (\ref{pct}) for pair creation is exceeded. It is usually assumed that the pairs are produced in very high harmonics, where a synchrotron-like formula \citep{E66} applies. However, the absorption coefficient has square-root singularities \citep{DH96,M13} at the threshold for each $n,n'$ $n,n'=0,1,\ldots$ in equation (\ref{pce}), and for $B\gtrsim0.1B_{\rm cr}$ with $B_{\rm cr}=m^2c^2/e\hbar=4.4.\times10^9\,$T the photon is absorbed at the lowest harmonics, or evolves into a bound state of positronium \citep{SU84,UM96}. Irrespective of the initial Landau levels in which the pairs are created, they quickly relax to the lowest Landau state, $n=0$, resulting in a one-dimensional (1D) distribution with $p_\perp=0$.

\subsubsection{Secondary pair plasma}

The properties of this pair plasma include the number density, with associated plasma frequency $\omega_p$, the mean Lorentz factor associated with the outward streaming and the spread in Lorentz factors about this mean. The number density can be expressed in terms of the Goldreich-Julian number density, $n_{\rm GJ}=|\rho_{\rm GJ}|/e$, given by equation (\ref{rhoGJ}). The associated plasma frequency,
\be
\omega_{\rm GJ}=(\omega_*\Omega_e)^{1/2},
\label{omegaGJ}
\ee
where factors of order unity are ignored, may be interpreted as the plasma frequency associated with the primary particles. The value of $\omega_{\rm GJ}$ at the stellar surface is proportional to $({\dot P}/P)^{1/4}$, which is relatively insensitive to the properties of the pulsar, and it decreases with radial distance $\omega_{\rm GJ}\propto(R_*/r)^{3/2}$.

The number density of secondary pairs in the pulsar magnetosphere may be written as $\kappa n_{\rm GJ}$, where $\kappa$ is the multiplicity. The characteristic plasma frequency in the magnetosphere is then $\omega_p=\kappa^{1/2}\omega_{\rm GJ}$. Recent estimates suggest a multiplicity of order $10^5$ \citep{TH15}. Numerical models suggest Lorentz factors in the range tens to hundreds \citep{ZH00,HA01,AE02}. However, considerable uncertainty remains concerning the generation and resulting properties of the secondary pair plasma. In particular, models based on primary particles from the stellar surface are now regarded as questionable because they result in a charge-separated, dome-disk model \citep{KM85,S04}, referred to as an electrosphere, rather than the widely-accepted polar-cap model. There are also arguments for an {\br ion-electron} plasma \citep{jones}, rather than a pair plasma.  An alternative model in which charges from the surface play no role, with pair creation being the only source of plasma in the magnetosphere \citep{T10b}, seems plausible. {\br In the following discussion of wave dispersion in a pulsar plasma it is assumed that no ions are present, and that the only particles are electrons and positrons. The most important effect of ions on the model is the contribution of their inertia to the Alfv\'en speed, which may be included by redefining $\beta_A^2$ appropriately.}

\subsubsection{Plasma inhomogeneities}

One expects plasma generated in a pair cascade to be highly structured in both space and time. Quasi-stationary gaps are unstable, and $E_\parallel\ne0$ is more plausibly described in terms of large-amplitude electrostatic oscillations \citep{Letal05,BT07} propagating outwards \citep{LM08}. Pair creation is then time-dependent, depending on the phase of the oscillation, and the distribution of outwardly streaming plasma is expected to be highly structured along the field lines. Any structure in the pair-creation across field lines is preserved as the plasma propagates outward, suggesting that strong gradients of the plasma properties across field lines is also to be expected.

\subsection{Wave dispersion in a pulsar plasma}
\label{sect:wave_dispersion}

The properties of a pulsar plasma differ from most other plasmas in a number of ways, each of which can affect the properties of the plasma. These properties include: the super-strong magnetic field,  the resulting 1D ($p_\perp=0$) distribution, the presence of both electrons and positrons {\br (with no ions)}, the net charge density, relativistic streaming of the bulk plasma, relative streaming of different plasma components (electrons, positrons and primary particles) and relativistic spread in parallel momentum.

\subsubsection{Cold-plasma models}

{\br Some early models \citep{HR76,HR78,MS77,AM82} for the wave dispersion were based on a cold electron gas, also called magnetoionic theory, with three notable changes. First, it is assumed that there are no ions, and the two magnetoionic parameters, $X$ and $Y$,  are complemented by an additional cold-plasma parameter, $\epsilon$ say, with $\epsilon=-1$ in the absence of positrons and $\epsilon=0$ in a charge-neutral pair plasma. Second, the approximations $\omega,\omega_p\ll\Omega_e$ are made, with the ratio $\Omega_e/\omega_p$ interpreted as $\beta_A=v_A/c$, which interpretation needs to be modified to take account of relativistic effects. Third, the relativistic streaming is taken into account either by Lorentz transforming the response tensor to the pulsar frame and solving for the wave properties in this frame \citep{HR76,HR78}, or by solving for the wave properties in the rest frame and Lorentz transforming these properties to the pulsar frame in which streaming is present \citep{MS77,AM82}. The latter approach is assumed in more detailed treatments that take into account the relativistic spread in energies in the rest frame, and is adopted here to treat the cold-plasma case.}

The cold plasma dielectric tensor is then given by equation (\ref{cp1}) with $S$ and $P$ unaffected, and with $D$ replaced by $-\epsilon XY/(1-Y^2)$. For radio waves in a pulsar magnetosphere, one has $\omega\ll\Omega_e$, allowing one to approximate by expanding in powers of $1/Y$. The parameter $S$ may be approximated according to
\be
S\approx1+\frac{X}{Y^2}=1+\frac{1}{\beta_A^2}=\frac{1}{\beta_0^2},
\qquad
\beta_A=\frac{\Omega_e}{\omega_p},
\quad
\beta_0=\frac{\beta_A}{(1+\beta_A^2)^{1/2}},
\label{Sapprox}
\ee
where $v_{\rm A}=\beta_Ac$ is the Alfv\'en speed, with $\beta_A\gg1$ in a pulsar plasma. The MHD speed becomes $\beta_0c\lesssim c$.  The two modes in this limit are conventionally referred to as the O~and X~modes {\br \citep{AB86}} in the pulsar literature.

If the multiplicity is large, $\kappa\gg1$, then $\epsilon=1/\kappa$ is small, and may be neglected to a first approximation. The  cold-plasma dispersion equation with $D=0$ becomes
\be
(S-n^2)(PS-P\cos^2\theta+S\sin^2\theta)=0.
\label{cold1}
\ee 
The solution $n^2=S$ corresponds to the X~mode. For $\sin\theta=0$ the second factor in equation (\ref{cold1}) reduces to $P(S-n^2)=0$. The solution $P=0$ implies $\omega=\omega_p$ and longitudinal polarization, and the other solution is for the O~mode. The dispersion curves $\omega=\omega_p$ and $n^2=S$ cross, and for $\sin\theta\ne0$ these reconnect and separate, forming two distinct branches, with the lower-frequency branch identified as the Alfv\'en mode and the higher-frequency branch called the LO~mode. The resulting dispersion relations are approximated using equation (\ref{Sapprox}), giving
\be
n_{\rm O}^2=\frac{1}{\beta_0^2}\frac{\omega^2-\omega_p^2}{\omega^2-\omega_p^2\cos^2\theta},
\qquad
n_{\rm X}^2=\frac{1}{\beta_0^2},
\label{MHDapprox}
\ee
At low frequencies, $\omega^2\ll\omega_p^2\cos^2\theta$, the O~mode dispersion relation reduces to $n^2_{\rm O}=1/\beta_0^2\cos^2\theta$, which corresponds to the Alfv\'en mode, which has a resonance at $\omega^2=\omega_p^2\cos^2\theta$. There is a stop band at $\omega_p^2\cos^2\theta<\omega^2<\omega_p^2$. The branch of the LO~mode at $\omega^2>\omega_p^2$ is strongly modified by dispersive effects due to the spread in velocities, which is not included in equations (\ref{cold1}) and (\ref{MHDapprox}). The X~mode may be interpreted as the magneto-acoustic mode in this approximation. 

{\br The foregoing cold plasma model can be misleading due to the important role played by the relativistic spread in energies in a pulsar plasma. The form of $S$ is changed in only a minor way by the inclusions of the relativistic spread, the the cold plasma form $P=1-\omega_p^2/\omega^2$ is strongly modified when the relativistic spread in energies is taken into account, with $P$ replaced by $K_{33}$ given by equation (\ref{7f.11b}) below. }

\subsubsection{Relativistic plasma dispersion function}

A more detailed treatment of wave dispersion in a pulsar plasma needs to be based on kinetic theory, {\br as first recognized in the early 1970s, cf.\  \citet{TK72}  and \S17 of \citet{KT73}, and developed in more detail in the 1980s \citep{GM83,VKM85,Letal86,AB86,BGI88}. In order to take account of the relativistic spread in energies one needs to introduce relativistic plasma dispersion functions (RPDFs), which depend on the assumed form of the distribution of electrons and positrons.} 

A general form of the dielectric tensor in a magnetized plasma involves an expansion in Bessel functions.\footnote{The anti-hermitian part of this tensor is used in the derivation of the gyromagnetic absorption coefficient (\ref{gammaM}) with (\ref{DV}).} For a 1D distribution, the argument of the Bessel functions is zero, and only $s=0$, $s=\pm1$ contribute in the sum over harmonic numbers. It is convenient to write the gyroresonance condition $\omega-k_\parallel v-s\Omega_e/\gamma=0$ with $s=0,\pm1$ as $\beta=z$, $\beta=z_\pm$, with $z=\omega/k_\parallel c$, $y=\Omega_e/k_\parallel c$, and 
\be
z_\pm=\frac{z\pm y(1+y^2-z^2)^{1/2}}{1+y^2}.
\label{zpm}
\ee
For radio waves in a pulsar plasma, the strong-field limit, $y/z\to\infty$, corresponds to $z_\pm\to\pm1$. 

Three plasma dispersion functions are required in general \citep{MGKF99,MG99,KML00}. For a 1D distribution function $f(u)$ with $p=mcu$, $u=\gamma\beta$, $\beta=v/c$, $\gamma=(1-\beta^2)^{-1/2}$, it is convenient to define the average value of any function $M(\beta)$ as
\be
\langle M\rangle=\int_{-\infty}^\infty du\,f(u)\,M(\beta),
\qquad \int_{-\infty}^\infty du\,f(u)=1,
\label{Wz}
\ee
where $f(u)$ is the distribution function, which may be for electrons alone, positrons alone or for the sum of the electrons and positrons. {\br Three RPDFs are defined by the averages}
\be
W(z)=\left\langle\frac{1}{\gamma^3(\beta-z)^2}\right\rangle,
\quad
R(z)=\left\langle\frac{1}{\gamma(\beta-z)}\right\rangle,
\quad
S(z)=\left\langle\frac{1}{\gamma^2(\beta-z)}\right\rangle.
\label{WRS}
\ee

In terms of these three RPDFs, the components of the dielectric tensor for either an electron or a positron gas ($\epsilon=\mp1$) are \citep{KML00}
\bea
K_{11}&=&K_{22}=1-
\frac{\omega_p^2}{\omega^2}\frac{1}{1+y^2}
\left[\left\langle\frac{1}{\gamma}\right\rangle+\frac{(z-z_+)^2R(z_+)-(z-z_-)^2R(z_-)}{z_+-z_-}
\right],
\nonumber
\\
K_{33}&=&1-{\omega_p^2\over\omega^2}
\left\{
z^2W(z)+
\frac{\tan^2\theta}{1+y^2}
\left[\left\langle\frac{1}{\gamma}\right\rangle+\frac{z_+^2R(z_+)-z_-^2R(z_-)}{z_+-z_-}
\right]\right\},
\nonumber
\\
K_{13}&=&K_{31}=-\frac{\omega_p^2}{\omega^2}\frac{\tan\theta}{1+y^2}
\left[\frac{(z-z_+)z_+R(z_+)-(z-z_-)z_-R(z_-)}{z_+-z_-}\right],
\nonumber
\\
K_{12}&=&-K_{21}=-i \epsilon\frac{\omega_p^2}{\omega^2}\frac{y}{1+y^2}
\left[\frac{(z-z_+)S(z_+)-(z-z_-)S(z_-)}{z_+-z_-}
\right],
\nonumber
\\
K_{23}&=&K_{32}=i \epsilon\frac{\omega_p^2}{\omega^2}\frac{y\tan\theta}{1+y^2}
\left[\frac{z_+S(z_+)-z_-S(z_-)}{z_+-z_-}
\right].
\label{7f.11a}
\eea
If the electrons and positrons have identical distributions with different number densities, then equation (\ref{7f.11a}) applies, with $\omega_p^2$ proportion to the sum of the number densities, and $\epsilon$ equal to the difference divided by the sum.

\subsubsection{1D relativistic distribution functions}

{\br The RPDFs (\ref{WRS}) depend on the form of 1D relativistic distribution function, and several different forms have been considered. For the RPDF $W(z)$, which is the only relevant one at low frequencies,  \citet{KT73} assumed a power-law distribution, and made approximations in treating the dispersion, rather than introducing a RPDF explicitly; \citet{LM79} discussed this and other analyses critically and also suggested a gaussian distribution in $u$.} The RPDFs can be evaluated in terms of elementary functions for the water-bag \citep{AB86}, hard-bell and soft-bell \citep{GMG98} distributions, which correspond to $f_0(u)$, $f_1(u)$ and $f_2(u)$, respectively, with $u=\gamma\beta$, $u_m=\gamma_m\beta_m$ in
\be
f_n(u)=\frac{(u_m^2-u^2)^n}{A_n}\,H(u_m^2-u^2),
\qquad 
A_m=\int_{-u_m}^{u_m}du\,(u_m^2-u^2)^n,
\label{fnu}
\ee
where $H$ denotes the step function. The RPDFs can also be evaluated for the 1D relativistic thermal (J\"uttner) distribution \citep{MG99,AR00,M13}, specifically for
\be
f(u)=\frac{e^{-\zeta\gamma}}{2K_1(\zeta)},
\label{Juttner}
\ee 
with $\zeta=mc^2/T$ an inverse temperture and $K_n$ is the Macdonald function of order $n$. In this case, the RPDFs are transcendental functions that may be expressed in terms of a RPDF defined by \citet{GNT75}. The resulting expression for $W(z)$ is
\be
W(z)={T'(z,\zeta)\over2K_1(\zeta)},
\qquad
T(z,\zeta)=\int_{-1}^1d\beta\frac{e^{-\zeta\gamma}}{\beta-z},
\label{Tzrho}
\ee
where the prime denotes the derivative with respect to $z$.

\begin{figure}
\centering
% \psfragfig[width=1.0\textwidth]{z2W_rho_50_10_1}
% \vspace{4mm}
%       \psfragfig[width=0.49\textwidth]{z2W_rho_0_1}
%        \psfragfig[width=0.49\textwidth]{z2W_rho_0_01}
\vspace{-50mm}
\includegraphics[width=1.0\textwidth]{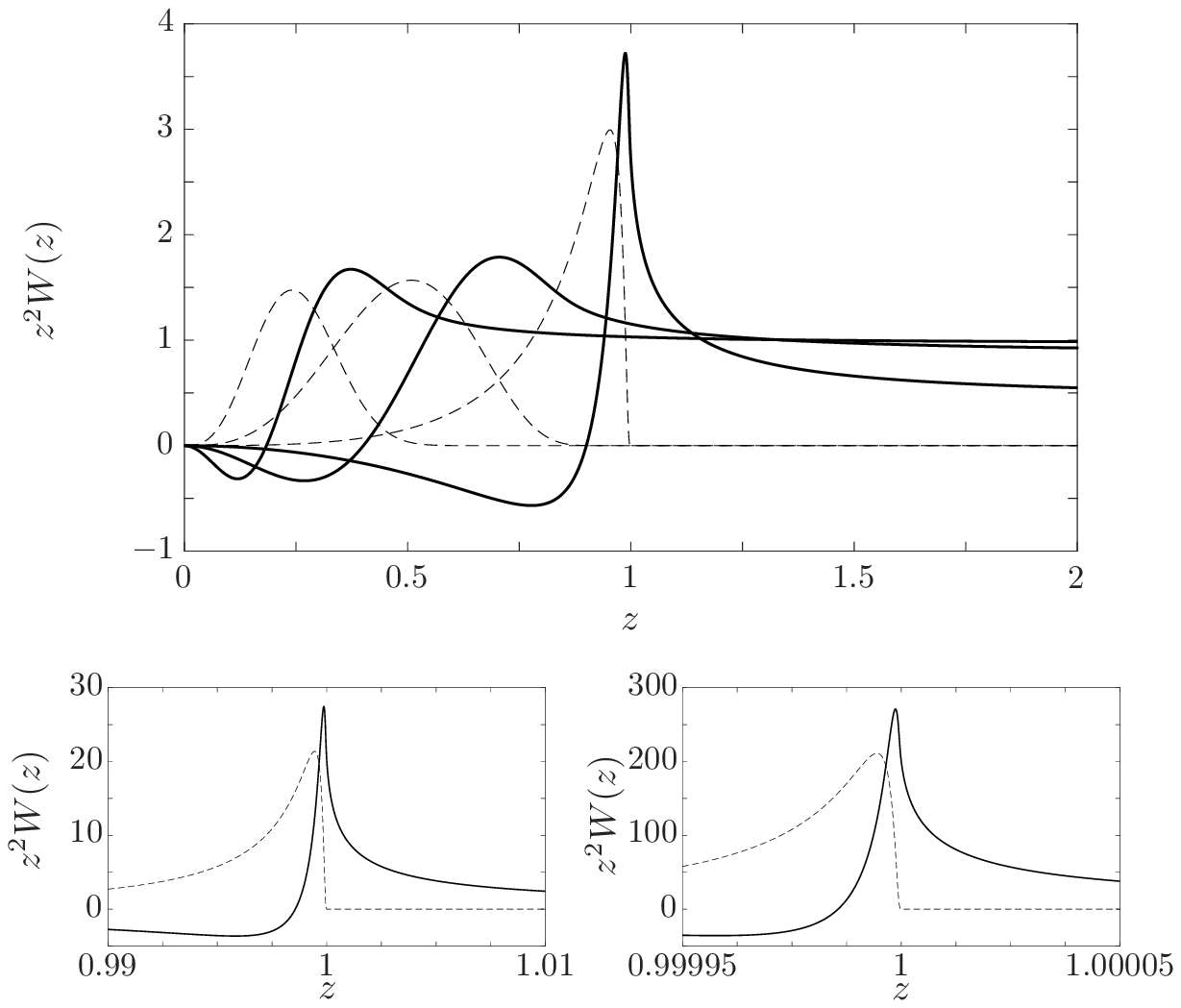}
\vspace{-60mm}
\caption{{\br The RPDF $z^2W(z)$ is plotted as a function of $z$ for 1D J\"uttner distributions: upper figure: leftmost $\zeta=50$, center $\zeta=10$ and rightmost $\zeta=1$; lower figures: left $\zeta=0.1$, right $\zeta=0.01$. The dashed curves correspond to the imaginary parts, which are identically zero for $z\ge1$.}}
\label{fig:MGKF3}  
\end{figure}

The function $z^2W(z)$ is plotted in Figure~\ref{fig:MGKF3} for five values of the inverse temperature, $\zeta=50,10,1,0.1,0.01$, with $\zeta=1$ corresponding to a temperature $\approx5\times10^9\,$K. The RPDF becomes increasingly sharply peaked as $\zeta$ decreases. The general form being relatively insensitive to the choice of distribution function \citep{MGKF99}. The specific values of $z^2W(z)$ for  $z\to\infty$ and $z=1$ are determined by moments of the distribution function:
\be
\lim_{z\to\infty}z^2W(z)=\left\langle\frac{1}{\gamma^3}\right\rangle,
\qquad
\lim_{z\to1}z^2W(z)=2\langle\gamma\rangle-\left\langle\frac{1}{\gamma}\right\rangle.
\label{Wlim}
\ee
For an ultrarelativistic distribution, $\langle\gamma\rangle\gg1$, $z^2W(z)$ is sharply peaked just below $z=1$, as illustrated in Figure~\ref{fig:MGKF3}. {\br The maximum value of $z^2W(z)$ at this peak is somewhat greater than the value $\approx2\langle\gamma\rangle$ at $z=1$, and occurs at $1-z$ less than $1/\langle\gamma\rangle^2$. The form of the RPDF is similar for the different choices of distribution function with the same value of $\langle\gamma\rangle\gg1$ \citep{MGKF99}. }

In the radio range, one has $\omega\ll\Omega_e$ and then the approximations $z_\pm\to\pm1$ lead to
\be
R(\pm1)=\mp\langle\gamma\rangle-\langle\gamma\beta\rangle,
\qquad
S(\pm1)=\mp1-\langle\beta\rangle.
\label{yinfty}
\ee
The components (\ref{7f.11a}) then simplify to
\be
K_{11}=K_{22}=1+
{\omega_p^2\over\omega^2}\,{k_z^2c^2\over\Omega_e^2}
\big(z^2\left\langle \gamma\right\rangle 
-2z\left\langle \gamma \beta\right\rangle 
+\left\langle \gamma \beta^2\right\rangle 
\big),
\nonumber
\ee
\be
K_{33}=1-{\omega_p^2\over\omega^2}
\bigg[
z^2W(z)
-{k_\perp^2\over\Omega_e^2}\,
\left<\gamma \beta^2\right>
\bigg],
\qquad
K_{13}=-{\omega_p^2\over\omega^2}\,
{k_\perp k_zc^2\over\Omega_e^2}
\big(z\left\langle \gamma \beta\right\rangle -\left\langle \gamma \beta^2\right\rangle \big),
\nonumber
\ee
\be
K_{12}=-i \epsilon\,{\omega_p^2\over\omega^2}\,
{k_z c\over\Omega_e}\,
\big(z-\left\langle \beta\right\rangle \big),
\qquad
K_{23}=i \epsilon\,{\omega_p^2\over\omega^2}\,
{k_\perp c\over\Omega_e}\,
\left\langle \beta\right\rangle ,
\qquad
\label{7f.11b}
\ee
with $K_{31}=K_{13}$, $K_{21}=-K_{12}$, $K_{32}=-K_{23}$. If the electrons or positrons have identical distributions, then in the rest frame of the plasma one has $\left\langle \beta\right\rangle=0$ (and $\left\langle\gamma\beta\right\rangle=0$) implying $K_{23}=0$.  However, there is a net current in a pulsar plasma, requiring a net streaming of electrons relative to positrons, and $K_{23}$ can be neglected only for $|J_\parallel|\ll enc$.

\subsection{{\br Dispersion relations in the rest frame of} pulsar plasma}
\label{sect:pulsar_plasma}

An idealized model for the pulsar plasma, assumed to be created in pair cascades, has a relativistic outward streaming motion, a relativistic spread in velocities about the bulk streaming velocity, and electron and positron distributions that are the same to first order in an expansion in $1/\kappa$, where $\kappa$ is the multiplicity. It is convenient to discuss the wave dispersion in the rest frame of the plasma. Another relevant approximation is that the plasma inertia is small, corresponding to $\beta_A\gg1$. Wave dispersion in such an idealized pulsar plasma may be treated as follows.

To lowest order in an expansion in $1/\kappa$, one has $K_{12}=0$ and $K_{23}=0$. The remaining components  of $K_{ij}$ lead to the following components of $\Lambda_{ij}$ {\br \citep{MG99,MGKF99}}:
\be
\Lambda_{11}=\frac{1}{\beta_0^2}-\frac{1-b}{\beta_A^2z^2},
\qquad
\Lambda_{22}=\frac{1}{\beta_0^2}-\frac{1-b}{\beta_A^2z^2}+\frac{\sin^2\theta}{z^2},
\nonumber
\ee
\be
\Lambda_{33}=1-\frac{\omega_p^2}{\omega^2}z^2W(z)-\frac{1-b}{\beta_A^2z^2}\tan^2\theta,
\quad
\Lambda_{13}=\frac{1-b}{\beta_A^2z^2}\tan\theta,
\label{wpp1}
\ee
with $\beta_0^2=\beta_A^2/(1+\beta_A^2)$, $\beta_A^2=\Omega_e^2/\omega_p^2\langle\gamma\rangle$ and $b=\left\langle\gamma^{-1}\right\rangle/\langle\gamma\rangle$. The resulting dispersion equation factorizes, as in the cold-plasma case, into $\Lambda_{22}=0$ and $\Lambda_{11}\Lambda_{33}-\Lambda_{13}^2=0$.

{\br The dispersion equation $\Lambda_{22}=0$ gives the dispersion relation for the X~mode, which reproduces the cold-plasma dispersion relation (\ref{MHDapprox}) for $b\ll1$. Except for $\tan^2\theta\lesssim1/\beta_A^2$ the dispersion curve for the X~mode is in the region $z>1$.}

\begin{figure}
\centering
%   \psfragfig[width=0.7\columnwidth]{dispersion_rho_20_th_0n1}
\vspace{-70mm}
\includegraphics[width=1.0\textwidth]{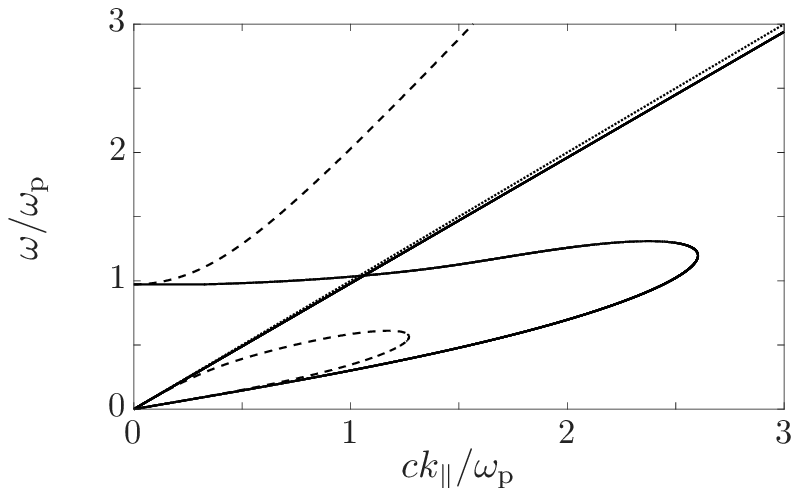}
\vspace{-75mm}
        \caption{{\br Dispersion curves  are shown for the parallel A~mode and parallel L~mode (solid lines, $\theta=0$), and for the oblique Alfv\'en mode and the LO~mode (dashed lines, $\theta=1$). The value of $z_A$ is close to unity and the dispersion relation $z=z_A$ is just to the right of $z=1$ (the dotted line). The calculation is for the distribution function (\ref{Juttner}) with $\zeta=20$.}
}
\label{fig:dp20}  
\end{figure}

\subsubsection{Parallel {\br A~mode and} L~mode}

For parallel propagation the dispersion equation for the LO~mode factorizes into $\Lambda_{11}=0$, {\br which corresponds to the parallel Alfv\'en (A) mode, and $\Lambda_{33}=0$, which corresponds to the parallel longitudinal (L) mode. The parallel A~mode has dispersion relation
\be
z^2=z_A^2, 
\qquad 
z_A^2=\frac{b}{a}\approx1+\frac{1}{\beta_A^2}=\beta_0^2,
\label{Amode}
\ee 
} 
The parallel L~mode has dispersion relation
\be
\omega^2=\omega_L^2(z)=\omega_p^2z^2W(z).
\label{Lmode}
\ee
The L~mode has a cutoff at $\omega=\omega_{\rm c}$ given by
\be
\omega_{\rm c}^2=\omega_L^2(\infty)=
\omega_p^2\left\langle \gamma^{-3}\right\rangle .
\label{7f.20}
\ee
{\br The frequency $\omega_{\rm c}$ is sometimes referred to as the relativistic plasma frequency.} The dispersion curve crosses the {\br line $z=1$} at $\omega=\omega_1$, given by
\be
\omega_1^2=\omega_L^2(1)=\omega_p^2\left(2\langle\gamma\rangle-\left\langle\frac{1}{\gamma}\right\rangle\right),
\label{7f.21}
\ee
where equation (\ref{Wlim}) is used. The ratio $\omega_1/\omega_{\rm c}$ is of order $\langle\gamma\rangle$ ($\langle \gamma^{-3}\rangle$ is of order $\langle \gamma\rangle^{-1}$), which is assumed to be much greater than unity in a pulsar plasma. 

{\br In the highly relativistic case, the dispersion curves are strongly concentrated near the line $z=1$. Following \citet{MGKF99} it is convenient to discuss this dispersion first for the nonrelativistic case, and then to indicate how the curves become strongly distorted as the plasma becomes increasingly relativistic. As illustrated in Figure~\ref{fig:dp20}, which is for a mildly relativistic thermal plasma ($\zeta=20$),  the dispersion curve for the L~mode is the solid tongue-like curve, which starts at the cutoff at $\omega_{\rm c}$ slightly less than $\omega_p$ and increases slowly with $k_\parallel=\omega/cz$ for $0<1/z<1$ until it crosses the line $z=1$ at $\omega_1$ slightly greater than $\omega_p$. The dispersion curves for the L~mode crosses that of  the A~modes $z=z_A$ at a crossover frequency $\omega_{\rm co}=\omega_L(z_A)$. The parallel L~mode  reaches a maximum frequency near where the phase speed is equal to the thermal speed of electrons, where it turns over, and continues as a lower-frequency retunr branch in the region of strong Landau damping. As the temperature increases from nonrelativistic values $\zeta\gg1$ to highly relativistic values $\zeta\ll1$, the cutoff frequency $\omega_{\rm c}$ decreases to $\ll\omega_p$ and the tongue-like curve narrows and becomes strongly aligned with the line $z=1$; it crosses $z=1$ at $\omega_1\gg\omega_p$, crosses the line $z=z_A$ at $\omega_{\rm co}>\omega_1$ and reaches its maximum at a slightly higher frequency. The highly relativistic case cannot be shown clearly on a diagram like figure~\ref{fig:dp20} because nearly all the curves as very close to the line$z=1$; the curves can be distinguished in a alternative plot of $\log(\omega/\omega_p)$ as a function of $z^{-2}$, cf.\ Fig.~3 of \citet{MG99}.
}

\subsection{Oblique Alfv\'en and LO modes}

The dispersion equation for the oblique Alfv\'en and the LO~modes may be written in the form \citep{MGKF99}
\be
\frac{\omega^2}{\omega_p^2}=\frac{z^2W(z)(z^2-z_A^2)}{z^2-z_A^2-b\tan^2\theta}.
\label{LOdr}
\ee
The dispersion curves for the parallel A~and L~modes intersect at $\omega=\omega_{\rm co}$ and for slightly oblique propagation they reconnect to form the oblique Alfv\'en and LO~modes. For small $\tan\theta$, the dispersion equation for the oblique Alfv\'en mode is nearly the same as for the parallel A~mode ($z=z_A$) for $\omega<\omega_{\rm co}$ and nearly the same as for the branch of the parallel L~mode for $\omega>\omega_{\rm co}$; the turnover defines the maximum frequency for the oblique Alfv\'en mode. The LO~mode follows the dispersion curve for the L~mode for $\omega_{\rm c}<\omega\lesssim\omega_{\rm co}$ and continues as $z\approx z_A$ for $\omega\gtrsim\omega_{\rm co}$. {\br There is a small range of $\tan^2\theta\lesssim1/\beta_A^2$ for which the (reconnected) LO~mode is in the range $1>z>z_A$. For larger angles $\theta$ the dispersion curve is entirely in the range $z>1$, as illustrated by the dashed curve in Figure~\ref{fig:dp20}. }

\subsubsection{Lorentz transformation to the pulsar frame}

In the pulsar frame, the plasma is assumed to be streaming outward along open field lines, at speed $\beta_{\rm s}c$, say, with $\gamma_{\rm s}=(1-\beta_{\rm s}^2)^{-1/2}$. A Lorentz transformation between the rest (unprimed) and pulsar (primed) frame gives
\begin{align}
&\omega'=\gamma_{\rm s}(\omega+k_\parallel c\beta_{\rm s}),
\qquad\,
k'_\parallel=\gamma_{\rm s}(k_\parallel+\omega\beta_{\rm s}/c),
\qquad
k'_\perp=k_\perp,
\nn\\
\ms
&\qquad
z'=\frac{z+\beta_{\rm s}}{1+z\beta_{\rm s}},
\qquad
\tan\theta'=\frac{\tan\theta}{\gamma_s(1+z\beta_s)}.
\label{LT1}
\end{align}
For example, the parallel L~mode in the superluminal range between $\omega_{\rm c}, z=\infty$ and $\omega_1,z=1$ transforms into a superluminal range between $\gamma_{\rm s}\omega_{\rm c}, z'=1/\beta_{\rm s}$ and $\gamma_{\rm s}(1+\beta_s)\omega_1,z'=1$.

Escaping radio emission must be in either the X~or  LO~modes, both of which have approximately vacuum-like dispersion. Such emission is relativistically boosted, from $\omega$ in the rest frame to $\omega'$ of order $\gamma_{\rm s}\omega$ in the pulsar frame. Relativistic aberration implies that emission at nearly all angles $\theta$ in the rest frame is confined to a narrow forward cone $\theta'\lesssim1/\gamma_{\rm s}$ in the pulsar frame.

\begin{figure}[t]
\begin{center}
\includegraphics[width=0.5\textwidth]{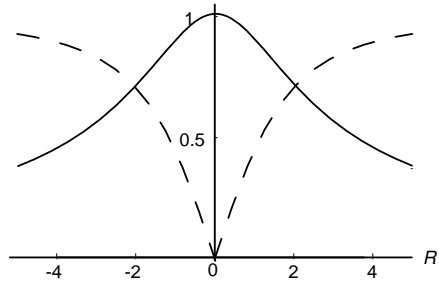}
\caption{The degrees of  circular (solid line) and linear (dashed line) polarization are plotted for one mode as a function of $R$, with $R=0$ corresponding to the cyclotron frequency [from \citet{ML04}]}
\label{fig:CP}
\end{center}
\end{figure}

\subsubsection{Elliptical polarization: cyclotron resonance}

Elliptical polarization of the natural modes of a pulsar plasma is relevant to the interpretation of observed elliptical polarization. The wave modes have a circularly polarized component if the distributions of electrons and positrons are different, with the simplest example being a net charge density, $\epsilon\ne0$. The transverse parts of the polarization vectors for the two modes{\br, labeled $\pm$, are then oppositely elliptically polarized with axial ratios $T=T_\pm$, with $T$ satisfying a quadratic equation that implies $T_-=-1/T_+$. This equation} may be written
\be
T^2-RT-1=0,
\qquad
T_\pm=\half R
\pm 
\half(R^2+4)^{1/2},
\label{SPP8}
\ee
with the parameter $R$ determined by the components of the dielectric tensor with $K_{12}$ and/or $K_{23}$ nonzero. For example, in the cold-plasma model one has
\be
R=-{Y\sin^2\theta\over(1-X){\epsilon}\cos\theta}(1-E),
\qquad
E={X(1-{\epsilon}^2)\over1-Y^2}.
\label{newT} 
\ee
The ellipticity of the natural modes may be described in terms of the degrees of circular and linear polarization. These are plotted as a function of $R$, given by equation (\ref{SPP8}), in Figure~\ref{fig:CP}.

Along its escape path pulsar radio emission encounters a cyclotron resonance region \citep{ML04}, where the wave modes become strongly elliptically polarized, and cyclotron absorption is possible. The wave dispersion associated with the cyclotron resonance is described by the RPDFs $R(z),S(z)$ in the dielectric tensor (\ref{7f.11a}). The elliptical polarization observed in the radio emission from some pulsars may be imposed on the escaping radiation as it propagates through the cyclotron resonance region.

\subsection{Relativistic beam instabilities}
\label{sect:instabilities}

One of the suggested pulsar radio emission mechanisms is relativistic plasma emission, that is, a relativistic form of the mechanism that operates in solar radio bursts. There are several difficulties with this {\br emission mechanism, one of which is that most discussions of it are based on unrealistic assumptions about the wave dispersion, as discussed in \S\ref{sect:emission}. This difficulty is ignored here in order to discuss another problem: assuming that Langmuir-like waves exist,}  the growth rates for beam instabilities appear to be too small to allow effective wave growth. Any instability must operate in a relativistically outflowing plasma, and there are gradients in the properties of the plasma along the magnetic field lines. Assuming the growing waves are stationary in the outflowing plasma, the time available for wave growth is limited by these gradients causing the waves to move out of resonance with the beam that drives their growth. The waves must grow fast enough for the instability to saturate before the plasma properties have changed significantly. 

\subsubsection{Maser version of beam instability}

A maser treatment of a beam instability requires that the waves and particles satisfy the Cerenkov resonance condition, which reduces to  $\omega-k_\parallel v_\parallel=\omega(1-n\beta\cos\theta)=0$ or $z-\beta=0$ in the 1D case. This condition can be satisfied only for waves with refractive index $n>1$, {\br or $z=\omega/k_\parallel c<1$ in the 1D case}. Negative absorption is possible when this resonance condition is satisfied provided  that the distribution of particles has a positive slope, {\br $df(u)/du>0$, in the streaming direction. This requires that the distribution function have a maximum, with  $df(u)/du=0$ at $\gamma=\gamma_{\rm peak}$ say, and that the growing waves have $z<\beta_{\rm peak}\approx1-1/2\gamma_{\rm peak}^2$. The growth rates for maser instabilities is too slow to be effective.} 

%Note that the parallel L~mode in a pulsar plasma would seem to be the most obvious counterpart of Langmuir waves in a nonrelativistic plasma, but this mode is superluminal between its cutoff and cross-over frequencies, given by equations (\ref{7f.20}) and (\ref{7f.21}), respectively. It is impossible to generate such L~waves through the Cerenkov resonance. Put another way, there are no Langmuir-like waves in a pulsar plasma that can be generated through a streaming instability. {\br This difficulty is ignored in the following discussion of relativistic beam instabilities.} 

Beam instabilities in a pulsar plasma are usually assumed to be reactive. This seems plausible when addressing the problem of identifying the fastest growing instability: reactive instabilities grow faster than analogous maser instabilities. However, reactive instabilities are usually treated assuming cold distributions, which involves neglecting the spread in energies, whereas the spread in energies is thought to be relativistic in a pulsar plasma. This inconsistency is ignored in the following summary of some possible reactive, relativistic beam instabilities.

\subsubsection{Relativistic reactive beam instability}

When the spread in energies of the beam is neglected, the beam is cold, and the only possible instability is reactive \citep{ELM83,GMG02}. For a cold beam with number density $n_{\rm b}$, velocity $\beta_b$ and Lorentz factor $\gamma_b$ propagating through a cold background plasma, the dispersion equation for parallel, longitudinal waves is
\be
1-{\omega_p^2\over\omega^2}
-{n_{\rm b}\over n_e}\,{\omega_p^2\over\gamma_b^3(\omega-k_z\beta_b)^2}=0. 
\label{8d.8a}
\ee
This dispersion equation is a quartic equation for $\omega$. We are interested in the case where the beam is weak, in the sense $n_{\rm b}/\gamma_b^3\ll n_e$. 

In the limit of arbitrarily large $k_z\beta_b$, the four solutions of (\ref{8d.8a}) approach $\omega=\pm\omega_p,k_z\beta_b\pm\omega_p(n_{\rm b}/n_e\gamma_b^3)^{1/2}$. The solution near $\omega=-\omega_p$ is of no interest here, and it is removed by approximating the quartic equation by the cubic equation
\be
(\omega-\omega_p)(\omega-k_z\beta_b)^2
-{n_{\rm b}\over2n_e\gamma_b^3}\,\omega_p\omega^2=0.
\label{8d.9}
\ee
The solutions of the cubic equation simplify in two cases: the ``resonant'' case $k_z\beta_b\approx\omega_p$, and the ``nonresonant'' case $\omega\ll\omega_p$. The approximate solutions for the growth rate in these two cases are \citep{GMG02}
\be
\omega\approx
\left\{\begin{array}{ll}
{\displaystyle\omega_p
+i{\omega_p\over\gamma_b}{\sqrt{3}\over2}\left(\frac{n_{\rm b}}{2n_e}\right)^{1/3}
 }, &{\rm resonant},
\cr
\ms
{\displaystyle k_z\beta_b
+i{\omega_p\over\gamma_b^{3/2}}\left(\frac{n_{\rm b}}{2n_e}\right)^{1/2} 
},
&{\rm nonresonant}.
\end{array}
\right.
\label{8d.10}
\ee
The nonresonant version of the beam instability applies only at frequencies below the resonant frequency, $\omega=k_z\beta_b$. As the resonant frequency is approached the nonresonant instability transforms into the faster-growing resonant instability. Despite having a lower growth rate, the nonresonant instability can result in greater growth due to a longer growth path. As the beam propagates along field lines, $\omega_p$ decreases, and a growing wave can continue to grow only until the value of $\omega_p$ moves it out of resonance; this depends on the bandwith of the growing waves, which is of order the growth rate. This limitation does not apply to the nonresonant form of the instability.

The problem of identifying a large enough growth rate is of long standing. A favored model is based on the assumption that the pair creation is highly structured in space and time. This results in localized clumps of enhanced pair density propagating outward. The counterstreaming is assumed to occur when the faster particles in a following clump overtake the slower particles in the preceding clump \citep{U87,UU88,L92,AM98}.

\begin{figure}[t]
\begin{center}
\includegraphics[width=8cm]{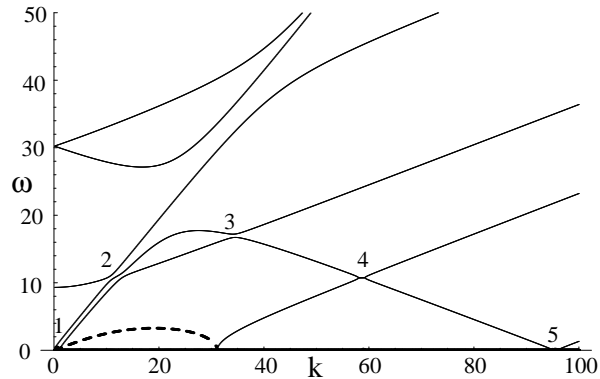}
\caption{Dispersion curves, $\omega$ vs. $k$ for nearly parallel propagation
($\theta=0.1\ {\rm rad}$) in a cold counter-streaming plasma with $\beta = 0.3$,
where $\omega_p=10$ and $\Omega_e=30$ (arbitrary units); dashed lines
are imaginary parts and solid lines are real parts; numbers label points mentioned in the text [from {\br \citet{VM08}}]}
\label{fig:almostparallel}
\end{center}
\end{figure}

\subsubsection{Multiple relativistically streaming cold distributions}

There are several possible counter-streaming motions that can lead to a beam instability: primary particles streaming relative to the background secondary plasma, electrons and positrons counter-streaming associated with a steady-state current, and counter-streaming electrons and positrons associated with large-amplitude oscillations in $E_\parallel$. In a cold-plasma multiple-streaming model, for parallel propagation there are two (real) beam modes (one backward one forward) associated with each streaming distribution, leading to multiple intersections between the various modes. For slightly oblique propagation two intersecting beam modes reconnect becoming two complex conjugate modes, as illustrated in Figure~\ref{fig:almostparallel}. One class of unstable mode is purely growing, with zero real frequency, as illustrated by the dashed curve in Figure~\ref{fig:almostparallel} and by cases 1~and 5~in Figure~\ref{fig:almostparallel}.  In other cases two real modes become a complex conjugate pair of modes, one of which is necessarily a growing mode, as in case~4 in Figure~\ref{fig:almostparallel}. {\br The results illustrated in Figure~\ref{fig:almostparallel} are only for cold distributions of electrons and positrons \citep{VM08}; some of the features survive in the generalization to include a relativistic spread in energies in the rest frame of each distribution \citep{VM11}.}

It should be emphasized that the foregoing discussion of reactive instabilities is based on the assumption that the spread in energies of all distributions of particles can be neglected. This assumption is not justified in conventional models for the particle distributions in a pulsar plasma. 

\subsection{Pulsar radio emission mechanisms}
\label{sect:emission}

Four pulsar radio emission mechanisms, each of which has been in the literature for several decades \citep{M95}, are referred to here as coherent curvature emission (CE), relativistic plasma emission {\br (RPE)}, linear acceleration emission (LAE)  and anomalous Doppler emission. (Free electron maser emission is a further possibility, regarded as a variant of LAE in the following discussion.) Two of these, CE and LAE, do not depend on wave dispersion in the plasma, in the sense that these mechanisms exist in vacuo, and the other two, {\br RPE} and anomalous Doppler emission, depend intrinsically on the dispersive properties of the plasma in the source region.

\subsubsection{Coherent curvature emission}

The 1D motion of a charge along a curved field line involves an acceleration, and the resulting emission by the accelerated charge is CE. The force that causes this acceleration, such that all charges (with $p_\perp=0$) move along the curved field line, is the Lorentz force associated with the curvature drift velocity \citep{CES75}. The emission is due to the accelerated motion, and its properties do not depend intrinsically on the properties of the wave modes in the plasma. 

There is an analogy between CE and synchrotron emission. Both are due to acceleration perpendicular to the direction of motion. Emission by a highly relativistic particle is confined to a narrow forward cone, and an observer sees radiation only if this cone sweeps across the line of sight. Let the radius of curvature of the field line at the point of emission be $R_{\rm c}$. The particle radiates in the direction of the observer for a fraction $\approx1/\gamma$ of the period $2\pi R_{\rm c}/\beta c$ of its motion around the circle. The pulse received by the observer is of duration $\delta t_{\rm rec}\approx\pi R_{\rm c}/c\gamma^3$, which contains frequencies $\omega<\omega_{\rm max}\approx c\gamma^3/\pi R_{\rm c}$.  Coherent CE is attributed to relatively low energy (relativistic) particles, such that $\omega_{\rm max}$ is higher than the observed radio frequencies. Two difficulties with coherent CE as a pulsar radio emission mechanism are its low frequency and the coherence mechanism.

In the polar-cap model, the emission is assumed to come from well inside the light cylinder, where $c/R_{\rm c}$ is very small. For a curve $r=g(\chi)$ in polar coordinates $r,\chi$, one has $R_{\rm c}={(g^2+g'^2)^{3/2}/|gg''-2g'^2-g^2|}$, where a prime denotes a derivative. For a dipolar field line, $r=r_0\sin^2\chi$, with $\chi\ll1$ for $r\ll r_{\rm L}<r_0$, one finds $R_{\rm c}\approx4(rr_0)^{1/2}/3>4(rr_{\rm L})^{1/2}/3$. The characteristic frequency of curvature emission is then $\omega_{\rm max}\approx(3\omega_*\gamma^3/4\pi)(r_{\rm L}/r)^{1/2}$. Models for the pair cascade suggests $\gamma$ between about 10 and 100 \citep{HA01,AE02}. For a slow pulsar with $\omega_*\approx1\rm\,s^{-1}$, it is difficult to explain observed radio emission at $\omega\approx10^{10}\rm\,s^{-1}$. In the early literature it was assumed that non-dipolar components lead to much smaller $R_{\rm c}$ than the dipolar model implies, and hence to a higher $\omega_{\rm max}$.

In early literature the coherence mechanism for CE was assumed to be emission by bunches \citep{BB76,BB77}. Although the assumption of emission by bunches has been strongly criticized \citep{M81,Letal98}, it remains implicit in more recent models that invoke coherent CE.  In the earliest models  the outflowing plasma was assumed to be confined to charge sheets, and the coherence was attributed to these sheets of charge \citep{L70a,L70b,L70c,S71,RS75}. However, the normal to such a sheet needs to be within an angle $\approx1/\gamma$ of the direction of the field line for the coherence to be effective, and this condition cannot be maintained due to the curvature of the field lines \citep{M81}. {\br  More recent models based on CE have invoked other types of bunches, e.g. solitons \cite{GLM04}. Another difficulty is that the particles in the bunch must be nearly mono-energetic. A spread $\Delta\beta c=\Delta\gamma c/\gamma^3$ in speed causes the linear extent of the bunch to increase at a rate $\Delta\beta c$, and coherent emission ceases once this exceeds about a wavelength. } These is no convincing argument that the difficulties associated with coherent CE by bunches can be overcome under the wide variety of conditions required to explain the radio emission from all pulsars.

Coherence due to maser emission is an alternative, but negative absorption is not possible for CE in the simplest approximation. The proof of this \citep{B75,M78} is similar to the proof that synchrotron absorption cannot be negative. \citet{T58} argued that the absorption coefficient can be written as the integral over the emissivity times a momentum-derivative of the distribution function, and that the absorption is negative only if the momentum-derivative is positive, corresponding to $df(u)/du>0$. The integral gives a net negative absorption only if the momentum-derivative of the emissivity, obtained by partially integrating, is negative. The latter condition is not satisfied for synchrotron absorption \citep{WSW63}, implying that synchrotron absorption is positive even if there is a range of $\gamma$ with $df(u)/du>0$. The proof that absorption due to CE cannot be negative is based on the derivative of the emissivity with respect to $\gamma$ being positive \citep{B75,M78}. Similar to synchrotron absorption, which can be negative under special circumstances \citep{McCray66,Z67}, maser CE is possible under special circumstances \citep{ZS79,CS88,LM92,LM92a,LM95}. The necessary condition $df(u)/du>0$ for negative absorption requires that the distribution function have a maximum, with $df(u)/du>0$ for $\gamma$ below the value at which this maximum occurs. {\br Such a maser would occur at too low a frequency, $(R_{\rm c}/c)\gamma^3$ due to the small  value of $\gamma$ implied by the requirement $df(u)/du>0$.} Although maser CE is possible in principle, the various conditions required for it to be an effective radio emission mechanism for all pulsars are not satisfied. 

\subsubsection{Linear-acceleration/free-electron maser emission}

LAE (and free-electron maser emission) also does not depend intrinsically on the dispersive properties of the plasma, and it may be treated assuming vacuum wave properties.

In LAE \citep{C73,M78,KM79} the accelerated motion is due to a parallel electric field, such that the acceleration is parallel to the velocity of the particle. In the simplest model $E_\parallel$ is assumed to be oscillating, at frequency $\omega_0$ say, and its effect on the motion of a particle is treated as a perturbation. The emitted radiation then satisfies $\omega-{\bi k}\cdot{\bi v}=\omega_0$, giving $\omega-{\bi k}\cdot{\bi v}\approx\gamma^2\omega_0/2$ for emission in vacuo by a highly relativistic particle. If the oscillation is associated with a propagating wave, then $\omega_0$ is replaced by $\omega_0-{\bi k}_0\cdot{\bi v}$, where ${\bi k}_0$ is the wave vector of the propagating wave. LAE usually refers to the case where the phase speed is superluminal, $\omega_0/k_0>c$ \citep{R92a,R92b,R95}, when the Cerenkov resonance condition cannot be satisfied.  In free-electron maser emission the motion of the relativistic particle is assumed to be perturbed by an electric or magnetic field with a spatial structure \citep{FK04}. The emission is then analogous to LAE for a subluminal wave, with $\omega_0\to k_0c$. Absorption associated with LAE (or free-electron maser emission) can be negative, at frequencies $\omega\ll\omega_0\gamma^2$, provided the particle distribution satisfies $df(u)/du>0$. 

The assumption that the acceleration by $E_\parallel$ can be treated as a perturbation is not satisfied for large-amplitude oscillations. Specifically, if $E_\parallel$ is sufficiently large to modify the motion of a particle substantially, the properties of LAE are correspondingly modified \citep{MRL09,ML09,RK10}. %In a sufficiently large-amplitude wave, the motion of electrons and positrons corresponds to relativistic counter-streaming, in which case the associated streaming instability needs to be taken into account, and may dominate any maser LAE. This suggests that relativistic plasma emission may be more favorable as an emission mechanism than maser LAE.
{\br Maser emission due to such LAE is possible, but no detailed model for it is available.}

\subsubsection{Relativistic plasma emission (RPE)}

The concept of plasma emission is based on the generation of Langmuir-like waves through a beam instability and production of escaping radiation from the resulting Langmuir-like turbulence. Any specific generalization of plasma emission to a pulsar magnetosphere involves two stages: the generation of Langmuir-like waves and partial conversion of the wave energy into escaping radiation. {\br There are major difficulties with both stages. Besides the difficulty with the growth rate of any beam instability, as discussed above, a less recognized difficulty concerns the very existence of ``Langmuir-like waves'' in pulsar plasma.

%\subsubsection{Cerenkov resonance in pulsar plasma}

The maser form (and the resonant reactive form) of a beam instability requires that the waves and particles satisfy the Cerenkov resonance condition, which reduces to {\br $\omega-k_\parallel v_\parallel=\omega(1-n\beta\cos\theta)=0$ or $z-\beta=0$} in the 1D case. This condition can be satisfied only for waves with refractive index $n>1$, or $z=\omega/k_\parallel c<1$ in the 1D case. In a pulsar plasma with a relativistic spread in energies, the only weakly-damped waves that have $n>1$ in the radio range have $n-1\ll1$.  This property severely restricts the conditions under which RPE can occur in a pulsar plasma \citep{MG99}.

For parallel propagation a beam instability is possible in principle only for the L~mode at $\omega>\omega_1$, where the dispersion curve is in the range $z<1$, cf.\ Figure~\ref{fig:dp20}. For the oblique LO~mode the dispersion curve is in the range $z<1$ only for very small $\tan\theta$. For $\tan\theta=0$, the resonance condition $\beta=z$ with $z_A<z<1$ requires  $\gamma=\beta_A$ at $z=z_A$ ($\omega=\omega_{\rm co}$) increasing to $\gamma\to\infty$ for $z\to1$ ($\omega\to\omega_1$); the constraint on the required $\gamma$ increases further for $\tan\theta\ne0$. RPE due to waves generated in the LO~mode though a beam instability would have the attractive feature that such waves can escape directly (without any second stage in the plasma emission), but the condition $\gamma>\beta_A$ for this to occur seems unrealistic. Specifically, negative absorption requires that the distribution of particle has a positive slope, $df(u)/du>0$, in the streaming direction, and this requires that $\gamma$ be below the peak in the distribution function, which is inconsistent with $\gamma\gg\beta_A$ for the large values of $\beta_A$ expected in a pulsar magnetosphere.

A beam instability is not possible for the X~mode, or for the parallel A~mode, because their polarization vector is orthogonal to the magnetic field, and hence to the current associated with the beam.  {\br Also, except at very small $\tan\theta$, the oblique X~mode and the LO~mode have $z>1$ implying that $\beta=z$ is impossible. }

%\subsubsection{Estimate of $\beta_A^2$ in pulsar magnetosphere}

The parameter $\beta_A^2=\Omega_e^2/\omega_p^2\langle\gamma\rangle$, which plays an important role in the foregoing discussion of RPE, is very large near the star and decreases with increasing $r$. Using equation (\ref{omegaGJ}) in the form $\omega_p^2=\kappa2\pi\Omega_e/P$, gives $\beta_A^2=\Omega_eP/2\pi\kappa\langle\gamma\rangle$. A dipolar magnetic field implies $\Omega_e\propto1/r^3$ and with the surface magnetic field given by equation (\ref{Bstar1}), one has $\Omega_e=5.6\times10^{26}(P{\dot P})^{1/2}\rm\,s^{-1}$ at the surface. For a normal pulsar with ${\dot P}=10^{-15}$ and $P=1$, assuming $\kappa=10^5$ and $\langle\gamma\rangle=10$  gives $\beta_A^2$ of order $3\times10^{12}$ at the stellar surface and of order $30$ at the light cylinder, varying $\propto(r_{\rm L}/r)^3$ in between.  The resonance condition cannot be satisfied for plausible values, except perhaps near the light cylinder.  

An implication of these estimates is that the first stage of RPE, that is, generation of Langmuir-like waves through a beam instability, cannot plausibly occur in a pulsar magnetosphere. The second stage involves nonlinear processes in a pulsar plasma \citep{I88}, and this presents additional difficulties, referred to by \citet{U00} as a ``bottle-neck'' in the emission process.

The assumption that RPE is due to beam-driven Langmuir-like waves ignores the intrinsically relativistic spread in energies of the electrons and positrons in their rest frame. As indicated in Section~\ref{sect:instabilities}, in treating beam instabilities it has been conventional to assume the spread in energies is negligible, corresponding to the particle distributions being cold in their respective rest frames. A generalization of this approach is to assume that the spread is nonrelativistic in the rest frame, and to adapt models for streaming instabilities in a nonrelativistic plasma to the pulsar case \citep{W94,W97,W98}. %However, a better understanding of the viability or otherwise of relativistic plasma emission as a pulsar radio mechanism will require resolution of this disconnect in the literature.
Such a model is not consistent with the relativistic spread in energies thought to apply in a pulsar plasma generated through a pair cascade \citep{HA01,AE02}.}

\subsubsection{Anomalous Doppler emission}

The anomalous Doppler resonance corresponds to $s=-1$ in equation (\ref{gyroresonance}). The relevant anomalous Doppler instability \citep{MU79,KMM91a,LMB99a,LMB99b} corresponds to particles (electrons or positrons) in their lowest Landau orbital jumping to the first excited state, that is from $n=0$ to $n'=1$ in equation (\ref{pce}).  {\br An} attractive feature of this mechanism is that the absorption is negative if the occupation number of the initial state, $n=0$, is greater than that of the final state, $n'=1$, and this is obviously the case for the 1D pair distribution in a pulsar magnetosphere. It seems plausible that this instability should occur in regions of the pulsar magnetosphere where the cyclotron frequency is sufficiently small. {\br However}, this mechanism depends on the dispersive properties of the plasma, and the difficulties discussed above for  {\br RPE also apply to} anomalous Doppler emission.

One may write the resonance condition for $s=-1$, $v_\parallel=\beta c$ in the form
\be
\omega(1-n\beta\cos\theta)+\Omega_e/\gamma=0,
\label{s=-1}
\ee
where $n$ is the refractive index. The resonance condition (\ref{s=-1}) can be satisfied only for $n\beta\cos\theta>1$, which requires $n>1$. As for the Cerenkov resonance, for X~and (nearly parallel) O~mode waves with $n-1\approx1/2\beta_A^2\ll1$, the resonance requires $\gamma>\beta_A$. Then equation (\ref{s=-1}) {\br  implies $\omega\approx2\beta_A^2\Omega_e/\gamma$. This mechanism requires emission by particles with very high $\gamma$ in order for the frequency to be in the radio range. To see this note that the order-of-magnitude estimates made above for $\Omega_e$ and $\beta_A^2$ imply an emission frequency of order $(10^{26}/\kappa\langle\gamma\rangle)({\dot P}/P^4)(r_{\rm L}/r)^6\rm\,Hz$. Assuming ${\dot P}/P^4$ of order $10^{-15}$ and $\kappa\langle\gamma\rangle$ of order $10^6$, gives a frequency is of order $(10^5/P)(r_{\rm L}/r)^6/\gamma\rm\,Hz$ in the rest frame. In the pulsar frame the frequency is boosted by a factor of $\gamma_{\rm s}$, thought to be of order $10^2$--$10^3$. For emission at a height $r=0.1r_{\rm L}$, even for emission by primary particles (with $\gamma$ of order $10^6$--$10^7$) it is questionable whether this mechanism can account for the lowest observed emission frequencies. As with RPE, the effect of a relativistic spread $\langle\gamma\rangle$ on the wave dispersion imposes a severe constraint on this emission mechanism.

}

\subsubsection{Discussion of the radio emission mechanism}

The foregoing discussion is inconclusive in that none of the possible emission mechanisms is obviously much more favorable than the others. It is reasonable to conclude that none of them seems plausible, allowing wide scope for differing opinions on possible emission mechanisms. 

There are several general questions that need to be answered in order to make progress in identifying the emission mechanism:
\begin{itemize}

\item {\it Are there multiple emission mechanisms?} There is a long-standing difference of opinion among observers as to whether there is only a single pulsar radio emission mechanism or whether two or more different emission mechanism operate in the same pulsar or in different pulsars. An argument for two emission mechanisms is based on a distinction between core and coronal emission \citep{R83b,R83a,R86,R90}, and a counter-argument that no such distinction exists \citep{LM88}. From a theoretical viewpoint, we are unable to clearly identify one viable emission mechanism, and appealing to two or more exacerbates the problem.

\item {\it Is there a characteristic emission frequency?} There is no definitive constraint on the emission frequency, analogous to $\omega_p,2\omega_p$ in plasma emission and $\Omega_e$ in ECME. Some of the suggested emission mechanisms involve a natural frequency but these frequencies depend on poorly determined parameters. Although a general radius-to-frequency mapping is implied by the broadening of the pulse window with decreasing frequency, implying that lower frequencies originate from larger radii, the range of frequencies that originates from a given height is not well determined.  {\br A remarkable feature of pulsar radio emission is that it occurs in roughly the same range for most pulsars. This suggests that if there is a natural frequency, then it is of the same order of magnitude for all pulsars. A frequency $\propto\omega_p\approx(\kappa2\pi\Omega_e/P)^{1/2}\propto({\dot P}/P^3)^{1/4}$ satisfies this requirement, but encounters a difficulty in accounting for the observed frequencies \citep{Ketal98}. }

\item {\it Where is the source of the emission located in the magnetosphere?} Besides the uncertainty in the height, or radius, of the emission at a given frequency, the location of the source within the polar-cap region is poorly determined, with a general preference for emission from near the last closed field line. One approach to constraining the source region is to appeal to retardation and aberration effects \citep{GG03}, but the results are not compelling. {\br The opinion that the emission comes from high in the magnetosphere is supported by direct evidence from the double-pulsar system \citep{L10,LL14}.}

\item {\it What drives the coherent emission?} The source of free energy that drives the coherent emission is well accepted for both plasma emission and ECME, but the source of free energy that drives pulsar radio emission is not known. It seems likely that the radio emission is related to the pair creation, as in the case of EAS discussed in the next section, but no causal relation has been suggested.  One could speculate that the causal relation involves the accelerating electric field, $E_\parallel$, as in LAE, but there is no compelling evidence that this is the case.

\end{itemize}

\section{Coherent Emission from Cosmic-Ray Showers}
\label{sect:EAS}

Although coherent radio emission in an extensive air shower (EAS) in the Earth's atmosphere does not fall into the category of a coherent emission in an astrophysical plasma, it is potentially relevant to pulsar emission in that both are associated with a pair cascade. 

\subsection{Radio emission from an EAS}

Incoherent optical Cerenkov emission has long been recognized as a diagnostic for high energy cosmic rays, but its usefulness is limited because the Cerenkov light can be detected only during moonless, cloudless nights. Coherent radio emission has become an important diagnostic, and several different experiments have established it usefulness {\br \citep{H13}, for example, the LOPES \citep{Fetal05}, CODALEMA \citep{Aetal05}, and  MIDAS  \citep{Wetal10} experiments}.

Coherent emission in an EAS includes amplified Cerenkov emission, but it was recognized in the early literature that it is not the dominant coherent radio mechanism \citep{KL66,C67}, and all relevant mechanisms are now incorporated in modelling EASs \citep{Jetal11,H13}. Any time-varying current gives rise to radio emission, and one can associate a specific emission mechanism with a specific current. Besides the current associated with streaming particles, which produces Cerenkov emission, another relevant current results from the changing charge density is the EAS, for example, due to preferential loss of positrons due to annihilation with ambient electrons. A further current results from the deflection of electrons and positrons in opposite directions by the Earth's magnetic field; this current is in the direction perpendicular to both the axis of the EAS and to the magnetic field.  While Cerenkov emission alone is inadequate in describing the actual coherent emission in an EAS, it is convenient to concentrate on Cerenkov emission to explain the different methods that have been used. In particular, three different models for coherent emission can be illustrated using Cerenkov emission as an example: negative absorption, coherent emission from a bunch  treated as a continuum, and coherent emission from a bunch treated as a collection of discrete particles. 

\subsubsection{Cerenkov resonance}
  
The Cerenkov resonance $\omega-{\bi k}\cdot{\bi v}=\omega(1-n\beta\cos\theta)=0$ is possible in air due to the refractive index, $n$, being slightly greater than unity in the radio range. It is conventional to write $n$ in terms of the radio refractivity, $N$,
\begin{equation}
n=1+N\times10^{-6}.
\label{refractivity}
\end{equation}
In the following discussion the value $N=N_0=315$ is assumed. Cerenkov emission is possible for a particle with speed $\beta c$ satisfying $n\beta>1$, which requires $\gamma>\gamma_0=10^3/(2N_0)^{1/2}\approx40$, corresponding to an electron (or positron) with energy $\varepsilon\gtrsim20\,$MeV. An electron with $\gamma>\gamma_0$ emits Cerenkov radiation at the Cerenkov angle, $\Theta$, satisfying $1-n\beta\cos\Theta=0$, which corresponds to
\be
\Theta\approx\frac{1}{\gamma\gamma_0}(\gamma^2-\gamma_0^2)^{1/2}.
\label{Cc1}
\ee
Emission and absorption occurs only at the Cerenkov angle, which increases from $\Theta=0$ at $\gamma=\gamma_0$ to $\Theta=1/\gamma_0$ for $\gamma\gg\gamma_0$.

\subsubsection{Cerenkov emission and absorption}

Emission by a charge $q$ moving along a trajectory, ${\bi x}={\bi X}(t)$, can be treated in a general way by including the current due to the charge as a source term in (the Fourier transformed) Maxwell's equations, and including the response of the medium in terms of the associated current induced by the field of the moving charge. The current density associated with a single particle (sp) is ${\bi J}_{\rm sp}(t,{\bi x})=q{\bi v}(t)\delta^3[{\bi x}-{\bi X}(t)]$, where ${\bi v}(t)=d{\bi X}(t)/dt$ is its instantaneous velocity. Cerenkov emission is due to a charge in constant rectilinear motion, and in this case the trajectory may be written as ${\bi X}(t)={\bi x}_0+{\bi v}t$, where ${\bi x}_0$ is the position of the charge at $t=0$. The Fourier transform of the current is
\be
{\tilde{\bi J}}_{\rm sp}(\omega,{\bi k})=\int dt\,d^3{\bi x}\,{\bi J}_{\rm sp}(t,{\bi x})e^{i(\omega t-{\bi k}\cdot{\bi x})}
=e^{-i{\bi k}\cdot{\bi x_0}}q{\bi v}\,2\pi\delta(\omega-{\bi k}\cdot{\bi v}),
\label{Pcoh1}
\ee
where the final expression applies for Cerenkov emission. The current (\ref{Pcoh1}) is identified as the source terms ${\bi J}_{\rm ext}$, in the wave equation (\ref{we1}), which is solved to find the electric field that it generates. The rate the current does work against the electric field that it generates is identified as the power emitted.

The power, $P({\bi k})$ in the range $d^3{\bi k}/(2\pi)^3$, is proportional to $|{\bi e}^*\cdot{\tilde{\bi J}}(\omega,{\bi k})|^2$, where ${\bi e}$ is the polarization vector, and $\omega$ and ${\bi k}$ are related by a dispersion relation, which is $\omega=kc/n$ in the case of air. Summing over the two states of polarization, the single-particle emission reduces to
\be
P_{\rm sp}({\bi k})=\frac{q^2c^2|\bkappa\times\bbeta|^2}{2\varepsilon_0\omega n^2}2\pi\delta(1-n\beta\cos\Theta),
\label{Pcoh2}
\ee
with $\bbeta={\bi v}/c$ and ${\bi k}=(n\omega/c)\bkappa$. The power emitted per unit volume by a distribution of electrons is given by multiplying by the distribution function, $f({\bi p})$, and integrating over $d^3{\bi p}$. The absorption coefficient is given by a similar formula with $f({\bi p})$ replaced by $-{\bi k}\cdot\partial f({\bi p})/\partial{\bi p}$, cf.\ equation (\ref{kbi3}). 

Negative absorption results in amplified Cerenkov emission for ${\bi k}\cdot\partial f/\partial{\bi p}=\omega\partial f/\partial\varepsilon>0$. The condition $\partial f/\partial\varepsilon>0$ is satisfied at energies below where the distribution peaks, which depends on the energy of the primary cosmic ray that triggers the EAS. Only the square of the charge appears in equation (\ref{Pcoh2}), and hence in the emission and absorption coefficients. It follows that electrons and positrons contribute in the same way, in the sense that the distribution function, $f$, may be interpreted as the sum of the electrons and positron contributions when evaluating the emission and absorption coefficients. 

{\br Negative absorption is also possible for three emission mechanisms that are related to Cerenkov emission: creation, annihilation and transition emissions. The current associated with creation or annihilation emission is of the form given by equation (\ref{Pcoh1}) with the $\delta$-function replaced by a Lorentzian line profile, and both can occur in vacuo. Transition emission may be regarded as the combination of annihilation and creation emission as a particle crosses the boundary between two semi-infinite media, one of which may be the vacuum. Negative absorption for transition emission was discussed by \citet{PF97}. 

}

\subsubsection{Coherent Cerenkov emission by a bunch}

In a continuum model for a bunch of particles, the distribution of particles is regarded as a single macro-charge with a spatial distribution described by its charge density, $\rho(t,{\bi x})$, implying the current density ${\bi J}(t,{\bi x})=\rho(t,{\bi x}){\bi v}$, where it is assumed that all the particles in the bunch have the same velocity ${\bi v}$. This corresponds to replacing the charge density, $\rho_{\rm sp}(t,{\bi x})=q\delta^3[{\bi x}-{\bi x}_0-{\bi v}t]$, for a single charge, by $\rho(t,{\bi x})={\bar q}n({\bi x}-{\bi x}_0-{\bi v}t)$ for the distribution of charge, with the mean charge ${\bar q}=-e,e$ and $0$ for a bunch of electrons, positrons and pairs, respectively. The current associated with the bunch gives
\be
{\tilde{\bi J}}(\omega,{\bi k})={\bar q}{\bi v}\,{\tilde n}({\bi k})\,2\pi\delta(\omega-{\bi k}\cdot{\bi v}),
\qquad
{\tilde n}({\bi k})=\int d^3{\bi x}\,n({\bi x})e^{-i{\bi k}\cdot{\bi x}}.
\label{Pcoh3}
\ee
The power in coherent emission from the bunch is 
\be
P_{\rm coh}({\bi k})=\frac{{\bar q}^2}{e^2}|{\tilde n}({\bi k})|^2P_{\rm sp}({\bi k}).
\label{Pcoh4}
\ee
A Gaussian model for the bunch of $N$ particles with velocity along the $z$ axis is
\be
n({\bi x})=N\frac{e^{-(x^2+y^2)/2R_\perp^2-z^2/2Z^2}}{(2\pi)^{3/2}R_\perp^2Z},
\qquad
{\tilde n}({\bi k})=Ne^{-(k_\perp^2R_\perp^2+k_z^2Z^2)/2},
\label{Pcoh5}
\ee
where $R_\perp$ and $Z$ are constants. In the limit $k_\perp^2R_\perp^2+k_z^2Z^2\to0$ the power emitted by the bunch is $N^2({\bar q}/e)^2$ times the power emitted by a single charge, so that the bunch emits like a single electron with charge $-N{\bar q}$.

\subsubsection{Coherent Cerenkov emission by a collection of charges}

Emission by a collection of $N$ individual charges may be treated by replacing the single-particle current by the sum of the currents due to each of the charges. Let the $i$th electron or positron, with charge $q_i=\mp e$, have velocity ${\bi v}_i$ and be at ${\bi x}={\bi x}_i$ at $t=0$. The current (\ref{Pcoh1}) becomes
\be
{\tilde{\bi J}}(\omega,{\bi k})=\sum_ie^{-i{\bi k}\cdot{\bi x_i}}q_i{\bi v}_i\,2\pi\delta(\omega-{\bi k}\cdot{\bi v}_i),
\label{Pcoh6}
\ee
where the sum is over all the charges. The solution of the wave equation for the electric field is then a sum over the electric field due to each charge. The power emitted, $\propto|{\tilde{\bi J}}(\omega,{\bi k})|^2$, involves a double sum, over $i,j$ say. The $N$ terms with $i=j$ correspond to incoherent (spontaneous) emission from each of the charges. In this model the coherent emission is described by the $N(N-1)$ terms with $i\ne j$. Such a model is not restricted to Cerenkov emission; for example, \citet{WM82} applied such a model to coherent gyromagnetic emission. For Cerenkov emission, the $\delta$-functions for $i$ and $j$ must be satisfied simultaneously, implying that the contribution to coherent emission from particles $i,j$ is nonzero only for ${\bi k}\cdot({\bi v}_i-{\bi v}_j)=0$. A geometric interpretation is that, for ${\bi x}_i\ne{\bi x}_j$ and ${\bi v}_i\ne{\bi v}_j$, the surfaces corresponding to the two Cerenkov cones associated with the two particles intersect only along specific curves, and the coherent emission observed from charges $i,j$ is attributed to the observer being located on the relevant curve. An alternative interpretation, that applies for any emission mechanism, involves inverting the Fourier transform to identify the electric field at $t,{\bi x}$ associated with each of the particles; the square of the total electric field far from the emission region includes the cross terms between the electric fields from particles $i,j$.

\subsection{Simulations of radio emission from EASs}

Coherent emission in an EAS is conventionally treated using various air shower codes, for example, ``ZHAireS'' \citep{ACZ12}. In these codes the emission by individual charges is described using the Li\'enard-Wiechert potentials, modified from their form in vacuo to apply to an isotropic medium with refractive index $n$. The individual charges are assumed to move along trajectories, called tracks, and the interference between the emission from different tracks is taken into account. Conceptually, the simulation models are similar to emission by a collection of charges, as discussed above in connection with Cerenkov emission, with constant rectilinear motion replaced by the actual motion along the track. The Li\'enard-Wiechert potentials give the electromagnetic field at an arbitrary point (identified as the position of the radio receiver here) as a function of time in terms of the position on the track at the retarded time. The algorithm used does not assume any specific emission mechanism and includes the effects of the charge imbalance between electrons and positrons and of the geomagnetic field. 

The simulations of EASs do not identify a specific emission mechanism directly. The Cerenkov effect is important in producing ring-like structures in the emission pattern, and synchrotron-like emission is clearly important in the simulations. An adaption of methods used in treating synchrotron emission in astrophysics to treat synchrotron emission in air \citep{RM15} implies that the emission should be synchrotron-like at lower frequencies and Cerenkov-like at higher frequencies.

The success of the Monte Carlo approach in modelling radio emission from EASs raises the question as to whether it could be adapted to apply to pulsar radio emission. The use of the Li\'enard-Wiechert potentials is only valid for a constant (frequency-independent) refractive index, and this condition is not satisfied for wave dispersion in a pulsar plasma. The method can be generalized to find the electric field ${\bi E}(\omega,{\bi k})$ due to each charge, with the trajectory of each charge including all relevant effects, such as curvature of the field lines and acceleration by an oscillating $E_\parallel$. The electric field ${\bi E}(t,{\bi x})$ of each charge is found by inverting the Fourier transform. The total electric field at $t,{\bi x}$ due to all the particles can then be modelled using the Monte Carlo approach. It is of interest to develop such a model, which describes a form of coherent emission that has not been explored in connection with pulsars. The feasibility of doing so has yet to be tested.

\section{Phase Coherence}
\label{sect:coherent}

The concept of coherence is not well-defined in radio astronomy. Coherent emission is identified only in terms of its brightness temperature, $T_B$, when it is too high to be explained in terms of any incoherent emission mechanism. This identification is not related directly to phase coherence. Indeed, in the language of quantum optics, specifying the brightness temperature is equivalent to specifying the photon occupation number, and there is an uncertainty relation between the occupation number and the phase. Hence specifying $T_B$ implies uncertainty in the phase, corresponding to the random phase approximation (RPA). Some more direct measure of phase coherence is desirable.

Before discussing potentially observable quantities related to the phase coherence, it is relevant to note that all observed forms of coherent emission seem to include fine structures that are inconsistent with the RPA. Specifically, the only form of coherent emission consistent with the RPA is maser emission, and this requires that the bandwidth (which is the phase-mixing rate) of the growing wave exceed the growth rate. For fine structures in a sufficiently narrow frequency range, this condition must be violated.

\subsection{Fine structures}

Fine structures in solar radio bursts are discussed in Section~\ref{sect:modified}, and most of these can be explained in terms of various modified forms of plasma emission. However, ECME and several suggested pulsar emission are direct emission processes, in the sense that the output of the maser is identified as the escaping radiation. For such direct maser emission, fine structure can imply phase-coherent wave growth, rather than the random-phase growth implicit in a maser theory.

Three relevant examples of fine structures are considered here:
\begin{itemize}
\item Fine structures in DAM were observed by \citet{E73}, who noted the similarity with discrete VLF emissions in the terrestrial magnetosphere.

\item  Observations by \citet{CR99} with very high time resolution led to the identification of phase-coherent features in DAM S bursts.

\item Observations of the Crab pulsar on nanosecond timescales \citep{HE07,EH16} led to the identification of fine structures called nano-shots.
\end{itemize}

Consider a fine structure on a dynamic spectrum, consisting of a curve with narrow widths in both frequency and time, $\Delta\omega$ and $\Delta t$ say. To produce such a fine structure requires growth of a narrow-frequency signal. Effective growth requires a large growth factor, requiring at least several tens of growth times. Assuming maser growth implies a growth rate $<\Delta\omega$ and equating the growth time to $\Delta t$ suggests that any fine structure with $\Delta\omega\Delta t\lesssim10$--100 is inconsistent with maser emission. This suggests that a phase-coherent form of wave growth is required. A possible archetype for such wave growth is provided by discrete VLF emissions in the terrestrial magnetosphere.

\begin{figure}
\begin{center}
 \includegraphics[scale=0.25, angle=0]{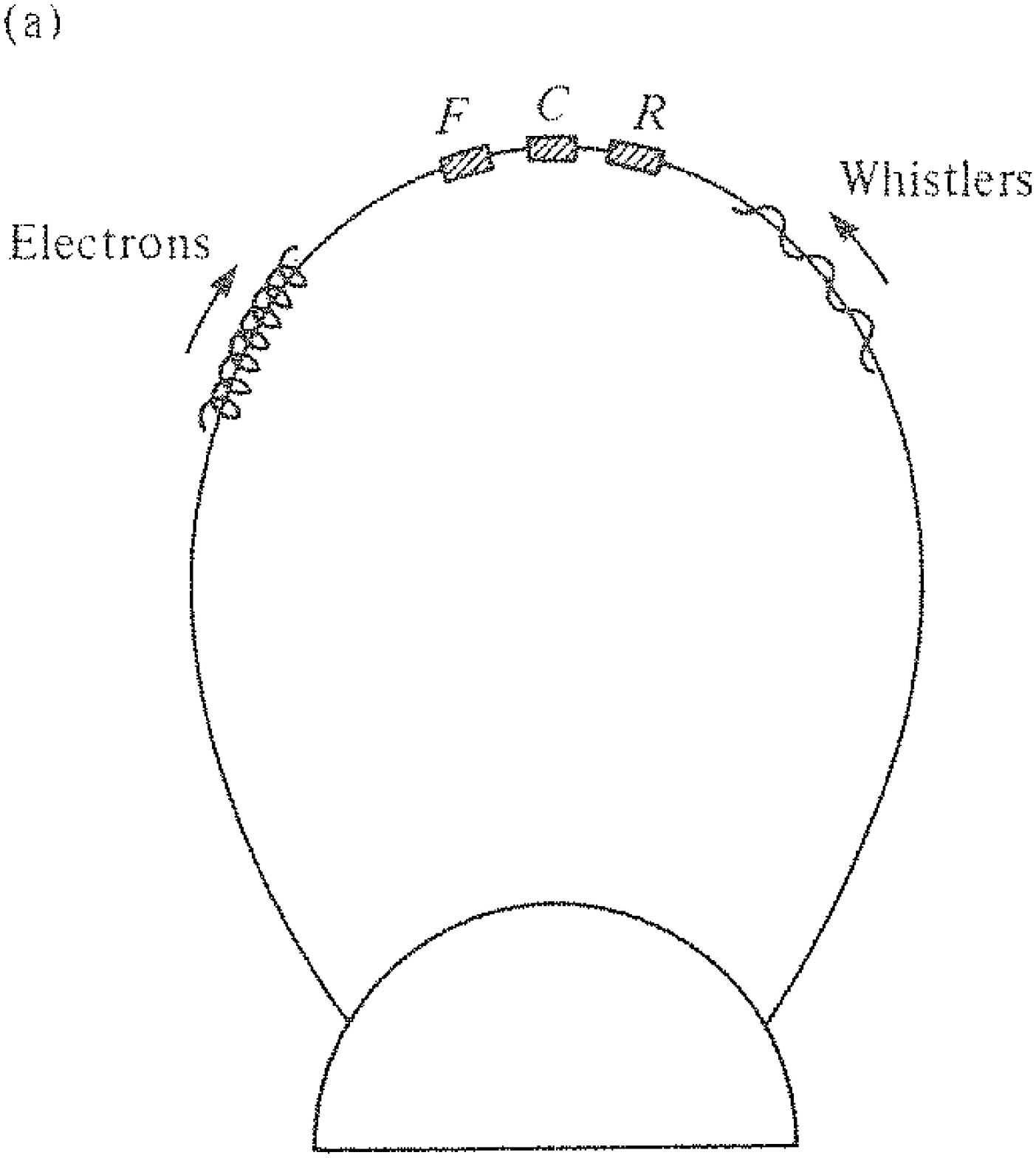}
 \includegraphics[scale=0.25, angle=0]{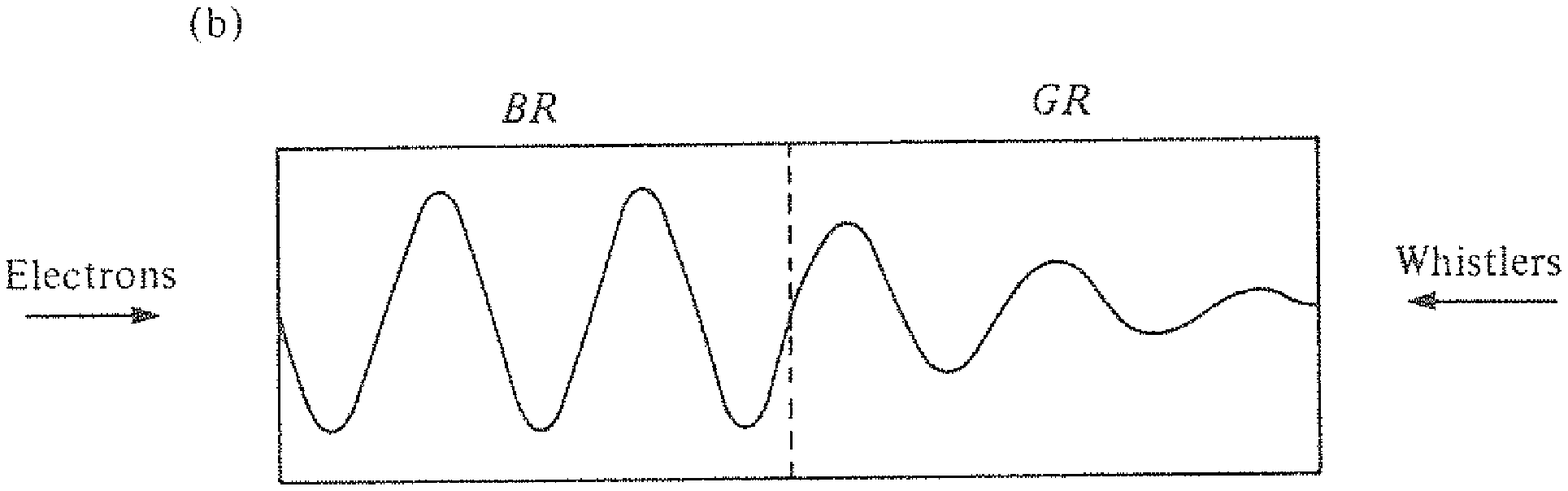}
\caption{The model of \citet{H67} for discrete VLF emissions [from \citet{M86}]}
\label{fig:VLF} 
\end{center}     
\end{figure}

\subsubsection{Discrete VLF emissions}

Discrete VLF emissions in the magnetosphere are whistler waves excited by resonant interactions with electrons \citep{H65}. These emissions can be triggered by radio emission from the ground, a notable feature being triggering by Morse code dashes (150$\,$ms) but not by Morse code dots (50$\,$ms). Individual narrow-band VLF emissions drift in frequency, and exhibit a rich variety of features. \citet{H67} developed a phenomenological model to account for these emission in terms of a resonant interaction between whistlers and electrons in an interaction region (IR). In this model the IR drifts along dipolar field lines such that the wave properties, including the frequency, are changing while preserving the resonance condition. The resonance condition, $\omega-\Omega_e-k_\parallel v_\parallel=0$ can be approximated by $\Omega_e+k_\parallel v_\parallel=0$, requiring that the resonant electrons and whistler waves propagate in opposite directions along the magnetic field lines, $k_\parallel v_\parallel<0$. Assuming parallel propagation, the resonant velocity, $v_\parallel=v_{\rm res}$, corresponds to $v_{\rm res}=-\Omega_e/k$ with $k^2=\omega_p^2\omega/\Omega_ec^2$. As illustrated in Figure~\ref{fig:VLF},  resonant electrons enter the IR from one side, $BR$, and their phase-bunching by the waves increases as they propagate to the right; the waves enter the IR from the other side, $GR$, and their amplitude grows spatially within the IR due to the resonant interaction with the phase-bunched electrons.

Analytic models for the growth of discrete VLF emissions involve a reactive instability \citep{N74,Oetal91,N15} and associated trapping of particles in the growing wave \citep{RP78}. Trapping of particles in a finite-amplitude electrostatic wave is described by equations (\ref{wt1})--(\ref{wt3}). An analogous effect occurs in a finite-amplitude wave growing through a reactive cyclotron instability. As already remarked, there are reactive counterparts of parallel-driven and perpendicular maser cyclotron instabilities, and these are associated with axial and azimuthal bunching, respectively \citep{SD77,W83}. In the application to whistlers, the relevant bunching is in the difference between the phase of the wave and the gyrophase of the electron, cf.\ \S11.6 and \S13.3 in \citet{M86}. The trapping frequency in a parallel-propagating whistler wave may be approximated by
\be
\omega_{\rm t}=\left(\Omega_ekv_\perp\frac{B_w}{B}\right)^{1/2},
\label{VLF1}
\ee
with $k=n_w\omega/c$, $n_w\approx(\omega_p^2/\omega\Omega_e)^{1/2}$, and where $B_w=n_wE_w/c$ is the magnetic field in a whistler wave with electric field $E_w$.

\subsubsection{Trapping models for fine structures}

Fine structures with sufficiently narrow bandwidths imply phase-coherence and growth of a phase-coherent wave suggests a reactive instability. As a reactive instability develops, the growing wave traps particles, and this trapping and associated phase bunching underlies the wave growth. In the case of ECME, such bunching is in azimuthal phase, similar to that in discrete VLF emissions. The superficial similarity between the fine structures observed in DAM \citep{E73} and discrete VLF emissions suggests that a modified form of the model of \citet{H67} for VLF emissions applies to fine structures in DAM \citep{M86b,W02}. In such a model, the growing wave is a large-amplitude x~mode wave that traps resonant electrons about a specific relative phase between the wave and the gyrating electron. The electron distribution needs to be in a marginally stable (or unstable) state to allow this trapping to transfer energy from the resonant electrons to the growing wave. The trapping modifies the local (in the region of wave growth) distribution function of the electrons, such that the tendency to growth is reduced.

It seems plausible that some form of trapping model might also be relevant to fine structures in pulsar radio emission. For example the model of \citet{W97,W98} for nanostructures in the Crab pulsar incorporates some of these ideas. More generally, imposing the requirement that the (unknown) pulsar radio emission mechanism must be capable of accounting for the nano-shots in the Crab pulsar is a potentially useful constraint on possible emission mechanisms, requiring that it allow phase-coherent wave growth and associated particle trapping in the growing wave.

\subsubsection{Compatibility with random-phase models}

Such trapping models suggest a different model for coherent emission than the maser-based models discussed above. The alternative model involves individual bursts of phase-coherent growth, and associated particle trapping in the growing wave, with a statistically large number of such individual bursts of growth occurring in the source region at any given time. In such a model, the particle distribution is maintained in a marginally stable state with the driving towards instability being balanced by a large number of individual bursts of  wave growth. This is similar to the model for intermittency in the growth of Langmuir waves discussed in Section~\ref{sect:IPM}, with the difference here being that the individual bursts of wave growth are phase-coherent. Superficially, this might seem inconsistent with a maser model, and the associated quasilinear relaxation of the particle distribution, whose derivation assumes the random phase approximation. However, there is a counter-argument \citep{MC89} that suggests that the maser/quasilinear treatment can remain valid even if the individual bursts of wave growth are phase coherent.

The argument is that the quasilinear equations for the waves and the particles can be derived by considering any statistical distribution of individual bursts of wave growth that involve transfer of energy (and momentum where relevant) between particles and waves. One can construct a transfer equation that describes the wave growth and a quasilinear equation (e.g., a Fokker-Planck equation) that describes the back reaction on the distribution of particles. Rather than requiring random phases in treating the wave growth, such a model requires only that the phases (and other properties) of the individual bursts of wave growth be uncorrelated.

\subsection{Measurement of coherence}

The only widely accepted criterion for ``coherent'' emission in astrophysics is that the brightness temperature is too high to be explained by any incoherent mechanism. The observation of phase-coherent features, such as in DAM S bursts \citep{CR99}, suggests that direct observation of phase-coherent features is possible. A measure that describes the degree of phase coherence is desirable. One such measure is familiar in quantum optics. Photon counting statistics are poissonian for a phase coherent signal, but are subject to photon bunching for a random-phase signal. A purely classical counterpart involves fluctuations in the intensity of the signal. One specific measure of coherence is the ratio $\langle I^2\rangle$ of the mean square intensity to the square of the mean intensity $\langle I\rangle$. More generally, one may write
\be
\langle I^N\rangle=g_N\langle I\rangle^N.
\label{IN}
\ee
For a phase-coherent signal one has $g_N=1$ and for a random-phase signal one has $g_N=N!$. Thus, for $N=2$, $1\le g_2\le2$ is a quantitative measure of the degree of coherence. For polarized emission, described by the Stokes parameters $I,Q,U,V$ one may generalize equation (\ref{IN}) with $N=2$ to the average of a product of any pair of $I,Q,U,V$ to the product of the averages, resulting in $g_2$ being generalized to ten polarization-dependent coefficients.

Consider the model suggested by \citet{CR99}, that the observed signal consists of ``a superposition of groups of pulses from closely spaced but independent short-lived emission centers that are nearly monochromatic but differ slightly in frequency''. Measurements with increasing resolution should show $g_2$ decreasing from 2 to 1 as a function of the sampling time and bandwidth, as the number of overlapping signals reduces from many to one. Any such measurement would provide new information on the coherence properties of the source, allowing specific models to be tested. As suggested by the foregoing discussion of discrete-VLF-like structures, one possible model for coherent emission involves a statistical distribution of phase-coherent structures that grow through a reactive instability and saturate due to trapping of particles in the wave, and this is consistent with the model suggested by \citet{CR99}.

\section{Conclusions}
\label{sect:conclusion}

Coherent emission is somewhat loosely defined as any emission that is too bright to be explained in terms of any incoherent emission mechanism. Each of the three forms of coherent emission discussed in this paper has been known for between five and seven decades, but over that time the theories developed to describe them have progressed at different rates and reached different levels of maturity. 

Plasma emission and its application to solar radio bursts is a mature field in which the underlying ideas are well established. The theory accounts well for general features of solar radio bursts and for many detailed features of emissions observed from the solar corona, the interplanetary medium and other sources of plasma emission. However, there  remain important aspects of solar radio bursts that are inadequately understood, notably type~I emission and various radio continua. It seems likely that future progress in understanding these aspects will involve modifications of the basic theory, rather than an intrinsically new theory. 

ECME is also a mature field but with some uncertainties concerning important details. In the application to AKR there are in situ data on the electron distribution that drives the ECME and on the plasma properties in the source region, and these data provide strong support for the horseshoe-driven version of ECME. However, in the application to DAM there is evidence (from the emission pattern and the elliptical polarization) that favors a loss-cone driven model. Which of these two versions of ECME operates in solar spike bursts and the emission from flare stars is uncertain. Despite such uncertainties, the importance of ECME as a coherent emission mechanism is well established. 

The pulsar radio emission mechanism remains an enigma. There is an enormous body of observational data on pulsars and a plethora of ideas relating to the interpretation of the radio emission, but no consensus has emerged concerning the emission mechanism. {\br There are ongoing arguments in favor of all four mechanisms, for example,  \citet{MGM09} in favor of coherent curvature emission (CE),  \citet{EH16} in favor of relativistic plasma emission (RPE) for nanoshots from the Crab pulsar (and also zebra patterns \citep{ZZZ12} similar to the solar counterpart),  and \citet{L07,LL14} for giant pulses from the Crab pulsar and emission from the double pulsar in terms of anomalous Doppler emission (ADE). However, as discussed in Section~\ref{sect:emission}, there are seemingly compelling arguments against each of these emission mechanisms.  

It may be that some important idea is missing in the discussion about emission mechanisms. For example, the strong argument \citep{EH16} in favor of RPE is negated by the seeming impossibility of generating Langmuir-like waves through a beam instability when the relativistic energy spread in the electrons is taken into account. Maybe, rather than the Langmuir-like waves being attributed to a beam instability, they should be identified with the oscillations \citep{Letal05,BT07,LM08} that develop as the plasma attempts to screen the $E_\parallel$ that arises from the inductive electric field. RPE based on this suggestion would requires a second stage to produce escaping radiation. A further suggestion that needs to be explored is that the conversion mechanism could be interpreted as a maser form of LAE  \citep{M78,MRL09,RK10}.}

{\br A more general aspect of coherent emission mechanisms is} that the instability involved operates only in a large number of localized, transient events. Some of the words used to describe this aspect include: intermittency, fine structures, discrete emissions, short-lived emission centers, microstructure and nanoshots; some of the ideas invoked to model it include: marginal stability, stochastic growth theory and wave trapping. We need both a basic theory to describe the appropriate instability and its local effect on the distribution of particles, and also a statistical model  to describe the radio source formed by the envelope of the localized, transient events.

The concept of coherence in radio astronomy needs further development, particularly through direct measurements of coherence. As noted here there is an observable quantity, $g_2=\langle I^2\rangle/\langle I\rangle^2$, that is measurable in principle, and provides a direct measure of coherence. In an idealized model, the value of $g_2$ should decrease from 2 to 1 as the time and frequency resolutions changes, providing information on the coherence properties of the source.

\begin{acknowledgements}
I thank Mike Wheatland {\br and Alpha Mastrano for helpful comments on the manuscript, Mohammad Rafat for help with figures~\ref{fig:MGKF3} and~\ref{fig:dp20}, and Gennady Chernov, another referee and Maxim Lyutikov for helpful critical comments an earlier version of the manuscript.}

\smallskip
\noindent
{\bf Disclosure of Potential Conflicts of Interest} \quad The author declares that he has no conflicts of interest.
\end{acknowledgements}

%\bibliography{CoherentRefs}   % name your BibTeX data base

\end{document}